\documentstyle[eqsecnum,aps,prb,multicol,epsfig]{revtex}

\begin{document}
\tighten

\title{Elasticity, fluctuations and vortex pinning in ferromagnetic superconductors:  \\
A ``columnar elastic glass''}

\author{A. M. Ettouhami$^1$, Karl Saunders$^2$, L. Radzihovsky$^3$ and  
John Toner$^4$}
\address{$^1$Department of Physics,
University of Florida, P.O. Box 118440, Gainesville, FL 32611-8440}

\address{$^2$Department of Physics, California Polytechnic State 
University, San Luis Obispo, CA93407}

\address{$^3$Department of Physics, University of Colorado, 390UCB, 
Boulder, CO 80309-0390}

\address{$^4$Department of Physics, Materials Science Institute and 
Institute of Theoretical Science, \\ University of Oregon, Eugene, OR 
97403}

\date{\today}
\maketitle

\begin{abstract}

We study the elasticity, fluctuations and pinning of a putative
spontaneous vortex solid in ferromagnetic superconductors.  Using a
rigorous thermodynamic argument, we show that in the idealized case of
vanishing crystalline pinning anisotropy the long-wavelength tilt
modulus of such a vortex solid vanishes identically, as guaranteed by
the underlying rotational invariance.  The vanishing of the tilt
modulus means that, to lowest order, the associated tension elasticity
is replaced by the softer, curvature elasticity. The effect of this is
to make the spontaneous vortex solid qualitatively more susceptible to
the disordering effects of thermal fluctuations and random pinning. We
study these effects, taking into account the nonlinear elasticity,
that, in three dimensions, is important at sufficiently long length
scales, and showing that a ``columnar elastic glass'' phase of
vortices results.  This phase is controlled by a previously unstudied zero-temperature
fixed point and it is characterized by elastic moduli that have
universal strong wave-vector dependence out to arbitrarily long length
scales, leading to non-Hookean elasticity.  We argue that, although
translationally disordered for weak disorder, the columnar elastic
glass is stable against the proliferation of dislocations and is
therefore a topologically ordered {\em elastic} glass. As a result,
the phenomenology of the spontaneous vortex state of isotropic
magnetic superconductors differs qualitatively from a conventional,
external-field-induced mixed state. For example, for weak external
fields $H$, the magnetic induction scales {\em universally} like
$B(H)\sim B(0)+ c H^{\alpha}$, with $\alpha\approx 0.72$.

\end{abstract}

\pacs{74.25.Qt,74.25.Ha,74.40.+k}

\begin{multicols}{2}

\section{Introduction}
\label{introduction}

It has been more than twenty years since superconducting materials
containing a periodic arrangement of magnetic rare earth ions were
discovered.\cite{Early-exp} These included the rare-earth compounds
$R$Rh$_4$B$_4$, R$_x$Mo$_6$S$_8$ and R$_x$Mo$_6$Se$_8$, with $x =$ 1.0
or 1.2 and $R$ a rare earth material (such as Lu, Y, Tm, Er, Ho and
Dy).  As temperature was varied, these materials displayed either a
superconducting or a magnetically ordered phase, but no phase where
both superconductivity and magnetism coexisted simultaneously. More
recent efforts, driven by the need to understand high-$T_c$
superconductivity and other strongly correlated materials, have led to
experimental discoveries of systems exhibiting coexistence of
superconductivity and magnetic order, such as the rare-earth nickel
borocarbides \cite{Stassis-et-al} $R$Ni$_2$B$_2$C with relatively high
superconducting transition temperatures ($T_c\simeq 16.5$ K for the Lu
and 15K for the Y compound). These materials exhibit a rich phase
diagram that includes superconductivity, antiferromagnetism,
ferromagnetism and spiral magnetic
order. \cite{PhysicsToday,experiments} While most of these materials
are antiferromagnets, there is now ample experimental evidence
\cite{Canfield-et-al} that below both the Curie ($\simeq$ 2.3 K) and
the superconducting transition temperatures ($\simeq$ 10.5 K),
superconductivity and ferromagnetism competitively coexist in
ErNi$_2$B$_2$C compounds. Other possible examples of ferromagnetic
superconductors (FS) are the recently discovered high temperature
superconductor Sr$_2$YRu$_{1-x}$Cu$_x$O$_6$ and the putative $p$-wave
triplet strontium ruthenate superconductor, Sr$_2$RuO$_4$, which
spontaneously breaks time reversal symmetry, as well as the recently
discovered compounds RuSr$_2$GdCu$_2$O$_8$,\cite{exp1}
UGe$_2$,\cite{exp2} ZrZn$_2$,\cite{exp3} and URhGe.\cite{exp4}

While there are many interesting unanswered microscopic questions
regarding the nature of such ferromagnetic superconductors, much of
their low-energy phenomenology is dictated by general symmetry
principles. These constrain the form of the Ginzburg-Landau theory
involving the local magnetization ${\bf M}$ and the superconducting
order parameter $\Psi$, that we expect to describe the low-energy
equilibrium thermodynamics. As we discuss in Sec. \ref{Phenom}, the
resulting theory predicts a rich
phenomenology,\cite{Blount,Tachiki1,Kuper,Tachiki2,Greenside,Ng1,Ng2} that
among other phases admits a very interesting spontaneous vortex (SV)
state driven by the spontaneous magnetization, rather than by an
external magnetic field ${\bf H}$.  The argument made in the context
of borocarbides \cite{Ng1,Ng2} (which we believe applies more
generally) implies that ferromagnetic superconductors are expected to
exhibit such a spontaneous ($H=0$) vortex state. Recent small angle
neutron scattering experiments\cite{Kawano-et-al} on
ErNi$_2$$^{11}$B$_2$C provide some evidence for the existence of a SV
state. The combined effects of soft elasticity, random pinning, and 
thermal fluctuations lead to a unique phenomenology for the SV solid;
this unique phenomenology is the subject of this paper.

We will show, in particular, that for $H=0$, and vanishing crystal
anisotropy pinning fields \cite{anisotropy} the elastic properties of
the resulting SV solid differ dramatically and qualitatively from
those of a conventional Abrikosov lattice. The key underlying
difference is the vanishing of the tilt modulus, which is guaranteed
by the underlying rotational invariance.\cite{anisotropy} This
invariance is of course broken by the magnetization ${\bf M}$, but
because this symmetry breaking is spontaneous, the tilt modulus
remains zero.  This is reflected by the invariance of the energy under
a simultaneous global rotation of the magnetization and vortex lattice
induced by it.  Formally, the vanishing of the tilt modulus
corresponds to the vanishing of the ``mass'' of the local field which
is a Goldstone mode associated with the rotational symmetry that is
spontaneously broken by the vortex lattice and magnetization. This
contrasts strongly with a conventional vortex solid, where the
rotational symmetry is explicitly broken by the applied field ${\bf
H}$ and crystalline anisotropy.  Of course, in a real sample there
will always be some degree of crystalline anisotropy.  We will show
that our results for the columnar elastic glass phase of the SV solid
are valid out to a long length scale that depends on the strength of
crystalline anisotropy pinning fields.  This length scale diverges
with vanishing anisotropy so that our results best describe materials
with weak crystalline anisotropy.

All of the conclusions that we draw about other distinctive properties
of the SV solid for ${\bf H}={\bf 0}$ are a direct consequence of this
important observation.  In particular, we find that this ``softness''
({\em i.e.}, vanishing tilt modulus) of the SV lattice dramatically enhances
the effects of quenched disorder and thermal fluctuations. As in
conventional vortex lattices,\cite{Larkin} any amount of disorder
$\Delta$, however weak, is sufficient to destroy translational order
in SV lattices.  Here we find that the Larkin lengths $R_c^\perp$ and
$R_c^z$, beyond which translational order is destroyed by random
forces and torques, are highly anisotropic both in their magnitude and
in the way they scale with pinning strength $\Delta$. We predict
\begin{mathletters}
\begin{eqnarray}
R_c^\perp\propto 1/\Delta^{2/3}\quad, \label{Rcperp}
\\
R_c^z	\propto 1/\Delta^{1/3}\quad, \label{Rcz}
\end{eqnarray}
\end{mathletters}
in $d=3$ dimensions, a prediction that can in principle be checked in
neutron scattering and transport measurements. However, as we will
show, for sufficiently weak pinning (such that dislocation loops
remain bound and the elastic description remains valid) the disorder
also qualitatively alters the SV solid's long-scale elastic
properties, leading to ``anomalous elasticity'': the universal scaling
of elastic moduli with wavevector ${\bf q}$ out to arbitrarily long
length scales (small $q$), with some elastic moduli vanishing and
others diverging as the wavevector ${\bf q}\rightarrow 0$. Other
related long length scale elastic properties include a universal
negative Poisson ratio and a non-Hookean elasticity.  This anomalous
elasticity is completely distinct from the well-known wavevector
dependence of the tilt and bulk moduli at {\em short} length scales
(less than the London penetration length $\lambda$, or, equivalently,
at wavevectors $q$ such that $q\lambda\gg 1$) in conventional vortex
lattices.\cite{Brandt} This {\em long-wavelength} anomalous behsvior
is characteristic of a new kind
of topologically ordered ``columnar elastic glass'' phase of vortices,
which can be shown \cite{SaundersThesis} to be stable, for weak
disorder, against the proliferation of dislocations. This spontaneous
vortex solid is another example of a class of ``soft'' elastic
systems\cite{softsolids}, where long-scale elasticity of the
strongly-interacting Goldstone modes (associated with the
spontaneously-ordered phase) is characterized by a nontrivial
low-temperature fixed point leading to universal anomalous elasticity.

\begin{figure}[ht]
\includegraphics[scale=0.6]{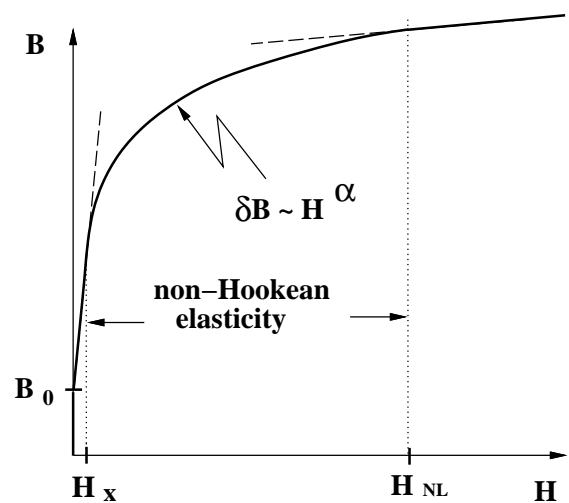}
\caption{
The nonlinear and universal power-law $B(H)$ scaling,
which at weakest fields $H<H_{cr.}$ and strongest fields $H>H_{NL}$
is cutoff by the crystal symmetry breaking anisotropy and $\xi^{NL}$,
respectively.
}\label{BvsH}
\end{figure}

The best way to experimentally probe the consequences of spontaneously
broken symmetry is to break it directly with a weak external
field. Here we predict that, as a consequence of the anomalous
elasticity, the increase in the magnetic induction $\delta B(H)\equiv
B(H)-B(0)$ over the spontaneous induction $B(0)$ as a function of weak
applied field $H$ obeys a {\em universal} ``non-Hookean'' scaling law:
\begin{equation}
\delta B(H) \propto H^{\alpha}\;,
\label{nonhook}
\end{equation}
with the {universal} exponent $\alpha=0.72 \pm 0.04$ (see Fig. \ref{BvsH}), a prediction
that should be experimentally testable. This non-Hookean scaling is in
contrast to conventional vortex lattices in which $\delta B(H)$ scales
approximately linearly with $H-H_{c1}$ for
$H>H_{c1}$.\cite{deGennes}

The rest of this article is organized as follows. In Sec.
\ref{Phenom}, we review the Ginzburg-Landau theory of magnetic

\medskip

\noindent
superconductors, considering the idealized case of vanishing
crystalline pinning anisotropy,\cite{anisotropy} and use it to derive
the effective interaction between vortex lines in such
materials. Using this result, we show rigorously, using a
thermodynamic argument, that as a consequence of the underlying
rotational invariance the tilt modulus at long wavelengths
$c_{44}(q_z=0)$ vanishes in the spontaneous vortex state, thereby
showing that the low-energy, long-wavelength tilt modes of the
spontaneous vortex lattice are controlled by curvature elasticity of
the form $\kappa(\partial_z^2{\bf u})^2$ (rather than by a
conventional tension type $(\partial_z{\bf u})^2$, with ${\bf u}({\bf
r})$ the transverse displacement field of the vortex lattice), and
derive the curvature modulus $\kappa$.  In Sec. \ref{thermal}, we
examine the consequences of the softness of the tilt modes for the
physics of the fluctuations and pinning of the spontaneous vortex
lattice. We show that while both are strongly enhanced quantitatively,
in {\em clean} samples we expect the spontaneous vortex lattice
phenomenology to remain unmodified. This is because the thermal
mean-squared fluctuations in the SV solid, in three dimensions, are
dominated by contributions from short length scales (at which the
spontaneous lattice elasticity should not differ qualitatively from
that of a conventional lattice.)  In particular, we expect only
quantitative changes in the melting curve of a spontaneous vortex
lattice in a clean sample. In contrast, we find that the effects of
pinning by random impurities differ qualitatively from that of a
conventional vortex lattice. Specifically, we show that for
$d<d^p_{uc}=9/2$, (in contrast to $d^p_{uc}=4$ in an ordinary vortex
lattice) random pinning disorders the spontaneous vortex lattice on
length scales larger than the highly anisotropic Larkin domains given
by Eqs. (\ref{Rcperp})-(\ref{Rcz}) (see also
Eqs. (\ref{LarkinR1})-(\ref{LarkinZ1}) below).  In Sec.
\ref{RandomPinning} we begin by ignoring the effects of nonlinear
elasticity and tilt disorder, and show that on length scales longer
than $R_c^{z,\perp}$, the power-law roughness of the vortex lattice
crosses over to the logarithmic growth of mean-squared phonon
distortions, as in conventional vortex lattices. Next, we include tilt
disorder (that is always present) and find that for $d<d_{uc}^t=7/2$
it leads to a far stronger power-law roughness of the vortex lattice,
thereby dominating over the logarithmic roughness caused by the vortex
positional pinning. In Sec. \ref{PT}, we show that as a consequence
of such strong power-law distortions, the nonlinear elasticity becomes
qualitatively important, as indicated by the breakdown of perturbation
theory in the presence of an (in principle) arbitrarily weak pinning.
These findings are consistent with similar results for other soft
solids in the presence of disorder, most notably from studies of
liquid crystals confined in a random matrix
\cite{softsolids,Radzihovsky-Toner}.
In Sec. \ref{RGanalysis} we use the momentum-shell renormalization
group method to determine the consequences of these effects for the
long-scale properties of the randomly-pinned spontaneous vortex
lattice. We show that for scales larger than an anisotropic nonlinear
crossover length scale, $\xi^{NL}_{z, \perp}$, the elasticity becomes
anomalous and scales universally with wavevector. It is controlled by
a nontrivial zero-temperature fixed point that is perturbative in
$\epsilon=7/2-d$.  In Sec. \ref{TopOrderApp} we analyze the topological stability
of the SV solid and argue that the resulting phase in the presence of disorder
is a ``columnar elastic glass"\cite{MSCprl,SaundersThesis} (CEG) 
that is stable against the proliferation of dislocations.
The term ``elastic glass" describes a phase that, while translationally
disordered, retains topological order. Other examples of elastic glasses
have been predicted for smectic
\cite{Radzihovsky-Toner,JSRT_smectic,JohnLeiming} and columnar
(discotic) \cite{SaundersThesis,RT_columnar} liquid crystals.
Associated with the loss of translational order is a broadening of the
Bragg scattering peaks, so that they now exhibit power-law decay
rather than delta-function divergences. 
The absence of even quasi-sharp Bragg peaks in these systems is the reason we
refer to them as ``elastic glasses'' rather than Bragg glasses (of
which the ordinary (field-induced) disordered vortex lattice is an
example\cite{GLD}). An elastic glass is distinguished from the liquid phase 
by being free of unbound dislocation loops, which proliferate in the liquid. 
The glass transition is identified as an ``unbinding" of the dislocation loops.
This transition is then, qualitatively, very similar to the melting of
the flux lattice in the absence of disorder, which can also be thought of as
an unbinding of dislocation loops. The only difference is that in the
glass problem, the flux lattice is translationally disordered {\em
both} above and {\em below} the transition. The absence of defects
below the transition, however, means that the low temperature solid
phase still has a finite shear modulus, leading to glassy behavior. The
experimental consequences of this behavior are discussed in Sec.
\ref{Exp_consq}, and Sec. \ref{Conclusion} contains a discussion of possible 
future theoretical directions.

\section{Ginzburg-Landau phenomenology and harmonic elastic theory of
a spontaneous vortex lattice}
\label{Phenom}

Irrespective of the microscopic mechanism that is ultimately
responsible for the stability of the spontaneous ferromagnetic
superconducting state, general symmetry principles dictate that the
long length scale phenomenology is described by an effective
Landau-Ginzburg free energy functional
\begin{eqnarray}
F &=& \int \!\! d{\bf r} \,\Big\{\frac{1}{2}a(T)|\psi|^2 +
\frac{b}{4}|\psi|^4 +
\frac{\hbar^2}{2m}|({\bf\nabla}-i\frac{2\pi}{\phi_0}{\bf a})\psi|^2
+ \nonumber \\
&+& \frac{{\bf b}^2}{8\pi}
+  \frac{1}{2}\,\alpha(T)\,|{\bf M}|^2 +
\frac{1}{4}\,\beta\,|{\bf M}|^4 +
\frac{1}{2}\,\gamma\,|{\bf \nabla}{\bf M}|^2 +
\nonumber\\
&-& {\bf b}\cdot{\bf M}- \frac{1}{4\pi}{\bf b}\cdot{\bf H}
\Big\} \, ,
\label{FGL}
\end{eqnarray}
where $\psi({\bf r})$ is the local superconducting order parameter,
${\bf M}({\bf r})$ is the magnetization, $\phi_0=hc/2e$ is the quantum
of flux, ${\bf a}$ is the vector potential, ${\bf
b}={\bbox\nabla}\times{\bf a}$ is the total flux density inside the
superconductor, ${\bf H}$ is the external magnetic field, and we have
(for now) ignored crystalline anisotropy.\cite{anisotropy} The
constants $a$, $b$, $\alpha$, $\beta$ and $\gamma$ are experimentally
accessible phenomenological parameters, with $a(T)$ and
$\tilde\alpha=\alpha(T)-4\pi $ changing sign at the superconducting
and ferromagnetic transition temperatures, respectively.\cite{FFLO}
The magnetic field ${\bf h}$ associated with the superconducting
currents satisfies Amp\`ere's law $\nabla\times{\bf
h}={4\pi}{\bf j}/c$ ($c$ is the speed of light in vacuum) and is
related to ${\bf b}$ by:
\begin{equation}
{\bf b}({\bf r})={\bf h}({\bf r}) + 4\pi {\bf M}({\bf r}) + {\bf H}.
\label{relbhm}
\end{equation}
The term $|{\bbox\nabla}{\bf M}|^2$ in Eq. (\ref{FGL}) stands for
$\sum_i({\bbox\nabla} {\bf M}_i)^2$. From Eq. (\ref{relbhm}) we
observe that the flux density ${\bf b}$ is determined by magnetization
of the local moments ${\bf M}$, the external magnetic field ${\bf H}$
and the screening magnetic field ${\bf h}$ generated by the
diamagnetic orbital currents ${\bf j}$.

The minimization of $F$ with respect to $\psi$, ${\bf a}$ and ${\bf
M}$ leads to the following three coupled equations:
\begin{mathletters}
\begin{eqnarray}
&\,&\frac{\hbar^2}{2m}\big({\bbox\nabla}-i\frac{2\pi}{\phi_0}
\,{\bf a}\big)^2\psi  =  - \frac{1}{2}\, a\psi
- \frac{1}{2}\,b|\psi|^2\psi  \,,
\label{glpsi}
\\
&\,&\frac{1}{4\pi}{\bbox\nabla}\times({\bbox\nabla}\times {\bf a})
= \frac{ie\hbar}{2 mc}\Big[\psi{\bbox\nabla}\psi^* -
\psi^*{\bbox\nabla}\psi\Big] +
\nonumber\\
& \,& \quad\quad -\frac{2e^2}{mc^2}|\psi|^2\,{\bf a} +
\nabla\times{\bf M}+\frac{1}{4\pi}\nabla\times{\bf H} \,,
\label{gla}
\\
&\,& {\bf b}({\bf r})  =  \alpha {\bf M} +
\beta |{\bf M}|^2{\bf M} - \gamma\nabla^2{\bf M} \,.
\label{glm}
\end{eqnarray}
\end{mathletters}
It has been shown many years ago by Kuper et al. \cite{Kuper} and by
Greenside et {al.}\cite{Greenside} that, depending on the actual
values of model parameters, the above mean-field equations admit a
variety of equilibrium phases, with transitions between them
controlled by the temperature, which enters model parameters, most
notably $\alpha(T)$ and $a(T)$.  In addition to the paramagnetic
normal phase, $M=\psi=0$ appearing for $\tilde\alpha(T)>0$, $a(T)>0$,
a pure superconducting state, $M=0$, $\psi\neq0$, a ferromagnetic
normal state ($M\neq0$, $\psi=0$) and a stable spiral phase (where
superconductivity coexists with spiraling magnetization) can
appear. Furthermore, the Ginzburg-Landau theory also predicts a
thermodynamically stable {\em spontaneous} vortex state for a range of
physically realistic parameters that give a large Abrikosov ratio
required for a robust mixed state ($\lambda/\xi
\sim {\cal O}(10)$, $\xi$ being the superconducting coherence length 
and $\lambda$ the London penetration depth), and a large exchange
$\gamma$ necessary to suppress the competing spiral phase.  In this
mixed state vortices are induced by the local magnetic moments that
are spontaneously ordered into a uniform ferromagnetic state. This is
in contrast to conventional vortex lattices that are induced by an
external magnetic field.

It is not our intention in this article to study the rich phase
diagram that follows from Eqs. (\ref{glpsi})-(\ref{glm}).  For this,
we refer the interested reader to previous work\cite{Kuper,Greenside},
where this has been done in great detail. Instead our focus here is on
the spontaneous vortex state. We use the above Ginzburg-Landau theory
to derive the vortex line interaction and elasticity of the resulting
spontaneous vortex solid phase and study it in the presence of thermal
fluctuations and random pinning.

\subsection{Interaction potential between vortices in the SV state}
\label{Vint}

The description of the vortex state can, as usual, be derived from the
Ginzburg-Landau theory by reexpressing the fields $\psi$ and ${\bf a}$
in terms of the vortex positions, thereby reexpressing energy in terms
of the vortex conformational degrees of freedom.  To this end, taking
the curl of equation (\ref{relbhm}), using Amp\`ere's law
$\nabla\times{\bf h}=\frac{4\pi}{c}{\bf j}$, and ${\bf
b}=\nabla\times{\bf a}$, we obtain
\begin{eqnarray}
\frac{1}{4\pi}\,{\bbox\nabla}\times({\bbox\nabla}\times{\bf a}) =
\frac{1}{c}\, {\bf j}({\bf r}) + \nabla\times{\bf M}({\bf r}) \
\end{eqnarray}
for zero external magnetic field.

Comparing this expression with Eq. (\ref{gla}) gives the standard
expression for the superconducting current:\cite{deGennes}
\begin{equation}
{\bf j}({\bf r})=\frac{ie\hbar}{2m}\Big[\psi{{\bbox\nabla}}\psi^* -
\psi^*{{\bbox\nabla}}\psi\Big] -\frac{2e^2}{mc}|\psi|^2\,{\bf a},
\end{equation}
which, in the London approximation $\psi=\psi_0\mbox{e}^{i\theta({\bf r})}$, 
(with $\psi_0$ a constant except inside the small vortex core
of size $\xi$, which is valid in the large $\lambda/\xi\gg 1$ limit), reduces
to\cite{deGennes}
\begin{eqnarray}
{\bf j}({\bf r})  =  -\frac{c}{4\pi\lambda^2}({\bf a}({\bf r})
-\frac{\phi_0}{2\pi}{\bbox\nabla}\theta) \,.
\label{eqj}
\end{eqnarray}
This relates superconducting currents to the vector potential and the
phase of the superconducting order parameter, with the London
penetration depth given by $\lambda=\big(mc^2/4\pi
e^2\psi_0^2\big)^{\frac{1}{2}}$.  Taking the curl of this current and
using Amp\`ere's law, together with Eq. (\ref{relbhm}) and
\begin{eqnarray}
{\bbox\nabla}\times{\bbox\nabla}\theta=2\pi\sum_\nu\delta_2({\bf
r}-{\bf r}_\nu(z))\frac{d{\bf R}_\nu}{dz} \,,
\label{rhsLondon}
\end{eqnarray}
where the sum runs over vortices, with
\begin{equation}
{\bf R}_\nu(z)=({\bf r}_\nu(z),z)
\label{R_trajectory}
\end{equation}
parametrizing the trajectory of the $\nu$th vortex line as it
traverses the superconducting sample (we choose the average
magnetization and vortex direction to be along the $\hat{\bf z}$
axis), we find:
\begin{eqnarray}
{\bf b}({\bf r}) - \lambda^2\nabla^2{\bf b}({\bf r}) & = &
\phi_0\sum_\nu\delta_2({\bf r}-{\bf r}_\nu(z))\frac{d{\bf R}_\nu}{dz}
\nonumber\\
&-& 4\pi\lambda^2\nabla^2{\bf M}({\bf r}) \,.
\label{eqb}
\end{eqnarray}

This London equation differs from its counterpart in ordinary
superconductors by the presence of the local magnetization term on its
right hand side. It needs to be solved simultaneously with the
constitutive relation, Eq. (\ref{glm}), relating flux-density ${\bf
b}$ and magnetization ${\bf M}$.

Using Eq. (\ref{glm}) and the ${\bf H}=0$ version of Eq.
(\ref{relbhm}), we can write the following equation for the screening
field ${\bf h}({\bf r})$:
\begin{equation}
{\bf h}({\bf r})  =  \tilde\alpha {\bf M} +
\beta |{\bf M}|^2{\bf M} - \gamma\nabla^2{\bf M}
\label{eq:hvsM}
\end{equation}
with $\tilde\alpha=\alpha-4\pi$.  For small {\bf h({\bf r})} and
$\tilde\alpha<0$ this equation has the spatially uniform solution
${\bf M}({\bf r}) = \hat{\bf h}\sqrt{|{\tilde\alpha}|\over
\beta} \equiv {\bf M}_0$.  Linearizing about this
solution by writing ${\bf M}({\bf r}) = {\bf M}_0 + {\bf \delta
M}({\bf r})$ and expanding to linear order in ${\bf \delta M}({\bf
r})$ gives, for ${\bf q} \neq {\bf 0}$,
\begin{equation} 
{\bf M}({\bf q}) = {\bf
\delta M}({\bf q}) = \chi({\bf q} ) {\bf h}({\bf q}),
\label{def_chi}
\end{equation}
where the second equality follows from the fact that ${\bf M}_0$,
being spatially uniform, contributes nothing to ${\bf M} ({\bf q})$
for ${\bf q} \neq {\bf 0}$ , and
\begin{eqnarray}
\chi(q) \equiv \frac{1}{2|\tilde{\alpha}(T)| + \gamma q^2}\ .
\label{chi(q)}
\end{eqnarray}
Combining this equation with Eq. (\ref{eqb}) gives, in Fourier space:
\begin{eqnarray}
(1 + \lambda^2q^2){\bf b}({\bf q}) & = &  \phi_0\sum_\nu\int dz\,
\mbox{e}^{-i{\bf q}\cdot{\bf R}_\nu(z)}
\,\frac{d{\bf R}_\nu}{dz}
\nonumber\\
& + & 4\pi\lambda^2q^2\chi({\bf q}){\bf h}({\bf q}) \,.
\label{eq-b-h}
\end{eqnarray}
Now, using Eq. (\ref{def_chi}) together with the ${\bf H}=0$ version of
Eq. (\ref{relbhm}) gives:
\begin{mathletters}
\begin{eqnarray}
{\bf b}({\bf q}) & = & \big( 1 + 4\pi\chi({\bf q}) \big)\,{\bf h}({\bf q}) \,,
\label{eq:bhchi}
\end{eqnarray}
\end{mathletters}
which allows us to solve for the magnetic field ${\bf h}({\bf q})$ and
flux density ${\bf b}({\bf q})$:
\begin{mathletters}
\begin{eqnarray}
{\bf h}({\bf q}) &\!=\!& \phi_0\sum_\nu\!\int\!\!dz
\frac{\mbox{e}^{-i{\bf q}\cdot{\bf R}_\nu(z)}}
{1+\lambda^2 q^2 + 4\pi\chi({\bf q})}\,\frac{d{\bf R}_\nu}{dz} ,
\label{result-h}
\\
{\bf b}({\bf q}) &\!=\!& \phi_0\sum_\nu\!\int\!\! dz
\frac{\big(1+4\pi\chi({\bf q})\big)\mbox{e}^{-i{\bf q}\cdot{\bf R}_\nu(z)}}
{1+\lambda^2 q^2 + 4\pi\chi({\bf q})}\,\frac{d{\bf R}_\nu}{dz} .
\label{result-b}
\end{eqnarray}
\end{mathletters}
It is easy to verify that, for one flux line, $\int d{\bf x}\;b_z({\bf x},z)=\phi_0$, 
giving the magnetic flux quantization for a single
vortex. On the other hand, the magnetic field ${\bf h}({\bf r})$ due
to the screening currents around the vortex cores satisfies $\int
d{\bf x}\,h_z({\bf x},z)\simeq\phi_0/(1+4\pi\chi(0))\neq \phi_0$ and
is therefore not quantized.

To proceed further we use the London approximation for the kinetic
energy density
\begin{eqnarray}
\frac{\hbar^2}{2m}|(\nabla-\frac{2i\pi}{\phi_0}\,{\bf a})\psi|^2
& = & \frac{2\pi\lambda^2}{c^2}\,j^2 \,,
\nonumber\\
& = & \frac{\lambda^2}{8\pi}\,(\nabla\times{\bf h})^2 \,,
\label{kineticE}
\end{eqnarray}
to reduce the ${\bf H} = 0$ Ginzburg-Landau free energy functional
(\ref{FGL}) to the London expression, omitting a constant associated
with the condensation energy:
\begin{eqnarray}
F_L &=& \int d{\bf r} \,\Big\{
\frac{\hbar^2}{2m}|({\bbox\nabla}-\frac{2i\pi}{\phi_0}{\bf
a})\psi|^2 + \frac{{\bf b}^2}{8\pi} + \nonumber \\
& + & \frac{1}{2}\,\alpha(T)\,|{\bf M}|^2 +
\frac{1}{4}\,\beta\,|{\bf M}|^4 +
\frac{1}{2}\,\gamma\,|{\bbox\nabla}{\bf M}|^2 - {\bf b}\cdot{\bf M}
\Big\} ,
\nonumber\\
& = & \int d{\bf r} \,\Big\{
\frac{\lambda^2}{2\pi}\,(\nabla\times{\bf h})^2 + \frac{{\bf 
b}^2}{8\pi} + \nonumber \\
& + & \frac{1}{2}\,\alpha(T)\,|{\bf M}|^2 + \frac{1}{4}\,\beta\,|{\bf M}|^4 +
\frac{1}{2}\,\gamma\,|{\bbox\nabla}{\bf M}|^2 - {\bf b}\cdot{\bf M}
\Big\} .
\nonumber\\
\label{FLondon}
\end{eqnarray}
Using the fact that ${\bf b}={\bf h}+4\pi{\bf M}$, the London free
energy reduces to a sum $F_L=F_s+F_m$, where (we henceforth use the
shorthand notation $\int_{\bf q}=\int\frac{d^3\bf q}{(2\pi)^3}$)
\begin{eqnarray}
F_s & = & \frac{1}{8\pi}\int\!d^3{\bf r}\, \Big[h^2({\bf r}) +
\lambda^2(\nabla\times{\bf h})^2
\Big] \,,
\nonumber\\
& = & \int_{\bf q}\frac{1+\lambda^2q^2}{8\pi}\,|{\bf h}({\bf q})|^2
\end{eqnarray}
is the free energy associated with superconducting currents, and
\begin{eqnarray}
F_m & = & \int d^3{\bf r}\,\Big[
\frac{1}{2}\,\tilde{\alpha}|{\bf M}|^2 + \frac{1}{4}\,\beta |{\bf M}|^4
+\frac{1}{2}\gamma|{\bf\nabla M}|^2
\Big] ,
\nonumber\\
& = & \int_{\bf q} \frac{1}{2}
\chi({\bf q})|{\bf h}({\bf q})|^2
\end{eqnarray}
is the ferromagnetic part. We thus obtain:
\begin{eqnarray}
F_L = \frac{1}{2}\int_{\bf q} \big[\frac{1+\lambda^2q^2}{4\pi} +
\chi({\bf q})\big]\,|{\bf h}({\bf q})|^2\,.
\end{eqnarray}
Now, using the expression (\ref{result-h}) for ${\bf h}({\bf q})$ we
finally obtain the London free energy of an arbitrary conformation of
interacting flux lines:
\begin{equation}
F_L  =  \frac{1}{2}\sum_{\mu\nu}\int dz\int dz'\,\,\frac{d{\bf R}_\mu}{dz}\cdot
V\big({\bf R}_\mu(z) - {\bf R}_\nu(z')\big)\,\frac{d{\bf R}_\nu}{dz'} ,
\label{free-energy}
\end{equation}
with the interaction potential:
\begin{eqnarray}
V\big({\bf R}_\mu(z) - {\bf R}_\nu(z')\big) & = &
\frac{\phi_0^2}{4\pi}\int_{\bf q}
\frac{\mbox{e}^{-i{\bf q}\cdot({\bf R}_\mu(z) - {\bf R}_\nu(z'))}}
{1+\lambda^2q^2 + 4\pi\chi({\bf q})}.
\label{result-V}
\end{eqnarray}
In the absence of the ferromagnetic component ($\chi({\bf q})=0$),
$V({\bf q})$ reduces to the usual London interaction between vortices
in a conventional isotropic superconductor.

Using our earlier result (\ref{chi(q)}) for $\chi({\bf q})$ implies
that the interaction potential $V$ is given in Fourier space by:
\begin{equation}
V({\bf q}) = \frac{\phi_0^2}{4\pi}
\frac{2|\tilde{\alpha}|+\gamma 
q^2}{(1+\lambda^2q^2)(2|\tilde{\alpha}|+\gamma q^2) + 4\pi}\,.
\label{V(q)}
\end{equation}
Expressions (\ref{chi(q)}) and (\ref{V(q)}) are the expressions we
shall use below to derive the elastic moduli of the SV lattice.

\subsection{Tilt and curvature moduli, and harmonic elastic
theory of a spontaneous vortex lattice}
\label{Sectionc44}

Having derived the interaction potential between flux lines, we
are now in a position to calculate the tilt modulus of the spontaneous
vortex lattice and to show that in the absence of an external field,
it indeed vanishes identically as dictated by rotational
symmetry.\cite{anisotropy} To this end, we shall write the vortex
trajectories as
\begin{equation}
{\bf r}_\nu(z) = {\bf X}_{\nu} + {\bf u}({\bf X}_{\nu},z),
\label{u_trajectory}
\end{equation}
with ${\bf X}_\nu$ the equilibrium lattice position of the $\nu$th
flux line and ${\bf u}({\bf X}_{\nu},z)$ the two component 
displacement at height
$z$ relative to ${\bf X}_\nu$.
Expanding the London free energy $F_L$ of the
spontaneous vortex system, Eq. (\ref{free-energy}), up to quadratic
order in the displacements ${\bf u}({\bf X}_{\nu},z)$ leads to
$F_L = F_0 + F_{el}$, where $F_0$ is the energy of an undistorted vortex
lattice:

\begin{eqnarray}
F_0 & = & \frac{1}{2}\sum_{\mu\nu}\int dz\int dz'\,\,
V\big({\bf X}_\mu - {\bf X}_\nu,z-z'\big) \,,
\nonumber\\
& = & \frac{1}{2} \Omega n^2\sum_{\bf Q}V({\bf Q}) \,,
\label{F0}
\end{eqnarray}
and $F_{el}$ is the elastic energy of the lattice, which in Fourier
space is given by:
\begin{eqnarray}
F_{el} = \frac{1}{2} \int_{\bf q} u_\alpha({\bf q})
\Phi_{\alpha\beta}({\bf q}) u_\beta(-{\bf q}) \,,
\label{F_el}
\end{eqnarray}
where $\Omega$ is the volume of the system, $n=B/\phi_0$ is the
average density of flux lines ($B$ being the magnetic induction),
the ${\bf Q}$'s are the reciprocal lattice vectors, and we have used Einstein's
implicit summation convention over the repeated indices $\alpha $ and
$\beta $, which run over the two directions ($x$ and $y$)
perpendicular to $z$, the mean direction of the flux lines.  The
elastic matrix $\Phi_{\alpha\beta}({\bf q})$ of the FLL derived from
using the expansion (\ref{u_trajectory}) in the elastic Hamiltonian
(\ref{free-energy}) has the usual form:\cite{Brandt}
\begin{eqnarray}
\Phi_{\alpha\beta}({\bf q}) & = &
n^2\sum_{\bf Q}
\big\{\big[(Q_\alpha  -  q_\alpha)(Q_\beta - q_\beta)
\nonumber\\
& + & q_z^2\delta_{\alpha\beta}\big]
V({\bf Q}-{\bf q})
  -   Q_\alpha Q_\beta V({\bf Q}) \big\}
\label{ElasticMatrix}
\end{eqnarray}
with $V({\bf q})$ the interaction potential of Eq. (\ref{V(q)}).

As discussed in the Introduction, in a SV state with no external field
to explicitly break rotational symmetry,\cite{anisotropy} the underlying
rotational invariance guarantees a vanishing of the vortex-line tilt
modulus $c_{44}$.  We now prove this explicitly for a spontaneous
vortex lattice, via a rigorous thermodynamic argument.

Given that the elastic matrix of a hexagonal vortex lattice (which is
the lattice type we expect for an isotropic interaction potential) is
of the general form\cite{Brandt} (in the following, and throughout the
rest of the paper, ${\bf q}_\perp$ stands for the projection of ${\bf q}$
onto the plane that is orthogonal to $\hat{\bf z}$, and greek indices
in $q_\alpha,\, q_\beta,\ldots$ run over components of ${\bf q}_\perp$ 
only):
\begin{equation}
\Phi_{\alpha\beta}({\bf q})=(c_{11}-c_{66})q_{\alpha}q_{\beta}
+\delta_{\alpha\beta}(c_{44}q_z^2+c_{66}q_\perp^2) \,,
\label{Phigeneral}
\end{equation}
we see that the tilt modulus $c_{44}({\bf q})$ can be extracted from
$\Phi_{\alpha\beta}(q_z\hat{\bf z})$, which, according to
(\ref{Phigeneral}), should be simply
\begin{equation}
\Phi_{\alpha\beta}(q_z\hat{\bf z}) = c_{44}(q_z\hat{\bf z}) q_z^2
\delta_{\alpha\beta} .
\label{c441}
\end{equation}
Setting ${\bf q} = q_z\hat{\bf z}$ in our general expression
(\ref{ElasticMatrix}) for $\Phi_{\alpha\beta}$ gives
\begin{eqnarray}
\Phi_{\alpha\beta}({\bf q}) & = & n^2\sum_{\bf Q} Q_{\alpha} Q_{\beta} 
\Big[ V({\bf Q}-q_z\hat{\bf z}) - V({\bf Q}) \Big]
\nonumber\\
& + & q_z^2\delta_{\alpha\beta}  V({\bf Q} - q_z \hat{\bf z}) \Big].
\label{c442}
\end{eqnarray}

Using the fact that $V({\bf q})$ depends only on the squared magnitude
$q^2$ of ${\bf q}$ , we can rewrite this as:
\begin{eqnarray}
\Phi_{\alpha\beta}(q_z\hat{\bf z}) & = &n^2\sum_{\bf Q}
Q_\alpha Q_\beta\Big[ \tilde{V}(Q^2+q_z^2) \!-\!
\tilde{V}(Q^2)\Big]
\nonumber\\
& + &
  \tilde{V}(Q^2 + q_z^2) q_z^2 \delta_{\alpha\beta}
\Big],
\label{c443}
\end{eqnarray}
where $\tilde{V}$ is the interaction potential such that $V({\bf q}) =
\tilde{V}(q^2)$.
Now, using the fact that for a hexagonal lattice
\begin{equation}
\sum_{\bf Q} Q_\alpha Q_\beta f(Q^2) = \frac{1}{2}\sum_{\bf Q} Q^2 f(Q^2)
\delta_{\alpha\beta}
\,,
\end{equation}
we can rewrite (\ref{c443}) as
\begin{eqnarray}
\Phi_{\alpha\beta}(q_z\hat{\bf z}) & =  \delta_{\alpha\beta} &n^2\sum_{\bf
Q}  \frac{Q^2}{2}\Big[ \tilde{V}(Q^2+q_z^2) \!-\!
\tilde{V}(Q^2)\Big]
\nonumber\\
& + &
  \tilde{V}(Q^2 + q_z^2) q_z^2
\Big],
\label{c444}
\end{eqnarray}
which is exactly of the form required by (\ref{c441}), enabling us to
identify
\begin{equation}
c_{44}(q_z) \!=\! n^2\sum_{\bf Q}
\frac{Q^2}{2}\Big[\frac{\tilde{V}(Q^2+q_z^2) \!-\!
\tilde{V}(Q^2)}{q_z^2}
+ \tilde{V}(Q^2 + q_z^2)
\Big].
\label{tildeVc44}
\end{equation}
This expression of $c_{44}(q_z)$ has the following limit as $q_z\to 0$:
\begin{eqnarray}
c_{44}(0) = n^2\sum_{\bf Q}\Big[\,\frac{Q^2}{2}\tilde{V}'(Q^2)
+ \tilde{V}(Q^2)
\,\Big] \,,
\end{eqnarray}
which can be rewritten as:\cite{Brandt}
\begin{eqnarray}
c_{44}(0) = B\,\frac{\partial}{\partial B}\,\big[
\frac{1}{2}\, n^2\sum_{\bf Q} V({\bf Q})
\big] \,.
\label{c44(0)}
\end{eqnarray}
To see this last step more explicitly we first note that keeping the
lattice structure hexagonal, increasing the magnetic field B must {\it
decrease} the lattice constant $a$ (since the flux carried by each
vortex line is a fixed quantum of flux $\phi_0$). Indeed, since for
hexagonal lattice $B = \phi_0/(a^2\sqrt{3})$, we see that $Q^2
\propto 1/a^2(B) \propto B$, for all reciprocal lattice vectors ${\bf
Q}$. Equivalently,
\begin{equation}
Q^2(B) = Q^2(B_0) B/B_0 \  ,
\label{BQ1}
\end{equation}
for all reciprocal lattice vectors ${\bf Q}$. It then follows that
\begin{eqnarray}
\frac{\partial \tilde V( Q^2)}{\partial B}\Big|_L &=& 
\tilde V'(Q^2)\frac{\partial  Q^2}{\partial B} \Big|_L,\nonumber\\
& = & \tilde V'(Q^2)\frac{Q^2(B_0)}{B_0},\nonumber\\
& = & \tilde V'(Q^2)\frac{Q^2(B)}{B},
\label{BQ2}
\end{eqnarray}
where $\frac{ \partial}{\partial B}\Big|_L$ denotes derivatives with
respect to $B$ keeping the lattice structure fixed, and we have used
(\ref{BQ1}) in the last two equalities.  Now using this result
(\ref{BQ2}) and $n=B/\phi_0$ , we have
\begin{eqnarray}
B\frac{\partial}{\partial B}\,\big[
\frac{1}{2}\, n^2\sum_{\bf Q} V({\bf Q})\big] &=& \sum_{\bf Q} \Big[
\frac{1}{2}\, n^2 \tilde V'(Q^2)Q^2
\nonumber\\
&+&\tilde V(Q^2)  Bn \frac{\partial n}{\partial B}\Big]\,\nonumber\\
&=& c_{44}(0)\,,
\label{BQ3}
\end{eqnarray}
as claimed earlier.

As can be seen from Eq. (\ref{F0}) above, the quantity in square
brackets in this last equation is nothing but the free energy density
$(F_0/\Omega)$ of a lattice of straight vortex lines, which leads to:
\begin{eqnarray}
c_{44}(q_z=0) = B\,\frac{\partial}{\partial B}\,(F_0/\Omega) \,.
\end{eqnarray}
In the presence of an external magnetic field ${\bf H}$,
thermodynamics dictates that
\begin{eqnarray}
\frac{\partial (F_0/\Omega)}{\partial B} = \frac{H}{4\pi} \,,
\end{eqnarray}
and hence we find
\begin{eqnarray}
c_{44}(q_z=0) = \frac{BH}{4\pi} \,,
\end{eqnarray}
which shows, that for $H=0$ the long-wavelength tilt modulus vanishes
identically
\begin{eqnarray}
c_{44}(q_z = 0) = 0 \,,
\end{eqnarray}
as required by rotational invariance.

Having established the vanishing of the tilt modulus at $q_z=0$, let
us now find the limiting form of $c_{44}(q_z)$ for small $q_z$.  Using
a Taylor expansion of the right hand side of Eq. (\ref{tildeVc44}) near
$q_z=0$ leads to the following expression
\begin{eqnarray}
c_{44}(q_z) = n^2 q_z^2\sum_{\bf
Q}\Big[\frac{Q^2}{4}\tilde V''(Q^2) + \tilde V'(Q^2)\Big] \,.
\end{eqnarray}
Neglecting the periodicity of the flux lattice and retaining only the
${\bf Q}=0$ term (which is usually dominant at high density) in the
sum gives:
\begin{mathletters}
\begin{eqnarray}
&c_{44}(q_z) &\simeq \kappa q_z^2 \,,
\label{c44qz0}
\\
\mbox{with}\quad &\kappa &\simeq \frac{B^2}{4\pi}\Big(
\frac{\pi\gamma - {\tilde{\alpha}}^2\lambda^2}{[2\pi + 
|\tilde{\alpha}|]^2}\Big)\,.
\label{kappa}
\end{eqnarray}
\end{mathletters}
Hence, the long-wavelength bending energy of the spontaneous vortex
lattice is characterized by a curvature (as opposed to tension)
elastic energy of the form $\kappa(\partial_z^2 {\bf u})^2$.  Note,
however, that expression (\ref{c44qz0}) and expression (\ref{kappa})
for the curvature modulus $\kappa$ are valid only at long wavelengths,
{\em i.e.} for $q_z\to 0$. At short wavelengths, we can restrict the
reciprocal lattice sum in Eq. (\ref{tildeVc44}) to the ${\bf Q}=0$
term (the so-called continuum approximation), upon which the tilt
modulus $c_{44}(q)$ takes the following expression:
\begin{eqnarray}
c_{44}({\bf q}) \simeq \frac{B^2}{4\pi}\,\frac{1}{1+\lambda^2q^2 +
4\pi\chi({\bf q})} \,,
\label{usualc44}
\end{eqnarray}
where we used the fact that $n=B/\phi_0$. In the absence of a magnetic
component, $\chi({\bf q})=0$ and the above expression reduces to the
usual result for the tilt modulus in conventional type-II
superconductors.

We can now combine this result for the tilt modulus with the standard
analysis\cite{Brandt} usually applied to obtain the compression and shear moduli and
thus write down the harmonic elastic energy for the spontaneous vortex
lattice. From the general expression of the elastic matrix, Eq.
(\ref{Phigeneral}), the compression and shear moduli are given by
$c_{11}(q)=\Phi_{xx}(q\hat{\bf x})/q^2$ and $c_{66}(q)=\Phi_{xx}(q\hat{\bf
y})/q^2$ respectively.  Hence, using Eq. (\ref{ElasticMatrix}) for
$\Phi({\bf q})$, we obtain:
\begin{mathletters}
\begin{eqnarray}
c_{11}(q) & = & \frac{n^2}{q^2}\sum_{\bf Q}\big[
(Q_x-q)^2V({\bf Q}-q\hat{\bf x}) \!-\! Q_x^2V({\bf Q})
\big] ,
\nonumber\\
\label{newc11}
\\
c_{66}(q) & = & \frac{n^2}{q^2}\sum_{{\bf Q}\neq 0}
Q_x^2\big[V({\bf Q}-q\hat{\bf y}) - V({\bf Q})
\big]\,.
\nonumber\\
\label{newc66}
\end{eqnarray}
\end{mathletters}
The compression modulus is readily evaluated in the continuum (${\bf 
Q}=0$) limit, with the
result:
\begin{eqnarray}
c_{11}(q) = \frac{B^2}{4\pi}\frac{1}{1+\lambda^2q^2+4\pi\chi(q)} \,,
\end{eqnarray}
(where we have again used the fact that $n=B/\phi_0$).  On the other hand,
in expression (\ref{newc66}) for the shear modulus, expanding $V({\bf
Q}-q\hat{\bf y})$ to second order in $q$ we obtain the following
expression:
\begin{eqnarray}
c_{66} = \frac{n^2}{2}\sum_{{\bf Q}\neq 0}
Q_x^2\frac{\partial^2V({\bf Q})}{\partial Q_y^2} \,,
\end{eqnarray}
a result which should be valid ($c_{66}$ being in general weakly
dispersive) over the entire Brillouin zone of the SV lattice. These
results are identical to the standard expressions for the ordinary
(field-induced) vortex lattices in conventional superconductors, apart
from the presence of the susceptibility $\chi(q)$ which reduces to a
constant at small wavevector and therefore should not qualitatively
alter the long-wavelength behavior of the vortex lattice.

Before closing this section, we note that in the expression of the
elastic Hamiltonian of the SV lattice, (Eq. (\ref{F_el})),
$\Phi_{\alpha\beta}({\bf q})$ for a triangular lattice can be written
in the form:
\begin{eqnarray}
\Phi_{\alpha\beta}({\bf q}) = \Phi_L({\bf q})P^L_{\alpha\beta}({\bf q})
+ \Phi_T({\bf q})P^T_{\alpha\beta}({\bf q}),
\end{eqnarray}
where $P^L_{\alpha\beta}({\bf q})=q_\alpha q_\beta/q_\perp^2$ and
$P^T_{\alpha\beta}({\bf q})=\delta_{\alpha\beta}-P^L_{\alpha\beta}({\bf q})$ are the
longitudinal and transverse projection operators onto the direction of
$\hat{\bf q}_\perp={\bf q}_\perp/q_\perp$ and the perpendicular
direction in the $(q_x,q_y)$ plane respectively, and where the
longitudinal and transverse parts of the elastic matrix are given by
\begin{mathletters}
\begin{eqnarray}
\Phi_L({\bf q}) & = & c_{11}({\bf q})\,q_\perp^2 + c_{44}({\bf q})\,q_z^2\,,
\\
\Phi_T({\bf q}) & = & c_{66}q_\perp^2 + c_{44}({\bf q})\,q_z^2 \,.
\end{eqnarray}
\end{mathletters}

Having established the harmonic elastic theory for the spontaneous
vortex lattice, we next use it to study the effects of thermal
fluctuations and quenched disorder. As discussed in the Introduction,
we anticipate these to be significantly enhanced relative to that of a
conventional vortex lattice, due to the softness (vanishing of
$c_{44}$) of the spontaneous vortex solid.

\section{Thermal fluctuations and melting of a ``soft'' spontaneous
vortex lattice}
\label{thermal}

As usual the effect of thermal fluctuations on vortex lattice
translational order is encoded in the vortex line mean-squared
fluctuation $\langle u^2\rangle$ relative to the ideal lattice
positions. This quantity determines the level of translational order,
namely the intensity of the Bragg peaks (in {\em e.g.}, neutron scattering)
through the Debye-Waller factor:
\begin{equation}
W=e^{-Q^2\langle u^2\rangle}\,,
\end{equation}
with $Q$ the length of the shortest reciprocal lattice vector.

Similarly the location of the melting transition $B_m(T)$ in the B-T
phase diagram can be roughly located by using $\langle u^2\rangle$
through the Lindemann criterion
\begin{eqnarray}
\langle u^2\rangle = c_L^2 a^2 \,,
\end{eqnarray}
corresponding to thermal root-mean squared fluctuation of vortex lines
becoming comparable to the average vortex spacing $a=\sqrt{\phi_0/B}$,
with the Lindemann constant $c_L$ conventionally taken to be on the
order of $0.1$.

Within the harmonic approximation (which, as we will find {\em a
posteriori}, remains valid at low T) $\langle u^2\rangle$ can be
easily computed using Eq. (\ref{F_el}) above. Equipartition ({\em i.e.},
simple Gaussian integration) gives the standard expression
\cite{Blatter-et-al} (we use
units such that Boltzmann's constant $k_B=1$):
\begin{eqnarray}
\langle u^2\rangle \simeq  \int_{\bf q}
\frac{T}{c_{66}q_\perp^2 + c_{44}({\bf q})q_z^2} \,,
\label{rms-thermal}
\end{eqnarray}
where we have assumed that $c_{11}\gg c_{66}$ (as is usually the case
for materials with a high Ginzburg ratio $\lambda/\xi$). It is clear
from Eq. (\ref{rms-thermal}) that (as in conventional vortex
lattices\cite{Blatter-et-al}) thermal {\em rms} vortex fluctuations
are dominated by the largest wavevectors near the Brillouin zone
boundary, {\em i.e.} near $q_\perp\simeq \Lambda=\pi/a$, $q_z\simeq 1/\xi$.
At these short length scales we expect the spontaneous vortex lattice
to be characterized by a finite tilt modulus, {\em i.e.}, for $c_{44}(q_z)$
to approach the expression given in Eq. (\ref{usualc44}), which should
not be too different from that of a conventional (external magnetic
field-induced) vortex lattice. We therefore conclude that thermal
vortex fluctuations $\langle u^2\rangle$ and all other physical
observables, such as the melting curve and the strength of Bragg
peaks, to be qualitatively (and roughly even quantitatively) unchanged
by the long-wavelength ``softness'' ($c_{44}(q_z\to 0)\approx \kappa
q_z^2$) of the spontaneous vortex lattice.  We can compare the SV
lattice fluctuations with the conventional vortex (CV) lattice
fluctuations ($\langle u^2\rangle_{CV}$ ) by treating $\chi(q)$
perturbatively (valid deep in the ferromagnetic phase where
$|\tilde{\alpha}|\gg 1$)
\begin{equation}
\langle u^2\rangle \simeq  \langle u^2\rangle_{CV}
+ \int_{\bf q}
\frac{Tc_{44}^{CV}({\bf q})q_z^2}{(c_{66}q_\perp^2 + c_{44}^{CV}({\bf 
q})q_z^2)^2}\frac{4\pi\chi({\bf q})}{1+\lambda^2q^2} \,,
\label{SVCVrms-thermal}
\end{equation}
where $c_{44}^{CV}$ is the tilt modulus for a convention vortex
lattice. This shows that to lowest order in $\chi({\bf q})$ the SV
lattice fluctuations are larger, as we would expect due to the
additional softness.

\section{Random pinning within the harmonic elastic theory}
\label{RandomPinning}

As in a conventional vortex lattice, random pinning couples to the
vortex density $\rho({\bf r})$ (positional disorder) and to the local
vortex tilt-field $\partial_z{\bf u}$ (tilt-disorder, which we discuss
in more detail in Sec.
\ref{RanTilt}), that are proportional to the $z$- and
$\perp$-components, respectively, of the local magnetic induction
vector field ${\bf B}$.  The effects of disorder on the spontaneous
vortex lattice can therefore be incorporated through the Hamiltonian
$H=H_{el}+H_{d}$, where the elastic energy is given by
Eq. (\ref{F_el}) and the pinning energy is given
by\cite{NattermanReview,MSCprl}
\begin{eqnarray}
H_{d} = \int d{\bf r}\;\Big[ V_d({\bf r}) \rho ({\bf r}) + {\bf h}({\bf
r})\cdot
\partial_z{\bf u}({\bf r})
\Big] \,,
\end{eqnarray}
where $V_d({\bf r})$ and ${\bf h}({\bf r})$ are the random pinning 
potentials for positional and tilt disorders respectively. 

As we will see in a moment, unlike the effects of thermal fluctuations
studied above, quenched disorder leads to vortex lattice distortions
that are dominated by {\em long} length scale fluctuations.  As a
result, we anticipate that the softness associated with the vanishing
of the long wavelength tilt modulus in a SV lattice will have
qualitatively important effects.  Hence, to study pinning effects,
that will be the focus of the rest of the manuscript, we specialize to
the long wavelength form of the elastic Hamiltonian. This is
characterized by the elastic matrix $\Phi_{\alpha\beta}({\bf q}) =
\Phi_L({\bf q})P^L_{\alpha\beta}({\bf q}) + \Phi_T({\bf
q})P^T_{\alpha\beta}({\bf q})$ with the following long wavelength
behavior of the longitudinal and transverse components:
\begin{mathletters}
\begin{eqnarray}
\Phi_L({\bf q}) & = & c_{11}q_\perp^2 + \kappa q_z^4 \,,
\label{newdef-PhiL}
\\
\Phi_T({\bf q}) & = & c_{66}q_\perp^2 + \kappa q_z^4 \,.
\label{newdef-PhiT}
\end{eqnarray}
\end{mathletters}
This elastic matrix is unique to the spontaneous vortex lattice, with
the curvature modulus $\kappa$ given by Eq. (\ref{kappa}).

\subsection{Perturbative treatment of random pinning: Larkin approximation}
\label{vmag_PT_Larkin}

The simplest (yet quite revealing) approximation in a study of random
pinning of a periodic medium ({\em e.g.}, a vortex lattice) is the Larkin
approximation. This approximation amounts to Taylor-expanding the
positional pinning energy (which is a nonlinear function of ${\bf u}$)
to linear order in the phonon displacement ${\bf }u$.  This can be
done by expressing the vortex density $\rho({\bf r})$ as an explicit
function of ${\bf u}$
\begin{equation}
\rho({\bf r})={\rm Re} \sum_{\bf Q} \tilde{\rho}_{\bf Q}e^{i {\bf Q} \cdot
({\bf r}+{\bf u}({\bf r}))} \,,
\label{rho}
\end{equation} 
which leads to a positional pinning energy:
\begin{eqnarray}
H_{d \rho}=\int d{\bf r} {\rm Re} 
\sum_{\bf Q} U_{\bf Q}({\bf r})e^{i {\bf Q} \cdot {\bf u}({\bf r})}\;,
\label{Hrho}
\end{eqnarray}
where the ${\bf Q}$'s are the reciprocal lattice vectors of the vortex
lattice and
\begin{eqnarray}
U_{\bf Q}({\bf r})=
\tilde{\rho}_{\bf Q} V({\bf r})e^{i{\bf Q}\cdot{\bf r}}\;.
\label{UG}
\end{eqnarray}
We expect correlations of $V_d({\bf r})$ to be short-ranged. 
Even if they are not, those of $U_{\bf Q}({\bf r})$ {\em are},
due to the $e^{i{\bf Q}\cdot{\bf r}}$ factor, as discussed in Ref. \cite{Radzihovsky-Toner}.
This means that one can accurately capture the
long-length scale physics of the problem by taking them to be
zero-ranged and Gaussian, characterized by a second cumulant:
\begin{eqnarray}
\overline{U_{\bf Q}({\bf r})U_{\bf Q^{\prime}}^*({\bf r'})} =
\Delta_{{\bf Q}} \delta^d({\bf r}-{\bf r'})\delta_{\bf Q, Q^{\prime}}\,,
\label{CU2}
\end{eqnarray}
where the set of phenomenological parameters $\Delta_{{\bf Q}}$'s
depend upon the microscopic natures of the impurities, the vortex
lattice and their interactions.  

In using the Larkin approximation we expand the periodic nonlinear
piece for small ${\bf }u$, valid only at short length scales such that
the typical induced vortex displacement ${\bf u}_{rms}=\langle {\bf u}^2\rangle$ is small
compared to the correlation length of the random potential.  With this
approximation we can write the pinning energy as:
\begin{eqnarray}
H_{d}\approx\int d{\bf r}\;\Big[ {\bf f}({\bf r})\cdot{\bf u}({\bf r}) +
{\bf h}({\bf r})\cdot\partial_z{\bf u}({\bf r})\Big] \,,
\label{def-Hd}
\end{eqnarray}
with ${\bf f}({\bf r})$ an effective random force with two-point
correlations:
\begin{eqnarray}
\overline{f_\alpha({\bf r})f_\beta({\bf r}') } =
\Delta_f\delta_{\alpha\beta}\delta({\bf r}-{\bf r}') \,,
\end{eqnarray}
where $\Delta_f=\sum_{\bf Q}Q^2\Delta_{{\bf Q}}$.  This leads to a total
Hamiltonian in this short-scale Larkin regime that is given by:
\begin{eqnarray}
H = H_{el} + \int d{\bf r}\; {\bf f}({\bf r})
\cdot{\bf u}({\bf r}) \,.
\label{HLarkin}
\end{eqnarray}
Above we have dropped the tilt disorder ${\bf h}({\bf r})$, since it
is clearly subdominant to the uncorrelated random force ${\bf f}({\bf
r})$ arising from the Larkin approximation to the positional pinning
disorder. Standard Gaussian integration then leads to the
disorder-averaged phonon correlation function
\begin{equation}
C^{f}({\bf r}) = C^{f}_L({\bf r}) + C^{f}_T({\bf r}) = 
\overline{\langle[{\bf u}({\bf r})- {\bf u}({\bf 0})]^2\rangle},
\end{equation}
where the longitudinal and transverse parts $C^{\Delta}_s({\bf r})$
(with $s=L,\;T$ labeling the longitudinal and transverse
polarizations, respectively) of $C^{f}({\bf r})$ are given by:
\begin{eqnarray}
C^{f}_s({\bf r}) & = &
\int_{\bf q} \frac{2T[1-\cos({\bf q}\cdot{\bf 
r})]}{B_{s}q_\perp^2+\kappa q_z^4}
+ \int_{\bf q}
\frac{2\Delta_f[1-\cos({\bf q}\cdot{\bf r})]}{\big[ B_{s}q_\perp^2+
\kappa q_z^4\big]^2} \,,
\nonumber\\
& \simeq & 2\Delta_f\int_{\bf q}
\frac{1-\cos({\bf q}\cdot{\bf r})}{\big[ B_{s}q_\perp^2+
\kappa q_z^4\big]^2} \,,
\label{defCsLarkin}
\end{eqnarray}
with $B_L=c_{11}$ and $B_T=c_{66}$ the bulk and shear moduli of the
vortex lattice, respectively, and where, in going from the first to
the second equation, we have neglected the thermal contribution to
$C^{f}_s({\bf r})$, that is subdominant to the disorder part at
small wavevectors.

Simple dimensional analysis shows that, for $d\leq 9/2$, the above
integrals are dominated by long length scales, justifying our use of
long-wavelength elastic theory derived in the previous section.  In
$d$ dimensions we find:
\begin{mathletters}
\begin{eqnarray}
C_s^{f}({\bf r} = {\bf R}_\perp) &\approx&
\frac{\Delta_f}{\kappa^{1/4}B_s^{7/4}}\,R_\perp^{\frac{9}{2}-d} \,,
\\
C_s^{f}({\bf r} =R_z\hat{\bf z}) & \approx &
\frac{\Delta_f}{\kappa^{(5-d)/2}B_s^{(d-1)/2}}\,R_z^{9-2d} \,,
\end{eqnarray}
\end{mathletters}
where we have dropped overall numerical factors of order unity, and
$z$ and $\perp$ denote directions along and perpendicular to the flux
lines, respectively. In the physically relevant case of $d=3$, the
above two expressions become:
\begin{mathletters}
\begin{eqnarray}
C_s^{f}(R_\perp) &\approx& \frac{\Delta_f R_\perp^{3/2}}{\kappa^{1/4}B_s^{7/4}} \,,
\\
C_s^{f}(R_z) &\approx& \frac{\Delta_f R_z^3}{\kappa B_s} \,.
\end{eqnarray}
\end{mathletters}
Equating these correlation functions to the square of the lattice
spacing, $a^2$, leads to highly anisotropic Larkin lengths in the
$(xy)$ and $z$ directions, given respectively by:
\begin{mathletters}
\begin{eqnarray}
R_c^\perp &\approx &
\Big(\frac{\kappa^{1/4}B_s^{7/4}a^2}{\Delta_f}\Big)^{2/3} \,,
\label{LarkinR1}
\\
R_c^z &\approx & \Big(\frac{\kappa B_s a^2}{\Delta_f}\Big)^{1/3} \,.
\label{LarkinZ1}
\end{eqnarray}
\end{mathletters}

These lengths characterize the dimensions of ordered Larkin domains
beyond which pinning dominates over elastic energy, disrupting the
translational order of the spontaneous vortex lattice. The finiteness
of these $3d$ Larkin domains demonstrates the absence of long-range
translational order even for arbitrarily weak disorder. By this last
criterion, the Larkin lengths also define the range of length scales
over which the above random-force perturbation theory, as defined by
the Hamiltonian (\ref{HLarkin}), is valid.  At longer length scales
the rms value of $u$ exceeds the lattice spacing, which invalidates
the approximation we used in obtaining Eq. (\ref{def-Hd}).  Also,
because the random force dominates over the random tilt disorder, it
is the positional pinning (rather than tilt disorder) that determines
the size of Larkin domains.

However, as is well-known for conventional vortex lattices and other
periodic media\cite{Natterman}, on longer length scales the Larkin
approximation highly overestimates the effect of disorder. On scales
beyond $R_c^\perp$ and $R_c^z$, a more sophisticated approach that
takes into account the nonlinearity of the positional disorder is
necessary.  Next we shall use the replica Gaussian variational
method\cite{Mezard,GLD} to treat the long-scale effects of the
(seemingly dominant) positional disorder alone. For now, ignoring tilt
disorder and nonlinear elasticity, we shall find, that, in spite of
the soft elasticity of the tilt modes in a spontaneous vortex lattice,
positional disorder leads to a logarithmic growth of the displacement
correlation function $C^{f}({\bf r})$ in the physical case of
$d=3$ dimensions, as it does in conventional vortex lattices in
ordinary superconductors.\cite{Natterman,GLD}

\subsection{Positional pinning on long length scales}
\label{vmag_replicas}

As discussed above, the perturbative (random-force) treatment of
positional disorder breaks down on length scales longer than the
Larkin lengths $R_c^\perp$ and $R_c^z$ of Eqs.
(\ref{LarkinR1})-(\ref{LarkinZ1}). In this Subsection we shall use a
replica variational analysis\cite{Remark1,commentVariational} similar
to the one carried out in Refs.\onlinecite{Mezard,GLD} to find the
contribution to the long distance behavior of the two-point correlation function for
relative vortex displacement from the seemingly dominant positional
disorder alone. Our starting point here is the Hamiltonian:
\begin{eqnarray}
H & = & \int_{\bf q}\frac{1}{2} \big[K q_\perp^2
+ \kappa q_z^4 \big]|{\bf u}({\bf q})|^2 
\nonumber\\
& + &\int d{\bf r} Re \sum_{\bf Q} U_{\bf Q}({\bf r})e^{i {\bf Q} \cdot
{\bf u}({\bf r})}\,,
\label{H-ptdis}
\end{eqnarray}
where for simplicity we have used isotropic elasticity with a single
in-plane elastic constant $K$ and the positional disorder correlations
are given by Eq. (\ref{CU2}).
Employing the standard replica ``trick''\cite{Anderson} to average
over the quenched disorder\cite{Natterman},
we obtain the following replicated effective Hamiltonian
\begin{eqnarray}
H_{eff} &\simeq&  \sum_{a=1}^m \int_{\bf q} \frac{1}{2}
\big[\, K q_\perp^2  + \kappa q_z^4 \,\big]
\,{\bf u}_a({\bf q})\cdot{\bf u}_a(-{\bf q})
\nonumber\\
& - & \sum_{a,b=1}^m\sum_{{\bf Q}\neq 0}\int d{\bf r}\;
\frac{\Delta_{\bf Q}}{2T}\;\cos\big[{\bf Q}\cdot\big(
{\bf u}_a({\bf r}) - {\bf u}_b({\bf r})\big)\big] ,
\nonumber
\\
\label{Heff-replicas}
\end{eqnarray}
with $a,b$ labeling the $m\rightarrow 0$ replicas, ${\bf Q}$ denoting
reciprocal lattice vectors, and $\Delta_{\bf Q}$ defined by Eq. (\ref{CU2}).
In the above we have also dropped the coupling of disorder to the
long-scale fluctuations in the vortex density
$-\rho_0{\bbox\nabla}\cdot{\bf u}$, that in $3d$ is subdominant to the
short-scale pinning that we have kept.

We now want to study the above nonlinear Hamiltonian via a Gaussian
variational approximation\cite{Mezard,GLD} with the following harmonic
trial Hamiltonian:
\begin{eqnarray}
H_v = \frac{1}{2}\sum_{ab}\int_{\bf q}
(G^{-1}({\bf q}))_{ab}{\bf u}_a({\bf q})\cdot{\bf u}_b(-{\bf q}) \,,
\end{eqnarray}
which is parameterized by a variationally-determined inverse
``propagator''
\begin{eqnarray}
\big[ G^{-1}({\bf q})\big]_{ab} = (Kq_\perp^2 + \kappa q_z^4)\, \delta_{ab}
- \sigma_{ab} \,,
\end{eqnarray}
characterized by an $m\times m$ self-energy matrix $\sigma_{ab}$ of
variational parameters, that encodes the average vortex positional
correlations.

Minimization of the variational free energy
\begin{eqnarray}
F_v = \langle H_{eff} - H_v \rangle_v - T\ln Z_v \,,
\end{eqnarray}
where $Z_v=\mbox{Tr}\big(\exp(-H_v/T)\big)$ and
$\langle\cdots\rangle_v$ denotes averaging with statistical weight
$\exp(-H_v/T)/Z_v$, leads to saddle-point equations that are very
similar to those derived in Ref.\onlinecite{GLD}.  As discussed in
more detail there, the long-scale properties of the pinned state are
characterized by a replica-symmetry broken matrix $A_{ab}$ with
``hierarchical'' structure, that in the $m\to 0$ limit can be viewed
as a single function $\sigma(v)$ of a real variable $v$ in the
interval $[0,1]$, that is the key quantity of the replica method. The
hierarchical matrix $\sigma(v)$ is a solution of the saddle-point
equation (here $Q_0$ is the magnitude of the smallest reciprocal
lattice vectors):
\begin{eqnarray}
\sigma(v)\int_{\bf q}
\frac{TQ_0^2}{\big[\kappa q_z^4 + Kq_\perp^2 +[\sigma](v)\big]^2} = 1 \,,
\label{eq-sigma-1}
\end{eqnarray}
where we denote by $[\sigma](v)$ the quantity $[\sigma](v) =
v\sigma(v) -\int_0^v du\;\sigma(u)$.  The solution of the saddle-point
Eq. (\ref{eq-sigma-1}) proceeds in much the same way as for a
conventional, field-induced vortex lattice.\cite{GLD} We refer the
reader to Appendix
\ref{App_vmag_replicas} where we give more details
on this solution, and present here only the final results for the
disorder-averaged correlation function $C^{p}({\bf r}) =
\overline{\langle[{\bf u}({\bf r}) - {\bf u}({\bf 0})]^2\rangle}$
(we use the superscript $p$ in $C^{p}({\bf r})$ to distinguish the
long-scale translational correlator in presence of the periodic density pinning
alone -- {\em i.e.}, in the absence of tilt disorder and elastic
nonlinearities -- from the quantity $C^{f}({\bf r})$ of the previous
paragraph which corresponds to the Larkin approximation of random
pinning forces and which is only valid inside the Larkin domains
defined by the length scales $R_c^\perp$ and $R_c^z$ of Eqs.
(\ref{LarkinR1})-(\ref{LarkinZ1})).

For a spontaneous vortex lattice in three spatial dimensions, we find
(see Appendix \ref{App_vmag_replicas}):
\begin{mathletters}
\begin{eqnarray}
C^{p}(r_\perp) & = &
\frac{T}{\sqrt{2}u_0K\kappa^{1/4}}\,\ln(\Lambda r_\perp) \,,
\label{CW(rperp)}
\\
C^{p}(z) & = & \frac{T}{\pi\sqrt{2}u_0K\kappa^{1/4}}
\,\ln(\Lambda z) \,,
\label{CW(z)}
\end{eqnarray}
\end{mathletters}
where (here $d_\perp=d-1$ and $c_{d_\perp}$ is a numerical constant
defined in Appendix \ref{App_vmag_replicas})
\begin{equation}
u_0 = \frac{3\sqrt{2}c_{d_\perp}TQ_0^2}{4(7-2d_\perp)\kappa^{1/4}K^{3/2}} \,.
\end{equation}

We thus see that the softness of the tilt modes of the spontaneous
vortex lattice does not affect the large distance behavior of the
correlation function $C^{p}({\bf r})$, which, like in ordinary
(field-induced) disordered vortex lattices, grows only {\em
logarithmically} at long length scales. This is expected on general
grounds, since (weak) disorder-induced logarithmic growth is a
consequence of the periodic form of positional pinning in Eq.
(\ref{Heff-replicas}), which is guaranteed by the identity-symmetry of vortex
lines. However, this result is valid {\em only} if we ignore the tilt
disorder and nonlinear elasticity, which, as we will show below become
important in $d=3<7/2$ dimensions. In what follows, we shall consider
the effect of random tilt disorder on the spontaneous vortex lattice,
and show that it leads to power-law growth of the vortex line
distortions in $d=3$ dimensions, and therefore, at long scales
strongly dominates over the logarithmic growth induced by positional
disorder found above.

\subsection{Random tilt disorder}
\label{RanTilt}

\begin{figure}[hb]
\includegraphics[scale=0.7]{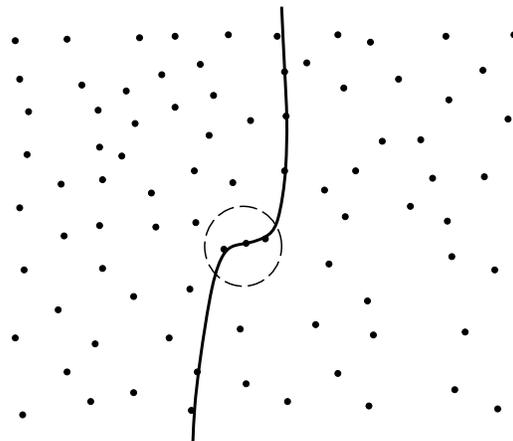}
\caption{
Schematic representation of the mechanism of random tilt
pinning. Randomly oriented anisotropic clusters of point-pins exert
random torques on vortex lines.  }
\label{Fig_random_tilt}
\end{figure}

In addition to the usual effects of pinning leading to a random
coupling to the vortex density, Eq. (\ref{H-ptdis}) (corresponding to
the pinning of the $z$-component of the magnetic field), it is
essential, in studying a spontaneous vortex solid, to include a random
tilt disorder, as appearing in $H_d$, Eq. (\ref{def-Hd}). As
schematically illustrated in Fig. \ref{Fig_random_tilt}, physically,
such a random term corresponds to local random torques being exerted
on the vortex lattice by random, short-ranged correlated clusters of
pinning sites.  The interaction with these randomly oriented clusters
of point pinners is similar to the interaction of vortex lines with a
randomly oriented local transverse magnetic field ${\bf h}({\bf r})$
(not to be confused with the screening field around vortices of
Sec.\ref{Phenom}), that couples to the transverse component of the
magnetic induction vector field ${\bf B}_\perp$, and therefore leads
to random coupling:
\begin{eqnarray}
H_t = \int d{\bf r}\; {\bf h}({\bf r})\cdot
\partial_z {\bf u}({\bf r}) \,.
\label{randomtilt}
\end{eqnarray}
For simplicity we take ${\bf h}({\bf r})$ to be a Gaussian random
field with zero mean and correlations:
\begin{eqnarray}
\overline{ h_\alpha({\bf r})h_\beta({\bf r}') } =
\Delta_t\,\delta_{\alpha\beta}\delta({\bf r}-{\bf r}') \,.
\end{eqnarray}

\medskip

Independent of the microscopic mechanism, it is simple to see, that
tilt disorder is always generated from the positional disorder upon
coarse-graining even if it is left out in the original model.  Such
disorder is generally left out in a treatment of conventional vortex
lattice, as in three dimensions it (as well as the aforementioned
coupling of disorder to the {\em long-scale} fluctuations in the
vortex density $-V({\bf r})\rho_0{\bbox\nabla}\cdot{\bf u}$) is
subdominant to the positional pinning that we have studied in the
previous Subsection.

\noindent
However, as we will show below, because of the
softness of the spontaneous vortex lattice
(more specifically, the
vanishing of its long wavelength tilt modulus $c_{44}$),
this tilt disorder
leads to power-law distortions of the spontaneous vortex lattice in three
dimensions. It is therefore the dominant form of disorder, that must be
taken into account on long scales.

To see this we treat tilt disorder alone. Because tilt disorder
couples only linearly to the phonon ${\bf u}$ field (in the absence of
positional disorder and nonlinear elasticity) it is straightforward to
study its effects on the vortex positional correlations. Using the
Hamiltonian $H = H_{el} + H_t$, a standard calculation gives for the
longitudinal and transverse contributions of the positional
correlation function $C^t({\bf r})=\overline{\langle[{\bf u}({\bf
r})-{\bf u}({\bf 0})]^2\rangle} = C^t_L({\bf r}) + C^t_T({\bf
r})$. Here we use the superscript $t$ to distinguish this correlator
associated with tilt disorder from the correlation functions
$C^{f}$ and $C^{p}$ of the previous two Subsections that were
associated with positional disorder. Keeping in mind that the
subscript $s$ denotes longitudinal ($L$) and transverse ($T$)
contributions we find that:
\begin{eqnarray}
C^t_s({\bf r}) & = & 2T\int_{\bf q}\frac{q_z^2\big[1 - \cos({\bf 
q}\cdot{\bf r})\big]}
{[\kappa q_z^4 + B_sq_\perp^2]}
\nonumber\\
& + & 2\Delta_t\int_{\bf q}
\frac{q_z^2\big[1 - \cos({\bf q}\cdot{\bf r})\big]}
{[\kappa q_z^4 + B_sq_\perp^2]^2} \,,
\nonumber\\
& \simeq & 2\Delta_t\,\int_{\bf q}
\frac{q_z^2\big[1 - \cos({\bf q}\cdot{\bf r})\big]}
{\big[\kappa q_z^4 + B_sq_\perp^2\big]^2} \,,
\label{Crantilt}
\end{eqnarray}
where, in going from the first to the second equation, we have 
dropped the subdominant thermal
part and only kept the $T=0$ random pinning contribution.
Asymptotic analysis of the above integrals gives (we again drop 
overall numerical factors of order unity):
\begin{mathletters}
\begin{eqnarray}
C^t_s(r_\perp) &\approx& \frac{\Delta_t r_\perp^{\frac{7}{2}-d}}
{\kappa^{3/4}B_s^{5/4}}\,,
\\
&\approx&\frac{\Delta_t \, r_\perp^{1/2}}{\kappa^{3/4}B_s^{5/4}} \,,
\label{CDelta(rperp)}
\\
C^t_s(z) &\approx& \frac{\Delta_t}{ \kappa^{(5-d)/2}B_s^{(d-1)/2}}
\, z^{7-2d} \,,
\\
&\approx&\frac{\Delta_t\,z}{\kappa B_s} \,,
\label{CDelta(z)}
\end{eqnarray}
\label{CDelta}
\end{mathletters}\noindent
where in Eqs. (\ref{CDelta(rperp)}) and (\ref{CDelta(z)}) we have
specialized to the physically relevant case of three
dimensions. Defining tilt-disorder spatial coherence lengths
$(\xi_t^\perp,\xi_t^z)$ as the lengths for which
$C^t_s(\xi_t^\perp,\xi_t^z)=a^2$, we find
\begin{mathletters}
\begin{eqnarray}
\xi_t^\perp & \simeq & \frac{B_s^{5/2}\kappa^{3/2}}{\Delta_t^2}
\; a^4 \,,
\label{xicperp}
\\
\xi_t^z & \simeq & \Big(\frac{\kappa B_s}{\Delta_t}\Big)
\, a^2 \,.
\label{xicz}
\end{eqnarray}
\end{mathletters}
We stress that (in the absence of positional disorder, elastic
nonlinearities, and dislocations) the power-law growth of the
correlation function $C^t_s({\bf r})$ (Eqs.
(\ref{CDelta(rperp)})-(\ref{CDelta(z)})) extends out to arbitrarily
long scales. This is in contrast to the Larkin random force
approximation of Sec.\ref{vmag_PT_Larkin}.  Hence at long length
scales, for $d=3<7/2$, $C^t({\bf r})$ quite clearly dominates over the
much slower, logarithmically growing distortions $C^{p}({\bf r})$
(Eqs. (\ref{CW(rperp)})-(\ref{CW(z)})) created by translational disorder
alone.  This is a very important point and is one that cannot be seen
just by comparing the coherence lengths ($\xi_t$) in the presence of
tilt disorder alone (given by Eqs. (\ref{xicperp}) and (\ref{xicz}))
with those ($\xi_p$) in the presence of positional disorder alone
(given by Eqs.  (\ref{LarkinR1}) and (\ref{LarkinZ1})). For equally
weak disorder strengths ({\em i.e.} small $\Delta_t$ and $\Delta_f$) one can
see that $\xi_t\gg\xi_p$ and Larkin length is determined by the
positional disorder, {\em i.e.}  $\xi_p=R_{c}$.  However, at length
scales beyond this Larkin length (where the Larkin approximation
breaks down) the vortex lattice distortions are controlled by the
random tilt (rather than positional) disorder.\cite{commentTilt}
Therefore, henceforth we shall ignore translational disorder
altogether and focus on tilt disorder only. As we will show below,
despite this considerable simplification, the asymptotic properties of
the spontaneous vortex lattice remain nontrivial and rich due to the
interplay of tilt disorder and anharmonic elasticity that we have
ignored so far.

\section{Perturbative treatment of elastic nonlinearities and tilt disorder}
\label{PT}

The elasticity of a pinned line-crystal can be most easily formulated
in terms of the (Lagrangian) left Cauchy-Green strain tensor
$v_{\alpha\beta}$,\cite{commentStrain} defined by
\begin{eqnarray}
v_{\alpha\beta}=\frac{1}{2}(\partial_i R_\alpha\partial_i
R_\beta-\delta_{\alpha\beta}).
\end{eqnarray}
For a line-vortex solid characterized by a conformation ${\bf R}({\bf
X}_{\mu},z)$, defined in Eqs. (\ref{R_trajectory}), (\ref{u_trajectory}), the
nonlinear Lagrangian strain tensor reduces to
\begin{eqnarray}
v_{\alpha\beta} = \frac{1}{2}\,\Big(\partial_\alpha u_\beta +
\partial_\beta u_\alpha +\partial_i u_\alpha\partial_i u_\beta\Big) \,,
\label{def-strain}
\end{eqnarray}
with $\alpha,\beta$ ranging over the $d_\perp$-dimensional subspace
($x,y$) transverse to the average spontaneous vortex line direction,
which we take to be $\hat{\bf z}$, and $i$ ranging over the full space
$(x,y,z)$. In conventional elastic solids, for weak disorder and low
temperature the gradients of the phonon field ${\bf u}({\bf r})$ are
usually small and it suffices to approximate the full nonlinear strain
tensor by its harmonic part
$v^0_{\alpha\beta}=\frac{1}{2}(\partial_\alpha u_\beta +
\partial_\beta u_\alpha)$. However, there exists a novel class of ``soft''
elastic systems\cite{softsolids} in which nonlinear elasticity plays
an essential qualitative role.  A unifying feature of solids in this
class is their underlying, spontaneously broken rotational invariance,
that strictly enforces a particular ``softness'' of the corresponding
Goldstone mode Hamiltonian. As a consequence, the usually small
nonlinear elastic terms are in fact comparable to harmonic ones, and
therefore must be taken into account.  Similar to their effects near
continuous phase transitions, but extending throughout an ordered
phase, fluctuations and disorder drive elastic nonlinearities to
qualitatively modify such soft states.  The resulting strongly
interacting ordered states at long length scales exhibit rich
phenomenology such as a universally nonlocal elasticity, a strictly
nonlinear response to an arbitrarily weak perturbation and a universal
ratio of wavevector-dependent singular elastic moduli, all controlled
by nontrivial infrared stable fixed points, illustrated for the problem at hand
in Fig. \ref{Fig_fixedpts}. As we have recently argued in a brief
communication\cite{MSCprl} a spontaneous vortex solid, with its
symmetry-enforced vanishing tilt modulus $c_{44}$ is a member of the
class of such soft solids. Hence, as we will show below, to understand
its pinned state the nonlinear part of the strain tensor
$v_{\alpha\beta}$, Eq. (\ref{def-strain}) must be taken into account.
In this Section we demonstrate the importance of nonlinear elastic
terms via a simple perturbation theory in these nonlinearities and
tilt disorder.

The full nonlinear elastic Hamiltonian of a spontaneous vortex
lattice pinned by the tilt disorder is given by (here the Lam\'e
elastic constant $\lambda$ should not be confused with the London
penetration depth):
\begin{equation}
H = \int \! d{\bf r}\,\Big[
\frac{1}{2}\kappa\,(\partial_z^2{\bf u})^2 +
\frac{\lambda}{2}\,v_{\alpha\alpha}^2 + \mu\, v_{\alpha\beta}^2 +
{\bf h}({\bf r})\cdot\partial_z{\bf u}({\bf r})
\Big],
\label{Hnonlinear}
\end{equation}
where $\kappa$ is the curvature modulus, and $\mu=c_{66}$ and
$\lambda=c_{11}-c_{66}$ are Lam\'e coefficients, respectively
characterizing the in-plane shear and in-plane bulk moduli of the
spontaneous vortex solid.  The key feature that distinguishes the
spontaneous vortex solid from a conventional field-induced one, and
requires us to keep nonlinear terms in $\partial_z u_\alpha$, is the
absence of the $\mu_{z\perp}v_{z\alpha}^2$ out-of-plane shear term.
This is because of the exact vanishing (enforced by
rotational-symmetry) of the tilt-shear modulus $\mu_{z\perp}=c_{44}$.
In contrast, because $\mu$ and $\lambda$ are finite, the corresponding
elastic nonlinearities in $\partial_\alpha u_\beta$ need {\em not} be
retained, as they are clearly subdominant at long length
scales. Consequently, for a spontaneous vortex solid the nonlinear
elastic strain that appears in $H$, Eq. (\ref{Hnonlinear}), is given
by:
\begin{eqnarray}
v_{\alpha\beta} \simeq \frac{1}{2}\,\Big(\partial_\alpha u_\beta +
\partial_\beta u_\alpha +\partial_z u_\alpha\partial_z u_\beta\Big).
\end{eqnarray}

The Hamiltonian $H$ above can be written as:
\begin{eqnarray}
H = H_0 + H_1 + H_d\; ,
\end{eqnarray}
where
\begin{eqnarray}
H_0 & = & \int d{\bf r}\;\Big[\,
\frac{1}{2}\kappa\,(\partial_z^2{\bf u})^2 +
\frac{1}{2}\,\mu(\partial_\alpha u_\beta)^2 +
\nonumber\\
&+&\frac{1}{2}\,\mu(\partial_\alpha u_\beta)(\partial_\beta
u_\alpha) +
\frac{1}{2}\,\lambda(\partial_\alpha u_\alpha)^2
\,\Big]
\end{eqnarray}
is the standard harmonic elastic part, while
\begin{eqnarray}
H_1  &=& \int \! d{\bf r}\;\Big[\,
\mu\partial_\alpha u_\beta\partial_z u_\alpha\partial_zu_\beta +
\frac{\lambda}{2}\,(\partial_\alpha u_\alpha)(\partial_zu_\beta)^2
\nonumber\\
&+& \frac{1}{8}\,(\lambda +
2\mu)\,(\partial_zu_\alpha)^2(\partial_z u_\beta)^2
\,\Big]
\end{eqnarray}
is the nonlinear elastic contribution, and
\begin{eqnarray}
H_d = \int d{\bf r}\; {\bf h}({\bf r})\cdot\partial_z{\bf u}({\bf r})
\end{eqnarray}
is the previously-introduced random tilt disorder.
The quenched tilt disorder can be most efficiently treated via the
standard replica trick,\cite{Anderson,commentReplica}
\begin{eqnarray}
\overline{Z^n} = \lim_{n\to 0} \frac{ \overline{Z^n} - 1}{n} \,,
\label{trick}
\end{eqnarray}
which allows us to formally average over disorder and therefore work
with a more convenient translationally invariant field theory.
Introducing $n$ replica fields labeled by $a$ and averaging the
resulting ``partition function'', $Z^n$ over the tilt disorder, ${\bf
h}({\bf r})$, we find
\begin{eqnarray}
\overline{Z^n}
& = & \int [d{\bf u}_1({\bf r})]\cdots[d{\bf u}_n({\bf r})]\;
\mbox{e}^{-\beta ({H}_{0n} + H_{1n})},
\end{eqnarray}
where $\beta=1/T$, and
\begin{equation}
H_{1n}=\sum_{a=1}^n H_1[{\bf u}_a],
\end{equation}
and we denote by ${H}_{0n}$ the following (quadratic) Hamiltonian:
\begin{eqnarray}
{H}_{0n} = \sum_{a,b}\int_{\bf q}\;\frac{1}{2}\,
\Gamma_{\alpha\beta}^{ab}({\bf q})
\,u_\alpha^a({\bf q})u_\beta^b(-{\bf q}) \,,
\label{Heff0}
\end{eqnarray}
with the kernel:
\begin{eqnarray}
\Gamma_{\alpha\beta}^{ab}({\bf q}) = \Phi_{\alpha\beta}({\bf q})
\delta_{ab} - \frac{\Delta_t}{T}\;q_z^2\delta_{\alpha\beta} \,,
\end{eqnarray}
where  $\Phi_{\alpha\beta}({\bf q}) =
\Phi_L({\bf q})P^L_{\alpha\beta}({\bf q}) + \Phi_T({\bf
q})P^T_{\alpha\beta}({\bf q})$ and $\Phi_L({\bf q})$ and $\Phi_T({\bf
q})$ are given by Eqs. (\ref{newdef-PhiL}) and (\ref{newdef-PhiT}).
From the above equation the noninteracting elastic propagator
$G_{\alpha\beta}({\bf q})=
\Omega^{-1}\langle u_{\alpha}({\bf q})u_{\beta}(-{\bf q})\rangle_0$
($\Omega$ being the system volume) can be easily obtained (see Appendix
\ref{App_elastic_propagator})
and is given in the limit $n\to 0$ by:
\begin{eqnarray}
G_{\alpha\beta}^{ab}({\bf q}) = G_L^{ab}({\bf q})P^L_{\alpha\beta}({\bf q}) +
G_T^{ab}({\bf q})P^T_{\alpha\beta}({\bf q}) \,,
\end{eqnarray}
where
\begin{mathletters}
\begin{eqnarray}
G_L^{ab}({\bf q}) & = & T\Gamma_L^{-1}({\bf q})\,\delta_{ab}
+ {\Delta_t} q_z^2\Gamma_L^{-2}({\bf q})\,,
\label{GL}
\\
G_T^{ab}({\bf q}) & = & T\Gamma_T^{-1}({\bf q})\,\delta_{ab}
+ {\Delta_t} q_z^2\Gamma_T^{-2}({\bf q}) \,,
\label{GT}
\end{eqnarray}
\end{mathletters}
with the kernels:
\begin{mathletters}
\begin{eqnarray}
\Gamma_L({\bf q})&=&(\lambda+2\mu)q_{\perp}^2 + \kappa q_z^4 \;,
\label{GLinverse}\\
\Gamma_T({\bf q})&=& \mu q_{\perp}^2 + \kappa q_z^4\;.
\label{GTinverse}
\end{eqnarray}
\end{mathletters}

To assess the importance of elastic nonlinearities we coarse-grain
$\overline{Z^n}$. To this end we decompose the displacement field
${\bf u}({\bf r})$ into low and high wavevector parts:
\begin{eqnarray}
{\bf u}({\bf r}) = {\bf u}^<({\bf r}) + {\bf u}^>({\bf r}) \,.
\label{shell_modes}
\end{eqnarray}
We then perturbatively (in the nonlinear elastic terms) integrate out the
quickly varying, short-scale components ${\bf u}^>({\bf r})$, and 
thereby obtain a
representation of $\overline{Z^n}$:
\begin{eqnarray}
\overline{Z^n} & = & \int\! [d{\bf u}_a^<({\bf r})]
\mbox{e}^{-\beta H_{0,n}^<} Z_0^>\Big\langle\mbox{e}^{-\beta
H_{1n}[u^<+u^>]}\Big\rangle_{0>} \,,
\nonumber\\
& = & \int [d{\bf u}_a^<({\bf r})]\;\mbox{e}^{-\beta H_{\rm eff}} \,,
\end{eqnarray}
in terms of a coarse-grained Hamiltonian $H_{\rm eff}$
\begin{equation}
H_{\rm eff}[{\bf u}] = H_{0,n}^<[{\bf u}] + \big\langle
H_{1,n}\big\rangle_{0>} - \frac{1}{2T}\big\langle
H_{1,n}^2\big\rangle_{0>}^c + \ldots,
\end{equation}
where $Z_0^>=\mbox{Tr}(-\beta H_{0,n}^>)$, $\langle\cdots\rangle_{0>}$
indicates an average with statistical weight $\exp(-\beta H_{0,n}^>)/Z_0^>$,
$\langle H_{1,n}^2 \rangle_{0>}^c$ denotes the connected cumulant
$\equiv\langle H_{1,n}^2 \rangle_{0>}-\langle H_{1,n} \rangle_{0>}^2$,
and where short-scale modes extend out to wavelengths of
size $L_z\times L_\perp$.

Restricting our attention here to the leading
order\cite{Radzihovsky-Toner} nonlinear part of $H_{1n}$:
\begin{equation}
H_{int} = \sum_{a=1}^n\int d{\bf r}\;\big[
\mu\partial_\alpha u_\beta^a\partial_z u_\alpha^a\partial_z u_\beta^a +
\frac{\lambda}{2}\,\partial_\alpha u_\alpha^a(\partial_z u_\beta^a)^2
\big],
\label{def-Hint}
\end{equation}
a standard calculation (with details relegated to Appendix
\ref{App_vmag_PT1}) shows\cite{linearH} that the resulting effective 
Hamiltonian,
$H_{\rm eff}[{\bf u}^<]$, can be put into the same form as the original
Hamiltonian, $H_{0n}[{\bf u}]$, but with effective elastic moduli
perturbatively corrected by the coarse-graining procedure. For 
example, as we show in great
detail in Appendix \ref{App_vmag_PT1}, the lowest order correction 
$\delta\mu$ to the bare
elastic shear modulus $\mu$ is given by:
\begin{eqnarray}
\delta\mu &\approx&
- \frac{\mu^2}{Td_\perp(d_\perp+2)}\;\int^{>}\frac{dq_z}{2\pi}
\int\frac{d^{d-1}{\bf q}_\perp}{(2\pi)^{d-1}}
q_z^4\;\Big[
2\big(G_L^{ab}({\bf q})\big)^2
\nonumber\\
&+& (d_\perp^2-2)\big(G_T^{ab}({\bf q})\big)^2
+ 2d_\perp G_L^{ab}({\bf q})G_T^{ab}({\bf q})
\Big] \,,
\end{eqnarray}
where $G_L^{ab}$ and $G_T^{ab}$ are the noninteracting propagators of
Eqs. (\ref{GL})-(\ref{GT}) and where the ``$>$" sign on the $q_z$
integral indicates that we impose the IR cutoff $1/L_z$, restricting
the integration range to the region $|q_z|>1/L_z$.  Evaluating the
above expression to leading order in the IR cutoff $L_z$ (and ignoring
the subdominant thermal parts) gives the perturbative relative
corrections to Lam\'e elastic moduli:
\begin{eqnarray}
\frac{\delta\mu}{\mu} \sim \frac{\delta\lambda}{\lambda} \sim \Delta_t\;
\mu^{\frac{3-d}{2}}\kappa^{\frac{d-7}{2}}\; L_z^{7-2d} \; ,
\label{new-del-mu}
\end{eqnarray}
that for $d<7/2$ diverge with $L_z,L_\perp$. As we will see in
subsequent Sections (with details given in Appendix
\ref{App_vmag_PT1}) similar analysis shows that the perturbative
corrections to the curvature modulus $\kappa$ and the tilt-disorder
strength $\Delta_t$ also diverge with system size like
$L_z^{7-2d}$. Thus, at long length scales perturbation theory breaks
down for spatial dimensions smaller than the upper critical dimension
\begin{eqnarray}
d_{uc} = \frac{7}{2} \; .
\end{eqnarray}

Note that this is the same as the critical dimension below which
positional fluctuations diverge. This is not a coincidence. Rather, it
is a consequence of the fact that, in power-counting terms, the
leading cubic and quartic anharmonicities that we have kept have the
same number of derivatives as the leading harmonic terms. In doing
this power counting, note that each $\perp$ derivative counts as two
$z$-derivatives, as can be seen from the strongly anisotropic form of
the harmonic terms. Given this, the only difference in scaling between
the anharmonic and the harmonic terms must come from the extra powers
of ${\bf u}$ in the anharmonic terms. Hence, the anharmonic terms only
become important when ${\bf u}$ scales up with increasing distance;
{\em i.e.}, when ${\bf u}$ fluctuations grow with length scale, as they do
below $d = d_{uc} = \frac{7}{2}$.

We define a nonlinear (crossover) length scale at which perturbation
theory breaks down by the value of $L_z$ at which the relative
corrections to the elastic moduli and disorder variance,
$\delta\lambda/\lambda$, $\delta\mu/\mu$, $\delta\kappa/\kappa$ and
$\delta\Delta_t/\Delta_t$, become of order $1$. Equation
(\ref{new-del-mu}) leads to the following estimate for this nonlinear
crossover length $\xi_z^{NL}$:
\begin{mathletters}
\begin{eqnarray}
\xi_z^{NL}&\simeq &
\Big(\frac{\kappa^{\frac{7-d}{2}}\mu^{\frac{d-3}{2}} }{\Delta_t}
\Big)^{\frac{1}{7-2d}} \;,\\
&\simeq& \Big(\frac{\kappa^2}{\Delta_t}
\Big)  , \quad \mbox{for d = 3}.
\end{eqnarray}
\end{mathletters}
A corresponding perpendicular nonlinear crossover length
$\xi_\perp^{NL}$ can also be defined by imposing a $1/L_\perp$
infrared cut-off on the ${\bf q}_\perp$ integrals in perturbation
theory, and is given by
\begin{eqnarray}
\xi_\perp^{NL} &\simeq& 
\Big(\frac{\mu}{\kappa}\Big)^{\frac{1}{2}}(\xi_z^{NL})^2 \,.
\end{eqnarray}
Hence, we find from the above perturbative coarse-graining procedure
that at sufficiently long length scales (greater than ${\rm
min}[(\xi_\perp^{NL},\xi_z^{NL})]$), the anharmonic elasticity becomes
qualitatively important, even for arbitrarily weak pinning, thus
invalidating the description of the pinned spontaneous vortex solid by
a conventional harmonic elasticity theory.  In order to describe the
pinned spontaneous vortex solid on scales longer than
$(\xi_\perp^{NL},\xi_z^{NL})$, we need to elevate the above
perturbation theory to a renormalization group analysis. We turn to
this task in the next Section.

\section{Nonlinear elasticity: RG analysis}
\label{RGanalysis}

Having established the importance of the nonlinear elasticity (as
indicated by the divergent perturbation theory of previous Section), we
now employ the standard Wilson momentum-shell renormalization-group
method\cite{Wilson} to study the pinned spontaneous vortex solid on
length scales longer than $\xi_{NL}$. The results we derive here were
first reported in Ref.\onlinecite{MSCprl}.

\subsection{Recursion relations and zero-temperature fixed points}

To this end, we use phonon mode decomposition,
Eq. (\ref{shell_modes}), with high wavevector modes ${\bf u}^>$
restricted to an infinitesimal cylindrical shell:
\begin{mathletters}
\begin{eqnarray}
\Lambda_z e^{-\ell} < &|q_z|& <\Lambda_z\,,
\label{shell1}
\\
0 < &|q_\perp|& < \infty\,,
\label{shell2}
\end{eqnarray}
\end{mathletters}
where $\ell$ is taken to be an infinitesimal positive number.  The
procedure is to integrate these short length scale modes ${\bf
u}^>({\bf r})$ out of the replicated partition function
$\overline{Z^n}$ and to interpret the result in terms of a Hamiltonian
of the same form but with effective $\ell$-dependent parameters. To
simplify the algebra that follows we introduce anisotropic rescalings
of lengths and fields:
\begin{mathletters}
\begin{eqnarray}
{\bf r}_\perp & = & \mbox{e}^{\omega\ell}{\bf r}_\perp' \,,
\label{resc1}\\
z & = & \mbox{e}^{\ell} z' \,,
\\
{\bf u}_\alpha^<({\bf r}) & = & \mbox{e}^{\phi \ell}{\bf
u}_{\alpha}'({\bf r}') \,.
\label{resc3}
\end{eqnarray}
\end{mathletters}
This leads to zero-th order (in elastic nonlinearities) recursion
relations
\begin{mathletters}
\begin{eqnarray}
\kappa(\ell) & = & \kappa\,\mbox{e}^{((d-1)\omega-3+2\phi)\ell} \,,
\label{rescaling1}
\\
\lambda(\ell) & = & \lambda\,\mbox{e}^{((d-3)\omega + 1 + 2\phi)\ell} \,,
\\
\mu(\ell) & = & \mu\,\mbox{e}^{((d-3)\omega + 1 +2\phi)\ell} \,,
\\
\frac{\Delta_t}{T}(\ell) & = &
\frac{\Delta_t}{T}\,\mbox{e}^{((d-1)\omega-1+2\phi)\ell} \,.
\label{rescaling4}
\end{eqnarray}
\end{mathletters}
It is convenient (but not necessary) to take advantage of the
rescaling freedom (\ref{resc1})-(\ref{resc3}), and to choose the
arbitrary exponent $\phi$ to be
\begin{equation}
\phi = 2 -\omega,
\label{omega}
\end{equation}
so as to keep fixed\cite{comment_phi} the form of the nonlinear strain
tensor $v_{\alpha\beta}=(\partial_\alpha u_\beta +
\partial_\beta u_\alpha + \partial_z u_\alpha\partial_z u_\beta)/2$.

Integrating out the high wavevector modes ${\bf u}^>$ to second
order\cite{linearH} in elastic nonlinearities leads to the following
corrections to effective Hamiltonian parameters (with calculation
details provided in Appendix \ref{App_vmag_PTkappa}):
\begin{mathletters}
\begin{eqnarray}
\delta D & = & g F_D(x) D\;d\ell \,,
\label{pert-lambda}\\
\delta\mu & = &  g F_\mu(x)\mu\;d\ell \,,\\
\label{pert-mu}
\delta\kappa & = & g F_\kappa(x)\kappa\;d\ell  \,,
\label{vmag_pert-kappa}\\
\delta\Big(\frac{\Delta_t}{T}\Big)
& = &  g F_{\Delta_t}(x)\Big(\frac{\Delta_t}{T}\Big) \;d\ell  \,,
\label{vmag_pert-Delta}
\end{eqnarray}
\end{mathletters}
where $D=\lambda+\mu$ and we have defined dimensionless couplings $x$ and $g$ such that:
\begin{mathletters}
\begin{eqnarray}
x & = & \frac{\lambda}{\mu} \,,
\\
g & = & \frac{\sqrt{2}C_{d-1}}{64}{\Delta_t}\;\kappa^{\frac{d-7}{2}}
\mu^{\frac{3-d}{2}}\Lambda^{2d-7} \,,
\end{eqnarray}
\end{mathletters}
the latter controlling the perturbative expansion; and where the scaling
functions $F_i(x)$ are given by:
\begin{mathletters}
\begin{eqnarray}
F_D(x) & = & -\frac{1}{30 (x+1)^2}\Big[
135x^3 + 351x^2 + 316x + 164 +
\nonumber\\
&+&\frac{90x^3 + 234 x^2 + 144 x - 64}{(2+x)^{5/4}}
\Big] \,,
\label{Flambda}
\\
F_\mu(x) & = & -\frac{2}{15(x+1)}\Big[
97 + 17 x - \frac{152+72x}{(x+2)^{5/4}}
\Big] \,,
\label{Fmu}
\\
F_\kappa(x) & \!=\! & \frac{1}{5(x+1)}\Big[
1686 \!+\! 366 x \!-\! \frac{3896 \!+\! 3156 x \!+\! 580 x^2}{(x+2)^{5/4}}
\Big],
\nonumber\\
\label{Fkappa}
\\
F_{\Delta_t}(x) & = & \frac{1}{10(x+1)}\Big[
543 + 159 x - \frac{614 + 230x}{(x+2)^{1/4}}
\Big] \,.
\label{FDelta}
\end{eqnarray}
\end{mathletters}

Combining these with the zero-th order rescalings, Eqs.
(\ref{rescaling1})-(\ref{rescaling4}), and Eq. (\ref{omega}), leads to
the differential flow equations:
\begin{mathletters}
\begin{eqnarray}
\frac{dD}{d\ell} & = & \big((d-5)\omega+ 5 +
g F_\lambda(x)\big)\,D(\ell) \,,
\label{vmag_reclambda}\\
\frac{d\mu}{d\ell} & = & \big((d-5)\omega + 5 +
gF_\mu(x)\big)\,\mu(\ell) \,,
\label{recmu}\\
\frac{d\kappa}{d\ell} & = & \big((d-3)\omega + 1 +
gF_\kappa(x)\big)\,\kappa(\ell) \,,
\label{reckappa}\\
\frac{d{\Delta_t}}{d\ell} & = & \big((d-3)\omega + 3 +
gF_{\Delta_t}(x)\big)\,{\Delta_t}(\ell) \,.
\label{recDelta}
\end{eqnarray}
\label{flow_final}
\end{mathletters}
It is easy to verify that the above differential flow equations,
Eqs. (\ref{vmag_reclambda})-(\ref{recDelta}), lead to the following
two closed flow equations for the two dimensionless couplings
$g(\ell)$ and $x(\ell)$:
\begin{mathletters}
\begin{eqnarray}
\frac{d g}{d\ell}
&=& \Big[(7-2d)+ g\big(F_{\Delta_t}(x) + \frac{3-d}{2}\,F_\mu(x)
\nonumber\\
&+& \frac{d-7}{2}\,F_\kappa(x)\big)\Big]g \,.
\label{recg1}\\
\frac{dx}{d\ell} & = & g\big[F_D(x)-F_\mu(x)\big]
\, \big(x(\ell) +1\big)\,.
\label{recx}
\end{eqnarray}
\end{mathletters}

\begin{figure}[t]
\includegraphics[width=8cm, height=5cm]{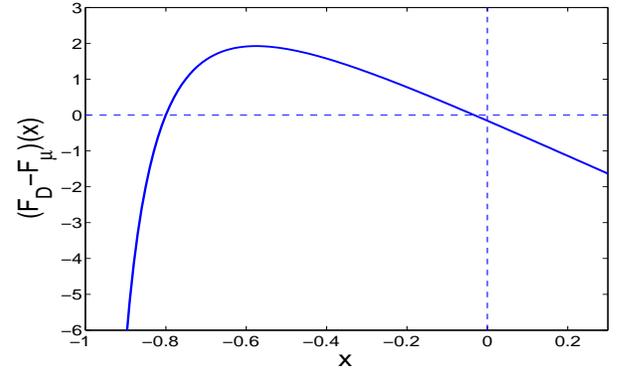}
\caption{
A plot of the function $\big(F_D(x)-F_\mu(x)\big)$, with its
zeros, $\big(F_D(x)-F_\mu(x)\big)=0$ at $x_1=-4/5$ and
$x_2\simeq -0.03272$ determining the fixed point solutions for the
Lam\'e coefficients ratio, $x_*=\lambda^*/\mu^*.$
}
\label{Fig_diff}
\end{figure}

The fixed point of Eq. (\ref{recx}) ({\em i.e.}, such that $(dx/d\ell)=0$)
can be easily computed by noting that for finite $g$ they are given by
the zeros of the function $\big(F_\lambda(x)-F_\mu(x)\big)$ plotted in
Fig. \ref{Fig_diff}. These are given by:

\begin{figure}[ht]
\includegraphics[scale=0.5]{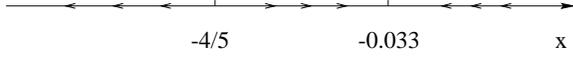}
\vspace{0.35cm}
\caption{
Renormalization group flow of the ratio of the Lam\'e coefficients
$ x = \lambda/\mu$. The fixed point at $x_2=-0.033$ controls the
long-scale properties (anomalous elasticity and associated
phenomenology) of the pinned spontaneous vortex solid. The critical
point at $x_1=-4/5$ controls the mechanical instability of the
spontaneous vortex solid.}
\label{Fig_fixedpts}
\end{figure}

\begin{mathletters}
\begin{eqnarray}
x_1&=&-4/5,\\
x_2&\simeq& -0.03272.
\end{eqnarray}
\end{mathletters}
As can be seen from Fig. \ref{Fig_diff},
$\big(F_D(x)-F_\mu(x)\big)$ is positive for $x_1<x<x_2$, and
negative otherwise. Therefore $x_1=-4/5$ is an unstable fixed point,
while $x_2\simeq -0.03272$ is the stable fixed point
of interest to us (see Fig. \ref{Fig_fixedpts}).

At the stable $x_2\simeq -0.03272$ fixed point, that we shall denote
hereafter by $x_*$, the recursion relation (\ref{recg1}) for the
dimensionless coupling $g(\ell)$ reduces to
\begin{eqnarray}
\frac{dg}{d\ell} = \big[\,2\epsilon -
12.8359\,g\,\big]\,g\,,
\end{eqnarray}
with $\epsilon=7/2 -d$ controlling the validity of the expansion. This
implies a stable fixed point (to ${\cal O}(\epsilon^2)$)
\begin{equation}
(x_*,g_*) \simeq (-0.03272,\, 0.1558\,\epsilon) \,,
\label{fixedpoint}
\end{equation}
that controls long-wavelength, weak-disorder and low-temperature
properties of the pinned spontaneous vortex solid.

The existence of this nontrivial fixed point has dramatic consequences
for the long-scale elastic properties of the pinned SV lattice. As
discussed above, it leads to the anomalous elasticity (common to soft
solids), that we explore below.

The critical point $x_1=-4/5$ and its instability correspond to the
stability limit, in $7/2$ dimensions, of the elastic constants,
$\lambda$ and $\mu$, beyond which the vortex lattice is unstable
against bulk distortions, {\em i.e.} ${\bf u}({\bf r}) \propto {\bf
r}_\perp$. Stability of the vortex lattice against bulk distortions
requires that the bulk modulus, $B=\lambda+(2\mu/d^*)$, (not to be
confused with magnetic flux; $d^*=d-1$ is the dimensionality of the
vortex {\em lattice}) be positive. At the stability limit, {\em i.e.} when
$B=0$,
\begin{eqnarray}
x \equiv {\lambda\over\mu} = {-2 \over d-1}\;,
\label{CBGgdefinition}
\end{eqnarray}

\medskip

\medskip

\noindent which for $d=7/2$ gives $x=-4/5$, the value of the unstable fixed
point. That the recursion relations reflect this stability limit is a
nontrivial check that the $\lambda$ and $\mu$ graphical corrections
were correctly evaluated.  For $x <-4/5$ the flows run away
negatively, corresponding to a phase transition from the stable vortex
lattice phase to one which, presumably, has different lattice
constants. The associated phenomenology would be an interesting
subject for future research.

\subsection{Long length-scale correlation functions: matching}
\label{subsec:matching}

Many of the basic properties (length scale-dependent elastic moduli
and the associated anomalous elasticity) of the spontaneous vortex
glass are encoded in the long-scale limit of the 2-point phonon
correlation function $G_{\alpha\beta}^{ab}({\bf
q})=\big(\Gamma_{\alpha\beta}^{ab}({\bf q})\big)^{-1}$, defined and
derived in Appendix \ref{App_elastic_propagator} for the harmonic
elastic case (valid on length scales shorter than $\xi_{NL}$).

We can utilize the renormalization group transformation and the harmonic
propagator to compute the correlation function at longer length scales
despite it being nonperturbative in the elastic
nonlinearities. To this end, recalling the definition of
$G_{\alpha\beta}^{ab}({\bf q})$,
\begin{eqnarray}
G_{\alpha\beta}^{ab}({\bf q}) = \frac{\langle u_\alpha^a({\bf q})u_\beta^b({\bf q}')\rangle}
{(2\pi)^d\delta({\bf q}+{\bf q}')} \,,
\label{eqn-def-G}
\end{eqnarray}
together with the rescalings, Eqs. (\ref{resc1})-(\ref{rescaling4}), 
it is easy to see that
\end{multicols}
\begin{eqnarray}
G_{\alpha\beta}^{ab}({\bf q})
& = &\mbox{e}^{((d-1)\omega+1+2\phi)\ell} \frac{\big\langle 
u_\alpha^a({\bf q}_\perp\mbox{e}^{\omega\ell},q_z\mbox{e}^{\ell})
u_\beta^b({\bf 
q}_\perp'\mbox{e}^{\omega\ell},q_z'\mbox{e}^{\ell})\,\big\rangle}
{(2\pi)^d\delta({\bf q}_\perp\mbox{e}^{\omega\ell}+{\bf 
q}_\perp'\mbox{e}^{\omega\ell})
\delta(q_z\mbox{e}^{\ell}+q_z'\mbox{e}^{\ell})} \,,
\nonumber\\
& = & \mbox{e}^{((d-3)\omega+5)\ell}\,G_{\alpha\beta}^{ab}({\bf q}_\perp
\mbox{e}^{\omega\ell},q_z\mbox{e}^{\ell}) \,,
\end{eqnarray}
where in going from the first to the second line we made the
convenient choice $\phi=2-\omega$. Finally, integrating out the
short-scale phonon modes $u^>$, gives a scaling relation satisfied by
the correlation function:
\begin{eqnarray}
G_{\alpha\beta}^{ab}({\bf q}_\perp,q_z,\lambda,\mu,\kappa,g)
&=& \mbox{e}^{((d-3)\omega+5)\ell}
G_{\alpha\beta}^{ab}\big({\bf q}_\perp\mbox{e}^{\omega\ell},
q_z\mbox{e}^{\ell},\lambda(\ell),\mu(\ell),
\kappa(\ell),g(\ell)\big) \,.
\label{trans-G}
\end{eqnarray}

\begin{multicols}{2}

This scaling (matching) relation demonstrates the real power of the
renormalization group transformation.\cite{Nelson_Rudnick} Choosing
$e^\ell$ large, we can relate the long-wavelength (small ${\bf q}$)
correlation function (that is impossible to compute perturbatively
because of the appearance of infrared divergences in the direct
perturbation theory) to the correlation function at large ${\bf q}$,
({\em e.g.} at the Brillouin zone boundary), that can be easily computed
using simple perturbation theory.

\subsubsection{Long wavelength behavior of $\kappa$ and ${\Delta_t}$}
\label{LongWaveKappaDelta}

To utilize the above scaling relation we first focus on the
$q_\perp=0$ case and take $\ell^*$ such that
\begin{eqnarray}
q_z\,\mbox{e}^{\ell^*} = \Lambda_z \,.
\label{def-lstar2}
\end{eqnarray}
We then consider sufficiently long length scales, $\Lambda_z/q_z\gg
1$, so that $\ell_*$ is large enough for the coupling constants
$(g(\ell_*),x(\ell_*))$ to approach their fixed point values
$(g_*,x_*)$.  Then using Eq. (\ref{trans-G}) together with the zero-th
order approximation for its right hand side, we find

\end{multicols}
\begin{eqnarray}
G_{\alpha\beta}^{ab}(q_\perp=0,q_z,\lambda,\mu,\kappa,g) & = &
\mbox{e}^{((d-3)\omega+5)\ell^*}
G_{\alpha\beta}^{ab}(0,\Lambda_z,\lambda(\ell^*),\mu(\ell^*),
\kappa(\ell^*),g_*)
\nonumber\\
& = & \mbox{e}^{((d-3)\omega+5)\ell^*}\Big[
\frac{T}{\kappa(\ell^*)\Lambda_z^4}\;\delta_{ab} +
\frac{{\Delta_t}(\ell^*)\Lambda_z^2}{[\kappa(\ell^*)\Lambda_z^4]^2}
\Big]\;\delta_{\alpha\beta} \,.
\label{G(kz)}
\end{eqnarray}
Using Eq. (\ref{def-lstar2}) to eliminate $\ell_*$ in favor of $q_z$, the
first term in square brackets gives:
\begin{mathletters}
\begin{eqnarray}
\mbox{e}^{((d-3)\omega+5)\ell^*}
\frac{T}{\kappa(\ell^*)\Lambda_z^4}\;\delta_{ab}\delta_{\alpha\beta}
& = & \frac{ T\mbox{e}^{((d-3)\omega+5)\ell^*} }
{\kappa q_z^4(\Lambda_z/q_z)^4\mbox{e}^{((d-3)\omega+1+
g_*F_\kappa(x_*))\ell^*} }
\delta_{ab}\delta_{\alpha\beta} ,
\\
& = &\frac{ T\big(\Lambda_z/q_z\big)^{((d-3)\omega+5)} }
{\kappa q_z^4\big(\Lambda_z/q_z\big)^{4+((d-3)\omega+1+
g_*F_\kappa(x_*))} } \delta_{ab}\delta_{\alpha\beta} ,
\\
& = &
\frac{T}{\kappa q_z^4(\Lambda_z/q_z)^{g_*F_\kappa(x_*)} }
\delta_{ab}\delta_{\alpha\beta} ,
\label{interm_kappa_resc}
\\
& = &
\frac{T}{\kappa(q_z)q_z^4}
\delta_{ab}\delta_{\alpha\beta}\,.
\end{eqnarray}
\end{mathletters}
\begin{multicols}{2}\noindent
We note that as required, the arbitrary rescaling exponent $\omega$
dropped out from the physical correlation function.  In the above
expression, we can identify $\kappa(q_z)$ as the effective
wavevector-dependent curvature modulus:
\begin{eqnarray}
\kappa(q_z) = \kappa \;\Big(\frac{q_z}{\Lambda_z}\Big)^{-\eta_\kappa},
\label{kappa-qz}
\end{eqnarray}
with the exponent
\begin{eqnarray}
\eta_\kappa = g_*\,F_\kappa(x_*) \,,
\label{def-etakappa}
\end{eqnarray}
clearly a universal number determined by the fixed point.  A similar
reasoning for the second term in (\ref{G(kz)}) leads to the
wavevector-dependent disorder variance:
\begin{eqnarray}
{\Delta_t}(q_z) & = & 
{\Delta_t}\;\Big(\frac{q_z}{\Lambda_z}\Big)^{-\eta_{\Delta_t}},
\label{Delta-qz}
\end{eqnarray}
that corresponds to effective long-range correlated tilt disorder,
with a universal exponent given by
\begin{eqnarray}
\eta_{\Delta_t} = g_*\,F_{\Delta_t}(x_*) \,.
\label{def-etaDelta}
\end{eqnarray}
From equations (\ref{def-etakappa}) and (\ref{def-etaDelta}) and the
definitions of the functions $F_\kappa(x)$ and $F_{\Delta_t}(x)$,
Eqs. (\ref{Fkappa})-(\ref{FDelta}), we obtain the following numerical
values of the exponents $\eta_\kappa$ and $\eta_{\Delta_t}$:
\begin{mathletters}
\begin{eqnarray}
\eta_\kappa & = & 1.478\epsilon \label{def__etakappa}\\
&\simeq& 0.739 \quad \mbox{in} \quad d=3 \quad \mbox{dimensions,}
\label{eq0.739}
\\
\eta_{\Delta_t} & = & 0.414\epsilon  \label{def__etaDelta}\\
&\simeq& 0.207 \quad \mbox{in} \quad d=3 \quad \mbox{dimensions.}
\label{eq0.207}
\end{eqnarray}
\end{mathletters}

\subsubsection{Long wavelength behavior of $\lambda$ and $\mu$}

Similarly, in order to determine the long wavelength behavior of the
effective Lam\'e moduli $\lambda$ and $\mu$, we again utilize the
scaling relation, Eq. (\ref{trans-G}). Here, quite clearly, we need to
keep $q_\perp$ finite. However, for convenience, we choose
$q_\perp\xi^{NL}_\perp\ll (q_z\xi_z^{NL})^\zeta$, so that (as in the
previous subsection) it is $q_z$ that controls the length scale
dependence of the elastic moduli, and therefore the rescaling
parameter $e^{\ell_*}$ is still determined by Eq. (\ref{def-lstar2}):
\begin{eqnarray}
\mbox{e}^{\ell^*} = \Big(\frac{\Lambda_z}{q_z}\Big)\,.
\label{elstar}
\end{eqnarray}
The longitudinal part of $G_{\alpha\beta}^{ab}\big({\bf q}\big)$ then
satisfies:
\begin{eqnarray}
G_L^{ab}({\bf q}_\perp,q_z) & = &
\mbox{e}^{((d-3)\omega+5)\ell^*}\,
G_L^{ab}\big(q_\perp e^{\omega\ell_*},\Lambda_z\big)
\nonumber\\
& = & \frac{T\mbox{e}^{((d-3)\omega+5)\ell^*}}
{\big[\lambda(\ell^*) + 2\mu(\ell^*)\big]\,q_\perp^2 e^{2\ell_*}
+ \kappa(\ell^*)\Lambda_z^4}.
\nonumber\\
\end{eqnarray}
Utilizing the elastic constants flow equations, Eq. (\ref{flow_final}),
the denominator of the expression on the right-hand side of this last
equation can be rewritten in the form
\begin{eqnarray}
&\mbox{e}^{((d-5)\omega+5)\ell^*}&\big[\lambda\mbox{e}^{g_*F_\lambda(x_*)\ell^*} 
+ 2\mu\mbox{e}^{g_*F_\mu(x_*)\ell^*}\big]
\,q_\perp^2 e^{2\ell_*}
\nonumber\\
&&+\kappa q_z^4\Big(\frac{\Lambda_z}{q_z}\Big)^4
\mbox{e}^{((d-3)\omega+1+ g_*F_\kappa(x_*))\ell^*}.
\nonumber
\end{eqnarray}
Using Eq. (\ref{elstar}) we obtain:
\end{multicols}
\begin{eqnarray}
G_L^{ab}({\bf q}_\perp,q_z) & = &  \frac{T}
{\big[\lambda\mbox{e}^{g_*F_\lambda(x_*)\ell^*} +
2\mu\mbox{e}^{g_*F_\mu(x_*)\ell^*}\big]\,q_\perp^2
+ \kappa\, q_z^4\,\mbox{e}^{g_*F_\kappa(x_*)\ell^*}} \,,
\end{eqnarray}
\begin{multicols}{2}\noindent
which shows that at long length scales the longitudinal part of the
correlation function $G_L^{ab}({\bf q}_\perp,q_z)$ takes on its
harmonic-elasticity form:
\begin{equation}
G_L^{ab}({\bf q}_\perp,q_z;\lambda,\mu,\kappa,g) =
\frac{T}{[\lambda(q_z) + 2\mu(q_z)]\,q_\perp^2 + \kappa(q_z)q_z^4},
\end{equation}
but with crucial wavevector-dependent elastic moduli, $\kappa(q_z)$,
Eq. (\ref{kappa-qz}), and
\begin{mathletters}
\begin{eqnarray}
\lambda(q_z) & = & \lambda\Big(\frac{q_z}{\Lambda_z}\Big)^{\eta_\lambda} \,,
\label{lambda-qz}
\\
\mu(q_z) & = & \mu\Big(\frac{q_z}{\Lambda_z}\Big)^{\eta_\mu} \,,
\label{mu-qz}
\end{eqnarray}
\end{mathletters}
where we have defined the anomalous exponents
\begin{eqnarray}
\eta_\mu  =  \eta_\lambda & = & - g_*F_\mu(x_*)  \,,
\label{def__lambda_mu}
\end{eqnarray}
that take the universal values:
\begin{mathletters}
\begin{eqnarray}
\eta_\mu  =  \eta_\lambda &\simeq& 0.6919\epsilon \,,
\\
& = & 0.346 \qquad\mbox{for}\quad d=3 \,,
\label{eq0.346}
\end{eqnarray}
\end{mathletters}
controlled by the stable fixed point.  The above results show that
both Lam\'e coefficients $\mu$ and $\lambda$ vanish at long
wavelengths with the same exponent $\eta_\mu$. However, their ratio
$x=\lambda/\mu$ flows to a stable fixed point $x_*=-0.03272$, implying
a universal negative Poisson ratio
\begin{eqnarray}
\sigma_P & = & \frac{\lambda}{\lambda+2\mu} = \frac{x_*}{x_*+2} \,,
\nonumber\\
& = & -0.0166 + O(\epsilon) \,,
\end{eqnarray}
characterizing the pinned spontaneous vortex solid.

The above calculations can clearly be extended to other directions of
${\bf q}$ by choosing a more general $\ell^*$ that is a function of
both $q_z$ and ${\bf q}_\perp$. To this end we now use a more general
choice:
\begin{eqnarray}
\kappa(\ell^*)\,\big(q_z\mbox{e}^{\ell^*}\big)^4
+ \mu(\ell^*)\,\big(q_\perp\mbox{e}^{\omega\ell^*}\big)^2 =
\kappa(\ell^*)\Lambda_z^4 \,,
\label{general-choice}
\end{eqnarray}
determined by the scaling of the denominator of a transverse part of
the phonon correlation function, that for $q_\perp=0$ reduces to the earlier
condition, Eq. (\ref{def-lstar2}).  On the other hand, for $q_z=0$ we
obtain the new condition (compare to equation (\ref{def-lstar2}))
\begin{eqnarray}
q_\perp\mbox{e}^{\omega\zeta\ell^*} =\Lambda_z^2\sqrt{\kappa/\mu} \,,
\label{lstar-qperp}
\end{eqnarray}
with the universal anisotropy exponent
\begin{eqnarray}
\zeta = 2 - \frac{\eta_\mu + \eta_\kappa}{2} \,.
\end{eqnarray}
Arguments analogous to the ones developed in the above paragraphs
show that the above condition, Eq. (\ref{lstar-qperp}),
leads to the following long-scale behavior of model parameters:
\begin{mathletters}
\begin{eqnarray}
\lambda(q_\perp) & = &
\lambda\,\Big(\frac{q_\perp}{\Lambda_z^2}
\sqrt{\frac{\mu}{\kappa}}\Big)^{\eta_\mu/\zeta} \,,
\label{lambda-qperp}
\\
\mu(q_\perp) & = & \mu\,\Big(\frac{q_\perp}{\Lambda_z^2}
\sqrt{\frac{\mu}{\kappa}}\Big)^{\eta_\mu/\zeta} \,,
\label{mu-qperp}
\\
\kappa(q_\perp) & = & \kappa\,\Big(\frac{q_\perp}{\Lambda_z^2}
\sqrt{\frac{\mu}{\kappa}}\Big)^{-\eta_\kappa/\zeta} \,,
\label{kappa-kperp}
\\
{\Delta_t}(q_\perp) & = & {\Delta_t}\,\Big(\frac{q_\perp}{\Lambda_z^2}
\sqrt{\frac{\mu}{\kappa}}\Big)^{-\eta_{\Delta_t}/\zeta} \,.
\label{Delta-qperp}
\end{eqnarray}
\end{mathletters}
Combining equations (\ref{kappa-qz})-(\ref{Delta-qz}),
(\ref{lambda-qz})-(\ref{mu-qz}) and
(\ref{lambda-qperp})-(\ref{Delta-qperp}), we finally see that the
general scaling behavior of the model parameters can be written in
the form:
\begin{mathletters}
\begin{eqnarray}
\lambda({\bf q}) & = & x_*\mu(q_z\xi_z^{NL})^{\eta_\mu}
f_\mu\big(q_\perp\xi_\perp^{NL}/(q_z\xi_z^{NL})^\zeta\big) \, ,
\\
\mu({\bf q}) & = & \mu(q_z\xi_z^{NL})^{\eta_\mu}
f_\mu\big(q_\perp\xi_\perp^{NL}/(q_z\xi_z^{NL})^\zeta\big) \, ,
\\
\kappa({\bf q}) & = & \kappa(q_z\xi_z^{NL})^{-\eta_\kappa}
f_\kappa\big(q_\perp\xi_\perp^{NL}/(q_z\xi_z^{NL})^\zeta\big) \, ,
\\
{\Delta_t}({\bf q}) & = & {\Delta_t}(q_z\xi_z^{NL})^{-\eta_{\Delta_t}}
f_{\Delta_t}\big(q_\perp\xi_\perp^{NL}/(q_z\xi_z^{NL})^\zeta\big) \, ,
\end{eqnarray}
\end{mathletters}
where we have taken $\Lambda_z=1/\xi_z^{NL}$ (which is a reasonable
choice since $\xi_z^{NL}$ is the shortest length scale one can go to before
perturbation theory can no longer be applied) and used the fact that
$\xi_\perp^{NL}=\big(\xi_z^{NL}\big)^2(\mu/\kappa)^{1/2}$.  The crossover
functions $f_\mu(x)$, $f_\kappa(x)$ and $f_{\Delta_t}(x)$
have the following asymptotic behavior:
\begin{eqnarray}
f_{\mu,\kappa,{\Delta_t}}(x)  =
\left\{\begin{array}{ll}
\mbox{const.}, \quad &\mbox{for} \quad x\ll 1\,,
\\
\nonumber
\\
x^{\eta_\mu/\zeta}, x^{-\eta_\kappa/\zeta},
\, x^{-\eta_{\Delta_t}/\zeta},\quad &\mbox{for} \quad x \gg 1\,.
\end{array}\right.
\end{eqnarray}

\subsubsection{Exact relation between exponents}

Since the recursion relation for $g$, Eq. (\ref{recg1}) involves
functions $F_{\mu,\kappa,{\Delta_t}}$ that respectively determined the
exponents ${\eta_\mu,\kappa,{\Delta_t}}$, it quite clearly leads to an
exact relation between these critical exponents.  To see this, we note
that at the fixed point $(x_*,g_*)$, $(dg/d\ell)|_{g_*}=0$. Combining
this with the definitions of the critical exponents,
Eqs. (\ref{def__lambda_mu}), (\ref{def__etakappa}) and
(\ref{def__etaDelta}), we obtain an exact exponent relation:
\begin{eqnarray}
(7-2d) + \eta_{\Delta_t} = \frac{3-d}{2}\;\eta_\mu +
\frac{7-d}{2}\;\eta_\kappa \,.
\label{exact}
\end{eqnarray}
From a more fundamental perspective, this exact scaling relation (Ward
identity) is also necessitated by the underlying rotational invariance
(that is spontaneously broken by the vortex lattice).   To see this, we
recall that rotational invariance requires that the form of the
nonlinear strain tensor $v_{\alpha\beta}$, Eq.  (\ref{def-strain}), be
preserved by the perturbation theory in these nonlinearities.
Performing the anisotropic rescaling in terms of the physical scaling
exponents and imposing this condition (contained in $\phi=2-\omega$),
we obtain the exact scaling relation, Eq. (\ref{exact}) above.

To order $\epsilon$, Eq. (\ref{exact}) is given by:
\begin{eqnarray}
2\epsilon + \eta_{\Delta_t} + \frac{1}{4}\eta_\mu =
\frac{7}{4}\,\eta_\kappa \,,
\end{eqnarray}
and is clearly satisfied by the exponents of equations
(\ref{eq0.739}), (\ref{eq0.207}) and (\ref{eq0.346}).  In $d=3$
it reduces to:
\begin{eqnarray}
\eta_{\Delta_t} = 2\eta_\kappa - 1 \,,
\end{eqnarray}
a relation that, in principle, should be experimentally testable.

\section{Topological Stability Of The Ramdomly Pinned Spontaneous Vortex Lattice}
\label{TopOrderApp}

We now turn our attention to the important question of whether
the new phase of the SV lattice predicted in the previous section, and that is
controlled by the low-temperature fixed point (\ref{fixedpoint}), is
stable against the proliferation of dislocations.
It has been shown that long-range translational order in the SV phase is 
destroyed by arbitrarily weak quenched disorder. This does not necessarily mean, 
however, that the topology of the phase is destroyed. It is the topology, or 
``connectivity'' of the phase, that is responsible for the phase's elasticity. 
If the topology is destroyed, then the elasticity is lost. Thus, if the SV
lattice remains topologically ordered, it will display SV solid-like
elasticity. It will, for example, still have a non-zero shear modulus and shear 
distortions in the $x\,y$ plane will still cost energy. We refer to a phase lacking long 
ranged order, but that still has topological order, as an elastic 
glass.\cite{GLD,DSFisher,GH} Of course, the topological order is eventually lost 
if the temperature or disorder strength become sufficiently 
large. For sufficiently large temperature $T$  or disorder strength $\Delta$,
the topological order of the putative elastic glass will be destroyed via
a thermodynamically sharp transition into a 
translationally {\em and} topologically disordered phase.

There are already examples of phase transitions between two phases which both
lack long-ranged order, the most famous being the Kosterlitz-Thouless 
transition in the $2d$ XY model.\cite{KT} In that problem, the transition  is 
associated not with the disappearance of long-ranged order, but rather with the 
unbinding of neutral pairs of topological defects with increasing temperature. 
The divergences associated with tilt disorder
that destroy the long-ranged translational order in the 
columnar elastic glass are of course much stronger (power-law) than in the $2d$ XY 
model where they are logarithmic. There are, however, examples of phases in much 
more strongly disordered systems in which topological defects nonetheless 
remain bound.\cite{Rokhsar} The possibility of such a transition from a 
translationally disordered smectic-A liquid crystal has also been considered by 
Radzihovsky and Toner. \cite{Radzihovsky-Toner}

\begin{figure}[t]
\includegraphics[width=8cm, height=5cm]{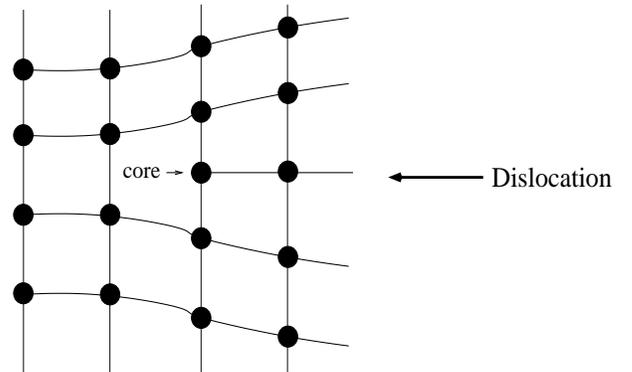}
\caption{
A square spontaneous vortex with an extra row of flux lines introduced from 
the right, along the $x$ axis. The core of the dislocation comes out of the page.
}
\label{disloc3}
\end{figure}

In general, the destruction of topological order in a phase occurs via the 
proliferation of defects. In the $2d$ XY model these defects are spin 
vortices. In the SV lattice the main mechanism for the destruction of 
topological order is through the   
proliferation of dislocations. An edge dislocation, which is an insertion 
of an extra row of flux lines is shown, for a square lattice, in 
Fig. \ref{disloc3}. The Burgers vector (the extra lattice vector) associated with this 
dislocation is ${\bf b}= a{\bf \hat{\bf y}}$. In the remainder of this Section,
we restrict ourselves to spatial dimensions $d=3$.

This edge dislocation will reduce the elasticity against shears in the  $yz$ 
plane. One could also have an edge dislocation corresponding to 
multiple extra inserted rows and in this case the Burgers vector will be a 
multiple of the lattice vector. If the rows of vortex lines are inserted in the $yz$ 
plane then the Burgers vector will be an integer number of 
lattice vectors in the ${\bf \hat x}$ direction. One could

\medskip

\noindent also have both rows inserted 
in the $yz$  and the $xz$ planes, and the corresponding Burgers vector would be 
a sum of lattice vectors. Of course, one can also consider dislocations in a 
hexagonal lattice, the only difference being that the lattice vectors do not 
simply point along ${\bf \hat x}$ and ${\bf \hat y}$. 

When the displacement field ${\bf u}({\bf r})$ is integrated around a closed 
curve enclosing the core of the edge dislocation shown in Fig. \ref{disloc3}, 
the result is not zero as it would be without a dislocation, but rather 
${\bf b} = {\bf a}_2 = a {\bf \hat y}$, {\em i.e.}, 
\begin{equation}
\oint d{\bf u} = {\bf a}_2\,.
\label{dislocation1}
\end{equation}
In general, a dislocation is a line or curve, with the property that an  
integral of $\bf {u}$ around any curve enclosing the dislocation core is some 
Burgers vector ${\bf b} = M {\bf a}_1 + N {\bf a}_2$, {\em i.e.},
\begin{equation}
\oint d{\bf u} = {\bf b} =  M {{\bf a}_1} + N {{\bf a}_2}\,,
\label{dislocation2}
\end{equation}
where $M$ and $N$ are integers. In the cases of square and hexagonal 
lattices, in which the lattice vectors are of equal length, 
Eq. (\ref{dislocation2}) can be rewritten as:
\begin{mathletters}
\begin{eqnarray}
\oint du_1 =  Ma \label{dislocation3a} \,,
\\
\oint du_2 =  Na \label{dislocation3b}\,,
\end{eqnarray}
\label{disl-3}
\end{mathletters}
where $u_i$ is the $i$'th component of ${\bf u}$, defined in the coordinate 
system with (not necessarily orthogonal) unit vectors 
$\hat{\bf a}_1= {{\bf a}_1}/a$, $\hat{\bf a}_2={{\bf a}_2}/a$, {\em i.e.}, 
${\bf u} = u_1 \hat{\bf a}_1 +u_2 \hat{\bf a}_2$. For a square lattice $i$ is 
either $x$ or $y$. 

Eqs. (\ref{disl-3}) apply for all types 
of dislocations. The edge dislocations, discussed above, have cores that run 
parallel to the vortex lines but in general dislocation lines can run in any 
direction. An extreme case is a screw dislocation which runs perpendicular to 
the vortices.

\subsection{The Defect Elastic Hamiltonian }
\label{CBG Defect Hamiltonian}

Eqs. (\ref{dislocation3a})-(\ref{dislocation3b}) imply that in the presence of 
a dislocation the ${\bf u}$ field is no longer single valued. Given that a single 
valued ${\bf u}$ field is obviously a prerequisite of a meaningful elastic 
theory it is necessary to introduce a ``cut'' in the following fashion. 
In order to remain single valued 
as the core of the dislocation is circulated, there must be points at which the 
value of ${\bf u}$ jumps discontinuously by an amount equal to the Burgers 
vector, ${\bf b}$. The cut is the locus of the points at which this jump occurs, 
and the value of ${\bf u}$ on either side of the cut differs by ${\bf b}$. In 
the absence of dislocations, the elasticity of the system can be described using 
a field
\begin{equation}
{{\bf v}_i} \equiv {\bf \nabla} u_i\;,
\label{vdefn}
\end{equation}
which is continuous and smooth almost everywhere. However, the presence of a cut means 
that ${{\bf v}_i}$ will be singular at the cut. The cut, however, does not 
represent anything {\em physical}, and therefore, nor does a singular ${\bf v}_i$. 
There is nothing singularly elastic happening in the real system as the 
(arbitrarily chosen) cut is crossed. Therefore, a modified ${{\bf v}_i}$, from 
which the unphysical, singular piece is removed, is used to construct an 
elastic theory in the presence of a dislocation. The modified ${{\bf v}_i}$ is 
equal to ${\bf \nabla} u_i$ everywhere except at the cut. 

Using Stokes law, Eqs. (\ref{dislocation3a})-(\ref{dislocation3b}) can be 
rewritten in differential form:
\begin{equation}
{\bbox\nabla} \times {\bf v}_i = {\bf m}_i\;.
\label{mdefn}
\end{equation}
Since ${{\bf v}_i}$ is no longer the gradient of a single valued field, it can 
have a non-zero curl. This is where the presence of the dislocation is manifest. In 
the absence of a dislocation the curl would be zero. The field ${\bf m}$ 
is defined as: 
\begin{equation}
{\bf m}_i({\bf r})=\sum_j\int a_i N_j {\bf t}_j({\bf r}_j(s_j))
\delta^3({\bf r}-{\bf r}_i(s_j)) d s_j\;,
\label{m}
\end{equation}
where $s_j$ parameterizes the $j$'th dislocation loop, ${\bf r}_j(s_j)$ is the 
position of that loop, ${\bf t}_j({\bf r}_j(s_j)$ is its local unit tangent, 
and $N_j$ the ``charge'' or number of excess lattice vectors along 
$\hat{\bf a}_i$ associated with the dislocation. Note that $N_j$ is independent of $s_j$, 
since the charge of a given line is constant along the line defect. It is 
important to note that ${{\bf v}_i}$ is {\em not} the $i$'th component of the 
three dimensional vector ${\bf v}$. Rather, it is one of the two vectors ${\bf v}_1$ 
and ${\bf v}_2$, constructed using Eq. (\ref{vdefn}). The same applies to 
${\bf m}_i$.

Eq. (\ref{mdefn}) implies that
\begin{equation}
{\bbox\nabla} \cdot  {\bf m}_i({\bf r})=0 \,,
\label{divm}
\end{equation}
which simply means that dislocation lines cannot end in the bulk of the sample:
they must either form closed loops or extend entirely through the system. It is 
useful to work with Fourier transformed fields:
\begin{mathletters}
\begin{eqnarray}
{{\bf v}_i}({\bf q}) =  i{\bf q} u_i({\bf q}) \,,
\label{FTv}\\
{{\bf m}_i}({\bf q}) =  i{\bf q} \times {{\bf v}_i}({\bf q})\,, \label{FTm}
\end{eqnarray}
\end{mathletters}
and work with the longitudinal and transverse components
${\bf v}_L({\bf q})$ and ${\bf v}_T({\bf q})$, 
defined by:
\begin{mathletters}
\begin{eqnarray}
{\bf v}_L({\bf q}) & = & i{\bf q} u_L({\bf q}) \,,
\label{vL}\\
{\bf v}_T({\bf q}) & = & i{\bf q} u_T({\bf q}) \,, \label{vT}
\end{eqnarray}
\end{mathletters}
where $u_L({\bf q})$ and $u_T({\bf q})$ are the longitudinal and transverse 
components of ${\bf u}({\bf q})$ given, in $d=3$, by
\begin{equation}
{\bf u}({\bf q}) =  u_L({\bf q}){\bf \hat q_\perp} + u_T({\bf q})({\bf 
\hat z}\times {\bf \hat q_\perp})
\label{uLT}\,. 
\end{equation}
The field ${{\bf v}_i}$ can be easily constructed from ${\bf v}_L({\bf q})$ and 
${\bf v}_T({\bf q})$ according to:
\begin{equation}
{{\bf v}_i}({\bf q}) =   v_L({\bf q})({\bf \hat q_\perp})_i 
+ v_T({\bf q})({\bf \hat z}\times {\bf \hat q_\perp})_i
\label{vLT}\,,
\end{equation}
where $({\bf \hat q_\perp})_i$ is the $i$'th component of ${\bf \hat q_\perp}$, 
and $({\bf \hat z}\times {\bf \hat q_\perp})_i$ is the $i$'th component of ${\bf 
\hat z}\times {\bf \hat q_\perp}$. Similarly,
\begin{equation}
{{\bf m}_i}({\bf q}) =  m_L({\bf q})({\bf \hat q_\perp})_i + m_T({\bf 
q})({\bf \hat z}\times {\bf \hat q_\perp})_i\label{mLT}\; . 
\end{equation}
In Fourier space the condition that dislocation lines must form closed loops or 
extend entirely through the sample, Eq. (\ref{divm}) gives: 
\begin{mathletters}
\begin{eqnarray}
{\bf q}\cdot{\bf m}_L({\bf q})=0 \label{FSdivmL} \,, 
\\
{\bf q}\cdot{\bf m}_T({\bf q})=0 \label{FSdivmT} \,,
\end{eqnarray}
\end{mathletters}
and the boundary condition, Eq. (\ref{mdefn}) corresponding to a topological 
defect is:
\begin{mathletters}
\begin{eqnarray}
i{\bf q}\times{\bf v}_L({\bf q})={\bf m}_L({\bf q}) \label{FSBCL} \,, 
\\
i{\bf q}\times{\bf v}_T({\bf q})={\bf m}_T({\bf q}) \label{FSBCT} \,. 
\end{eqnarray}
\end{mathletters}
The real space Hamiltonian to harmonic order in $d=3$ dimensions is given by Eq. (\ref{Hnonlinear})
that we rewrite here for definiteness: 
\begin{eqnarray}
H[{\bf u}({\bf r})]  =  \int d{\bf r}\; {1\over2}\Big[\Big(\lambda 
v_{\alpha\alpha}^2 &+& 2\mu v_{\alpha\beta}^2+ 
\kappa|\partial_z^2{\bf u}|^2\Big) 
\nonumber\\
&+& {\bf h}({\bf r})\cdot\partial_z{\bf u} \Big] \label{RSH}\;,
\end{eqnarray}
where the reader is reminded that ${\bf h}({\bf r})$ is the random tilt 
disorder, with short-ranged, isotropic correlations
\begin{equation}
\overline {h_\alpha({\bf r}) h_\beta({\bf r^{\prime}})} = \Delta_t \delta_{\alpha\beta}
\delta({\bf r}-{\bf r^{\prime}})
\label{RShcorr}\,.
\end{equation}
The above real space Hamiltonian can then be rewritten in Fourier space, where 
${\bf u}({\bf q})$ and ${\bf h}({\bf q})$ can both be decomposed into 
longitudinal and transverse pieces according to Eq. (\ref{uLT}). The longitudinal 
and transverse modes decouple and one obtains
\begin{equation}
H[{\bf u}({\bf q}),{\bf h}({\bf q})]=H_L[u_L({\bf q}),h_L({\bf q})] +
H_T[u_T({\bf q}),h_T({\bf q})]\label{HL+T}\;,
\end{equation}
with 
\end{multicols}
\begin{equation}
H_{L/T}[u_{L/T}({\bf q}),h_{L/T}({\bf 
q})]={1\over2}\sum_{\bf q} \left[ B_{L/T}\left(q_\perp^2 + 
{\lambda_{L/T}^2} q_z^4\right)|u_{L/T}({\bf q})|^2 - iq_z 
u({\bf q})h_{L/T}(-{\bf q})\right]
\label{HL/T}\,, 
\end{equation}
\begin{multicols}{2}\noindent
where $B_L=\lambda+2\mu$ , $B_T=\mu$ and  ${\lambda_{L/T}^2} = \kappa/B_{L/T}$. This 
Hamiltonian does not have any anomalous elasticity in it. It is instructive 
to perform the dislocation theory at the harmonic level first and then to 
generalize to a full anharmonic theory afterwards.

Using the decoupled Hamiltonians for $u_L({\bf q})$ and $u_T({\bf q})$ it is 
possible, using the constraints Eqs. (\ref{FSdivmL})-(\ref{FSdivmT}) and (\ref{FSBCL})-(\ref{FSBCT}) 
to obtain decoupled elastic Hamiltonians for the dislocation fields ${\bf m}_L({\bf q})$ 
and ${\bf m}_T({\bf q})$. The first step is to minimize $H_{L/T}[u_{L/T}({\bf q}),h_{L/T}({\bf q})]$ 
with respect to $u_{L/T}({\bf q})$. The Euler-Lagrange equation for $u_{L/T}({\bf q})$ is given by:
\begin{equation}
\left(q_\perp^2+{\lambda_{L/T}^2} q_z^4\right)u_{L/T}({\bf q})
-{iq_z h_{L/T}({\bf q})\over B_{L/T}} = 0\label{ELu}\;,
\end{equation}
and can be rewritten in terms of ${\bf v}_{L/T}({\bf q})$ as
\begin{equation}
{\bf q_\perp} \cdot {\bf v}_{L/T}^\perp({\bf q}) + 
{\lambda_{L/T}^2}q_z^3 v_{L/T}^z({\bf q})+{q_z 
h_{L/T}({\bf q})\over B_{L/T}} = 0\label{ELv}\;,
\end{equation}
where ${\bf v}_{L/T}^\perp({\bf q})$ and $v_{L/T}^z({\bf q})$ are the $\perp$ 
and $z$ components, respectively, of ${\bf v}_{L/T}({\bf q})$. The next step is 
to solve Eq. (\ref{ELv}) above and Eqs. (\ref{FSBCL})-(\ref{FSBCT}) for 
${\bf v}_{L/T}({\bf q})$. Doing this, one obtains:
\end{multicols}
\begin{equation}
{\bf v}_{L/T}({\bf q})={i{\bf q} \times {\bf m}_{L/T}({\bf q})\over q^2} 
+ {q_z {\bf q} \over \Gamma_q^{L/T}}
\left({i{({\bf m}_{L/T}}({\bf q}))_z\over q^2} (1-{\lambda_{L/T}^2}q_z^2) - {h_{L/T}({\bf q})\over 
B_{L/T} }\right)\label{v}\;,
\end{equation}
\begin{multicols}{2}\noindent
where $\Gamma_q^{L/T}$ is defined as
\begin{equation}
\Gamma_q^{L/T} \equiv q_\perp^2+{\lambda_{L/T}^2}q_z^4\label{Gamma}\;.
\end{equation}

These expressions for ${\bf v}_{L/T}^\perp({\bf q})$ can then be inserted back 
into the original elastic Hamiltonian of Eq. (\ref{HL+T}), which gives the defect 
Hamiltonian:
\begin{equation}
H_d=H_d^L[{\bf m}_L({\bf q}),h_L({\bf q})] +
H_d^T[{\bf m}_T({\bf q}),h_T({\bf q})]\label{HdL+T}\;,
\label{Hd}
\end{equation}
where
\end{multicols}
\begin{eqnarray}
H_d^{L/T}[{\bf m}_{L/T}({\bf q}),h_{L/T}({\bf q})]=
\sum_{\bf q} \bigg[ {\kappa q_z^2|{{\bf m}_{L/T}^\perp}({\bf q})|^2 + 
B_{L/T}|{{\bf m}_{L/T}^z}({\bf q})|^2 \over \Gamma_q^{L/T}} 
+ {{\bf m}_{L/T}}({\bf q})\cdot{\bf a}_{L/T}(-{\bf q})\bigg]
\label{HdL/T}\,,
\end{eqnarray}
\begin{multicols}{2}\noindent
and where ${\bf a}_{L/T}({\bf q})$ is related to $ h_{L/T}({\bf q})$ via: 
\begin{equation}
{\bf a}_{L/T}({\bf q}) = 
{ih_{L/T}({\bf q})({\bf \hat z} \times {\bf q}) \over 
\Gamma_q^{L/T}}\label{afield}\;. 
\end{equation}

\subsection{Duality Transformation}
\label{CBG Duality Transformation}

The defect Hamiltonian of Eq. (\ref{HdL/T}) makes it possible for us to study the 
statistical mechanics of the dislocations. We do this by performing a  
duality transformation. As will be seen below, to perform this transformation 
it is necessary to have a partition function in which the trace is taken over 
the real space dislocation fields. However, as will also be seen below, 
although this is possible, it is not necessary to rewrite the defect Hamiltonian 
above (which is expressed in terms of the fields ${\bf m}_{L/T}({\bf q})$) in 
terms of the fields ${\bf m}_1({\bf q})$ and ${\bf m}_2({\bf q})$. The reader is 
reminded that ${\bf m}_1({\bf r})$ and ${\bf m}_2({\bf r})$ are the dislocation 
loop fields associated with dislocations along ${{\bf a}_1}$ and ${{\bf a}_2}$ 
respectively, where ${{\bf a}_1}$ and ${{\bf a}_2}$ are basis 
vectors of the lattice.

Henceforth for simplicity a square lattice will be assumed, {\em i.e.}, ${\bf  a}_1 = a \hat{\bf x}$ 
and ${\bf a}_2= a \hat{\bf y}$. This assumption makes certain parts 
of the calculation more straightforward than they would be for arbitrary 
$\hat{\bf a}_1$ and $\hat{\bf a}_2$, {\em e.g.}, for a hexagonal lattice. This 
assumption may seem invalid given that the Hamiltonian that has been used thus 
far is that of a hexagonal lattice. However, the determination of topological 
stability is ultimately based upon {\em scaling} arguments. Since the scaling, 
{\em e.g.} of the propagators and correlation functions, is the same in the hexagonal 
and square lattices the answer, which depends {\em only} on this scaling,  
obtained using this innocuous simplification to a square lattice should be the 
same as that for the hexagonal lattice. This assumption would {\em not}, 
however, be so innocuous for a theory of the transition, where the structure of 
the lattice is important.

The first step in the duality transformation is to put the model on a simple 
cubic lattice (to make it well-defined at short distances); now, ${\bf 
m}_1({\bf r})$ and ${\bf m}_2({\bf r})$ are defined on the sites ${\bf r}$ of 
the lattice, and they take on values
\begin{equation}
{{\bf m}_i}({\bf r})={a_i \over d^2}\big(n_i^x({\bf r}),n_i^y({\bf r}),n_i^z({\bf 
r})\big)\;, \label{mdescrete}
\end{equation}
where $n_i^\alpha$ is the $\alpha$'th component of ${\bf n}_i$ and the Einstein 
summation convention has been suspended.  Each of $n_i^x$, $n_i^y$, and $n_i^z$ 
is an integer. For the square lattice considered here the lengths of the two 
basis  vectors are the same so that $a_1=a_2=a$. $d$ is the cubic lattice  
constant used in the discretization. The partition function for this model is 
then 
\begin{equation}
Z[\{{\bf h}\}]=\sum_i\sum_{\{{\bf m}_i({\bf r})\}}^\prime 
e^{-S[\{{{\bf m}_i}\}]}\;, \label{cbgZ}
\end{equation}
where 
\begin{equation}
S[\{{{\bf m}_i}\}]\equiv\beta\Big[H_d[\{{\bf m}_1,{\bf m}_2\}] + E_c\sum_{{\bf r},i}
|{{\bf m}_i}({\bf r})|^2\Big]\,,
\label{S}
\end{equation}
and the sum is over all integer-valued configurations of ${{\bf m}_i}$'s given by 
Eq. (\ref{mdescrete}), satisfying the dislocation line continuity constraint 
(which is denoted by the prime):
\begin{equation}
\nabla\cdot{{\bf m}_i}=0\;,
\label{continuity2}
\end{equation}
where the divergence now represents a {\it lattice} divergence, $H_d$ is 
obtainable from Eq. (\ref{HdL/T}), and a core energy term $E_c\sum_{{\bf 
r},i}|{{\bf m}_i}({\bf r})|^2$ has been added, to account for energies near the 
cores of the 
defect line that are not accurately treated by the above continuum elastic 
theory. The symmetry of the square lattice dictates that the core energies be the 
same for both ${\bf m}_1({\bf r})$ and ${\bf m}_2({\bf r})$. The reader's 
attention is directed to the fact that the partition function still depends 
implicitly on the configuration of the random fields $\{\bf h\}$.

To proceed, the constraints $\nabla\cdot{\bf m}_1=0$ and $\nabla\cdot{\bf m}_2=0$ 
are enforced by introducing new auxiliary fields $\theta_1({\bf r})$ 
and $\theta_2({\bf r})$, and rewriting the partition function Eq. (\ref{cbgZ}) as: 
\begin{eqnarray}
Z & = & \prod_{\bf r}\int d\theta_1({\bf r}) d\theta_2({\bf r})
\nonumber\\
&\times&\sum_i\sum_{\{{{\bf m}_i}({\bf r})\}} e^{-S[\{{{\bf m}_i}\}] + 
i\sum_{{\bf r},i}\theta_i({\bf r})\nabla\cdot{{\bf m}_i}({\bf r})d^2/a}\;,
\label{cbgZ2}
\end{eqnarray}
where the sum over $\{{\bf m}_i\}$ is now unconstrained (and no longer has a 
prime), the constraint $\nabla\cdot{{\bf m}_i}=0$ is enforced by integration over 
$\theta_i$, since 
\begin{equation}
\delta(\nabla\cdot{{\bf m}_i})=\int_0^{2 \pi} d\theta_i({\bf r})
e^{i\theta_i({\bf r})\nabla\cdot{{\bf m}_i}({\bf r})d^2/a}\;,
\end{equation}
where the $\delta$ is a Kronecker delta since ${\bf m}_i$, and, hence, 
$\nabla\cdot{{\bf m}_i}$, is integer valued. 

Now one can ``integrate'' (actually sum) by parts, and rewrite 
\begin{eqnarray}
\sum_{{\bf r},i}\theta_i({\bf r})\nabla\cdot{{\bf m}_i}({\bf r}) & = &
-\sum_{{\bf r},i}{{\bf m}_i}({\bf r})\cdot\nabla\theta_i({\bf r})
\nonumber\\
& + &\;\mbox{surface terms}\;.
\end{eqnarray}
The next step is to introduce dummy gauge fields ${\bf A}_1$ and ${\bf A}_2$ to 
mediate the long-ranged interaction between the defects loops ${\bf m}_1$ and 
${\bf m}_2$ in the Hamiltonian Eq. (\ref{Hd}). This is accomplished using the 
Hubbard-Stratonovich transformation and rewriting the partition function as:
\begin{eqnarray}
Z &=&\prod_{\bf r}\int d\theta_1({\bf r}) d\theta_2({\bf r})d{\bf A}_1({\bf 
r})d{\bf A}_2({\bf r})
\nonumber\\
&\times&\sum_i \sum_{\{{{\bf m}_i}({\bf r})\}} e^{-S[\{{\bf m}_i\},\theta_i,{\bf A}_i]} 
\delta(\nabla\cdot{\bf A}_i)\;,
\label{cbgZ3}
\end{eqnarray}
with
\end{multicols}
\begin{eqnarray}
S & = & {1 \over T}\sum_{{\bf r},i}\Big\{{{\bf m}_i}
({\bf r})\cdot\Big[-i{Td^2 \over a}{\bbox\nabla}\theta_i({\bf r})+d^3\big(
i{\bf A}_i({\bf r})+{\bf a}_i({\bf r})\big)\Big] + E_c|{{\bf m}_i}|^2\Big\}
\nonumber\\
&+&{1\over2}\sum_{\bf q}\Big[{\Gamma_q^L\over \kappa q_z^2}
|{\bf A}_L^\perp({\bf q})|^2 
+ {\Gamma_q^L\over B_L}|{\bf A}_L^z({\bf q})|^2 + 
{\Gamma_q^T\over \kappa q_z^2}|{\bf A}_T^\perp({\bf q})|^2 + 
{\Gamma_q^T\over B_T}|{\bf A}_T^z({\bf q})|^2\Big]\;, 
\label{S2} 
\end{eqnarray}
\begin{multicols}{2}\noindent
where ${\bf A}_L({\bf q})$ and ${\bf A}_T({\bf q})$ are defined in the same 
way as ${\bf v}_{L/T}({\bf q})$ and ${\bf m}_{L/T}({\bf q})$, {\em i.e.}, 
\begin{mathletters}
\begin{eqnarray}
{\bf A}_L({\bf q}) &=& {\bf \hat q}_1 {\bf A}_1({\bf q}) + {\bf \hat q}_2 
{\bf A}_2({\bf q})
\label{AL} \,, 
\\ 
{\bf A}_T({\bf q}) &=& ({\bf \hat z}\times{\bf \hat q})_1 {\bf A}_1({\bf q}) + 
({\bf \hat z}\times{\bf \hat q})_2 {\bf A}_1({\bf q})
\label{AT} \,,
\end{eqnarray}
\end{mathletters}
where $({\bf \hat q_\perp})_i$ is the $i$'th component of ${\bf \hat q_\perp}$ 
and $({\bf \hat z}\times {\bf \hat q_\perp})_i$ is the $i$'th component of ${\bf 
\hat z}\times {\bf \hat q_\perp}$, and where ${\bf a}_i({\bf r})$ is the Fourier 
transform of 
\begin{equation}
{\bf a}_i({\bf q}) =  {\bf a}_L({\bf q})(\hat{\bf q}_\perp)_i + 
{\bf a}_T({\bf q})(\hat{\bf z}\times \hat{\bf q}_\perp)_i
\label{aLT}\,.
\end{equation}

The two goals of all of these manipulations have now been achieved: the sum on 
$\{{{\bf m}_i}({\bf r})\}$ is now unconstrained, and the sum on each site over 
${{\bf m}_i}({\bf r})$ is now decoupled from that on every other site. 
Furthermore, this sum is readily recognized to be nothing more than the 
``periodic Gaussian'' first used by Villain.\cite{Natterman} The partition 
function Eq. (\ref{cbgZ3}) can thus be rewritten: 
\end{multicols}
\begin{eqnarray}
Z & = &\prod_{\bf r}\int d\theta_1({\bf r})d\theta_2({\bf r}) d{\bf A}_1({\bf r})
d{\bf A}_2({\bf r}) \delta(\nabla\cdot{\bf A}_1({\bf r}))
\delta(\nabla\cdot{\bf A}_1({\bf r})) \nonumber\\
&\times&\exp\Bigg\{-\sum_{{\bf r},i,\alpha} V_p[\theta_i({\bf r + \hat{x}_\alpha})-
\theta_i({\bf r})-A_i^\alpha({\bf r})+i a_i^\alpha({\bf r})]
\nonumber\\
& - &{1\over2}\sum_{\bf q}\bigg[{\Gamma_q^L\over \kappa q_z^2}|{\bf A}_L^\perp({\bf 
q})|^2+ {\Gamma_q^L\over B_L}|{\bf A}_L^z({\bf q})|^2 
+ {\Gamma_q^T\over \kappa q_z^2}|{\bf A}_T^\perp({\bf q})|^2 + {\Gamma_q^T\over B_T}
|{\bf A}_T^z({\bf q})|^2\bigg]\Bigg\}\,,
\label{Z4}
\end{eqnarray}
\begin{multicols}{2}\noindent
where $A_i^\alpha({\bf r})$ is the $\alpha$'th component of 
${\bf A}_i({\bf r})$ and the well-known $2\pi$-periodic Villain 
potential $V_p(x)$ 
\begin{equation}
e^{-V_p(x)}\equiv\sum_{n=-\infty}^\infty e^{-n^2 E_c/T + i x n}\;
\label{Vp}
\end{equation}
has the usual property that the {\it smaller} $E_c/T$ is ({\em i.e.}, the {\it 
higher} the temperature in the original random-tilt smectic model), the sharper 
the potential minima. This can be seen by looking at two extremes. When $E_c/T$ 
is zero, the Villain potential is a just a periodic set of delta functions. When 
it is very large, then the $n=0,\pm 1$ terms of the series dominate and the 
potential is a constant plus a cosine of $x$ which is obviously not as sharp as 
a set of delta functions. Furthermore, the amplitude of the cosine is 
proportional to $e^{-E_c/T}$ and hence gets smaller as $t$ is {\em reduced}. 
Thus {\it raising} the temperature in the original model is like {\it lowering} 
the temperature in the dual model Eq. (\ref{Z4}). It is precisely this familiar 
temperature inversion associated with duality that leads to an {\it inverted} XY 
transition for bulk smectics \cite{Toner} and three-dimensional disorder-free 
charged superfluids \cite{HLM}. It also plays an important role here, as we 
shall see in a moment.

Standard universality arguments imply that replacing the periodic potential 
$V_p(x)$ in Eq. (\ref{Z4}) by {\it any} other non-singular periodic function 
should not change the universality class of the transition. In particular, we 
could replace $V_p(x)$ by $\cos(x)$. The resultant model would be precisely 
the ``fixed length'' version of the ``soft spin'', or Landau-Ginsburg-Wilson 
model, with the {\em complex} ``action''
\end{multicols}
\begin{eqnarray}
S &=&\sum_{\{{\bf r}\}}\Big\{{c\over 2}\Big[
\nabla+{ad\over T}\big(i{\bf A}_i + {\bf a}_i\big)\Big]\psi_i
\cdot\Big[\nabla-{ad\over T}\big(i{\bf  A}_i + {\bf a}_i\big) \Big]\psi_i^* +t|\psi_i|^2+u |\psi_i|^4
+ w|\psi_1|^2|\psi_2|^2\Big\}
\nonumber\\
& + &{1\over2}\sum_{\bf q}\Big[
{\Gamma_q^L\over \kappa q_z^2}|{\bf A}_L^\perp({\bf q})|^2 
+ {\Gamma_q^L\over B_L}|{\bf A}_L^z({\bf q})|^2 + {\Gamma_q^T\over \kappa q_z^2}
|{\bf A}_T^\perp({\bf q})|^2 
+ {\Gamma_q^T\over B_T}|{\bf A}_T^z({\bf q})|^2\Big]\;,
\label{S3}
\end{eqnarray}
\begin{multicols}{2}\noindent
where $\psi_i({\bf r})$ is a complex field whose phase is $\theta_i({\bf r})$, 
and the reduced temperature $t$ and quartic coupling $u$ are parameters of the 
model (and are the same for $\psi_1({\bf r})$ and  $\psi_2({\bf r})$, because of 
the symmetry of the square lattice). Because of the duality transformation's 
inversion of the temperature axis, the reduced temperature $t$ is a 
monotonically {\it decreasing} function of the temperature $T$ (of the 
original dislocation loop model), which vanishes at the mean-field transition 
temperature $T_{MF}$ of the fixed length model Eq. (\ref{Z4}).

Universality also implies that this ``soft-spin'' model should be in the same 
universality class as the fixed length model Eq. (\ref{Z4}). Therefore  model 
Eq. (\ref{S3}), shall henceforth be used because it is more straightforward to 
analyze perturbatively.

Before undertaking that analysis, it is worth noting another consequence of 
duality inversion of the temperature axis: the {\it ordered} phase of the dual 
model Eq. (\ref{S3}) corresponds to the {\it disordered} ({\em i.e.}, dislocation loops 
unbound) phase of the original dislocation loop gas model. That is, the low 
dual-temperature phase described by
\begin{equation}
\langle\psi({\bf r})\rangle\neq 0\;
\end{equation}
corresponds to the topologically {\em disordered}, dislocation unbound phase of 
the SV lattice.

\subsection{Analysis Of The Dual Model}
\label{CBG Analysis Of The Dual Model}

Disorder is included in Eq. (\ref{S3}) through the quenched gauge-fields 
${\bf a}_1({\bf r})$ and ${\bf a}_2({\bf r})$, which are related to the random tilt 
field ${\bf h}({\bf r})$ by Eqs. (\ref{afield})-(\ref{aLT}). The partition 
function 
\begin{eqnarray}
Z[\{{\bf h}\}] & = &\int [d\psi_1d\psi_2][d{\bf A}_1d{\bf A}_2]
e^{-S[\psi_i,{\bf A}_i,{\bf h}]}
\nonumber\\
&\times&\delta(\nabla\cdot{\bf A}_1)\delta(\nabla\cdot{\bf A}_2)\;,
\label{Z5}
\end{eqnarray}
with $S$ given by Eq. (\ref{S3}), is thus an implicit function of the random tilt 
field configuration $\{{\bf h}({\bf r})\}$. 

The dependence of the partition function $Z$ on the quenched field  $\{{\bf 
h}\}$ is dealt with by using the replica trick, Eq. (\ref{trick}). Doing this leads to the 
replicated partition function:
\begin{eqnarray}
\overline{Z^n}&=&\int[d{{\bf a}_1}][d{{\bf a}_2}]\prod_{a=1}^n 
[d\psi_{1a}][d\psi_{2a}][d{\bf A}_{1a}][d{\bf A}_{2a}]
\nonumber\\ 
&\times& e^{-S_r[\psi_{ia},{\bf A}_{ia},{\bf a}_i]}
P[{\bf a}_1]P[{\bf a}_2] 
\delta(\nabla\cdot{\bf A}_1)\delta(\nabla\cdot{\bf A}_2)\;, 
\nonumber\\
\label{Zr}
\end{eqnarray}
with:
\end{multicols}
\begin{eqnarray}
S_r & = &\sum_{\bf r}\sum_{a=1}^n\Big\{ {c\over2}\Big[{\bbox\nabla}+{ad\over T} 
\big(i{\bf A}_{ia} + {\bf a}_i\big)\Big]\psi_{ia}
\cdot
\Big[{\bbox\nabla} -{ad\over T}\big(i{\bf A}_{ia} + {\bf a}_i\big)\Big]\psi_{ia}^* 
+ t|\psi_{ia}|^2+u |\psi_{ia}|^4 +w|\psi_{1a}|^2|\psi_{2a}|^2 \Big\}
\nonumber\\
& + &{1\over 2}\sum_{\bf q}\Big[{\Gamma_q^L\over \kappa q_z^2}|
{\bf A}^\perp_{La}({\bf q})|^2 + {\Gamma_q^L\over B_L}|{\bf A}^z_{La}({\bf q})|^2 
+ {\Gamma_q^T\over \kappa q_z^2}|{\bf A}^\perp_{Ta}({\bf q})|^2 + 
{\Gamma_q^T\over B_T}|{\bf A}^z_{Ta}({\bf q})|^2\Big]\,.
\label{Sr}
\end{eqnarray}
Note that ${\bf A}_{1a}$ is the $a$'th replica of the ${\bf A}_1$ field, not the $a$th component of ${\bf A}_1$.
The probability distribution $P[{\bf a}_i]$ of the field ${\bf a}_i$ in Eq. (\ref{Zr}) is 
Gaussian, defined by Eq. (\ref{RShcorr}). Thus the distributions of $P[{{\bf a}_1}]$ 
and $P[{{\bf a}_2}]$ are completely specified by the averages 
$\overline{a_1^i({\bf q})a_1^j(-{\bf q})}$ and $\overline{a_2^i({\bf q})a_2^j(-{\bf q})}$ respectively. 
These are easily evaluated using the 
relations  (\ref{afield}) and (\ref{aLT}). For ${\bf a}_1$:
\begin{eqnarray}
\overline{a_1^i({\bf q})a_1^j(-{\bf q})}&=&
\overline{a_L^i({\bf q})a_L^j(-{\bf q})}(({\bf \hat q_\perp})_1)^2
+\overline{a_T^i({\bf q})a_T^j(-{\bf q})}(({\bf \hat z}\times{\bf \hat q_\perp})_1)^2
+ 2\overline{a_L^i({\bf q})a_T^j(-{\bf q})}({\bf \hat q_\perp})_1({\bf 
\hat z}\times{\bf \hat q_\perp})_1
\nonumber\\
&=&\delta_{ij}^\perp\Delta_t q_\perp^2\left[{(({\bf \hat 
q_\perp})_1)^2\over(\Gamma_q^L)^2}+{(({\bf \hat z} \times {\bf \hat  
q_\perp})_1)^2\over(\Gamma_q^T)^2}\right]\;,
\label{aa1}
\end{eqnarray}
\begin{multicols}{2}\noindent
where the second equality was obtained using Eq. (\ref{RShcorr}),
and where we defined $\delta_{ij}^\perp =
\delta_{ij} - \delta_{iz}\delta_{jz}$.
Similarly,
\begin{equation}
\overline{a_2^i({\bf q})a_2^j(-{\bf q})}=\delta_{ij}^\perp\Delta_t 
q_\perp^2\left[{(({\bf \hat q_\perp})_2)^2\over(\Gamma_q^L)^2}+{(({\bf \hat 
z}\times{\bf \hat  q_\perp})_2)^2\over(\Gamma_q^T)^2}\right]\;.
\label{aa2}
\end{equation}
One now must consider the statistical mechanics of the model defined by 
Eqs. (\ref{Zr}) through (\ref{aa2}), in the limit $n\rightarrow 0$. 

Since it is only necessary to determine whether the topologically 
ordered columnar elastic glass phase is stable against disorder induced dislocation loop 
unbinding, a complete analysis of the critical properties of the
dislocation loop unbinding is unnecessary. Thus, it is sufficient to follow the 
example of Radzihovsky and Toner,\cite{Radzihovsky-Toner} and consider only the limit in 
which $u\ll 1$. As a result, the $\psi_i$  fluctuations are subdominant to to 
those of the gauge field ${\bf A}_i$ allowing  a mean field treatment of $\psi_i$. In this 
case the remaining gauge fluctuations may be treated {\em exactly}. The 
stability of the columnar elastic glass is assessed by studying the effect of 
diagrammatic corrections on the reduced dual temperature $t$. The lowest order 
contribution to the renormalized {\em dual} temperature $t_R$ comes from the 
average of the ``diamagnetic'' terms

\begin{figure}[t]
\includegraphics[width=8cm, height=2.5cm]{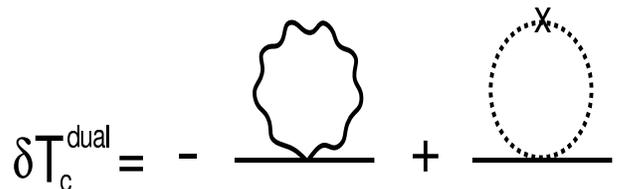}
\caption{
The Feynman diagrams that dominate the renormalization 
of the reduced dual temperature $t$.
}
\label{DualRenorm}
\end{figure}
\noindent
\begin{equation}
\delta S = {c a^2 d^2 \over 2 T^2} \sum_i\sum_{\bf q}\left(\langle|{\bf A}_i|^2\rangle
- \overline{|{\bf a}_i|^2}\right)|\psi_i|^2\;,
\label{deltaS}
\end{equation}
graphically illustrated in Fig. \ref{DualRenorm}.

Taking into account the symmetry of the square lattice, and first
angular averaging in the $\perp$ plane, the correction to the renormalized 
reduced dual temperature is, in $d=3$ dimensions,
\begin{eqnarray}
t_R & = & t_0 + {ca^2d^2\over T^2}\int_{\bf q}\Bigg[\kappa q_z^2\left(
{1\over\Gamma_q^L}+{1\over\Gamma_q^T}\right) + {B_L\over\Gamma_q^L} + 
{B_T\over\Gamma_q^T} 
\nonumber\\
& - & \Delta_t  
q_\perp^2\left({1\over(\Gamma_q^L)^2}+{1\over(\Gamma_q^T)^2}\right)\Bigg]\;,
\label{tR}
\end{eqnarray}
where the first four terms proportional to $1/\Gamma_q^{L/T}$ come from the 
first graph in Fig. \ref{DualRenorm}, with the wiggly line representing 
${\bf A}_i$ fluctuations, while the terms proportional to $\Delta_t$ come from the 
second graph  with the internal dotted line representing the quenched gauge 
field ${\bf a}_i$. The negative sign of the disorder contribution leads to an 
{\em increase} in the {\em dual} $T_c$ and can be traced back to the fact that 
the action $S_r$ in Eq. (\ref{Sr}) is complex-valued.

Performing this integral, one finds no infrared divergence
in spatial dimensions $d=3$ (the only dimensions in which the analysis
of this section applies), indicating that the 
renormalized $t_r$ remains finite as the system size diverges. A finite $t_r$ 
implies that there {\em is} a temperature regime in which the dual parameter is 
in its {\em disordered} phase, which in turn implies that there is a temperature 
regime in which dislocations in the columnar phase remain bound. Thus, in the 
harmonic model, there {\em is} a stable columnar elastic glass.

\subsection{Treating The Nonlinearities}
\label{CBG Treating The Nonlinearities}

Of course, one would like to also perform a full {\em anharmonic} theory of 
dislocations (since the anharmonic effects are obviously important) but 
unfortunately such a theory is simply intractable. In particular, the fact 
that, in an anharmonic theory, the interaction energy between dislocations {\em 
cannot} be written as a sum of pairwise interactions (since their fields do not 
simply add) makes it impossible to even write down a general expression for the 
energy of an arbitrary dislocation configuration. At best, one might hope to be 
able to write down the energy for a few simple, high symmetry configurations 
({\em e.g.}, a simple, straight dislocation line). Such specialized results would be 
of no use in a full statistical theory of defect unbinding, which requires 
consideration of very complicated, tangled configurations of dislocations, 
which for entropic reasons dominate the free energy near the dislocation 
unbinding transition.

Furthermore, even {\em if} one {\em could} write down the anharmonic energy for 
an arbitrary dislocation configuration, it would presumably be anharmonic in 
the dislocation fields ${\bf m}_i({\bf r})$, and hence, those fields would 
{\em not} be decoupled by a simple Hubbard-Stratonovich transformation, as they can 
in the harmonic model.

For all these reasons, a completely honest treatment of dislocations in the 
full, anharmonic model is extremely difficult. The next best, tractable, 
(but uncontrolled) approach, introduced by Radzihovsky and Toner,\cite{Radzihovsky-Toner} is to simply 
replace the elastic {\em constants} $\mu$, $\lambda$ and $\kappa$, and the tilt 
disorder variance $\Delta_t$ in Eq. (\ref{tR}) for the renormalized dislocation 
unbinding transition temperature, with the renormalized wavevector dependent 
moduli $\mu({\bf q})$, $\lambda({\bf q})$, $\kappa({\bf q})$, and $\Delta_t({\bf q})$, 
derived in Sec. \ref{subsec:matching}. One can hope to justify this procedure
by doing an RG matching calculation.

Doing this, simple power counting shows that in the IR limit ({\em i.e.}, at long 
length scales) it is the negative, disorder, contribution to the renormaliztion 
of the dual temperature which dominates, {\em i.e.}, 
\begin{equation}
\delta t=-{\cal O}(1)\times{ca^2d^2\over T^2}\int_{\bf q}{\Delta_t 
({\bf q}) q_\perp^2\over(q_\perp^2 + {\lambda_T^2}({\bf q})q_z^4)^2}\,,
\label{tRanom}
\end{equation}
where $\Delta_t({\bf q})$ is given by Eq. (\ref{Delta-qz}) and ${\lambda_T^2}({\bf 
q})=\kappa({\bf q})/\mu({\bf q})$. The contribution of the piece  
proportional to $1/(\Gamma_q^{L})^2$ is absorbed into the $O(1)$ coefficient, 
since in the IR limit $\Gamma_q^L$ and $\Gamma_q^T$ scale the same way with 
${\bf q}$ (because $\mu({\bf q})/\lambda({\bf q})$ is independent of ${\bf q}$). 
Imposing an infrared cutoff $q_z>L^{-1}$ on the wavevector integral in 
Eq. (\ref{tRanom}) one finds that the infrared piece scales with the linear size of the system 
$L$ like
\begin{equation}
{\delta}t_{IR} \approx -\left({L \over \xi_z^{NL}}\right)^\gamma{ca^2d^2\Delta_t 
\over {\lambda_T^2}}  \;,
\label{deltaIR}
\end{equation}
with
\begin{equation}
\gamma=\eta_\Delta-1\label{gammapower1}\;.
\end{equation}
If $\gamma>0$ the {\em negative} correction to $t$ diverges, corresponding to 
an infinite {\em downward} renormalization of the dual temperature, even for 
arbitrarily weak disorder, implying that dislocations in the SV lattice are 
{\em always} unbound. On the other hand, if $\gamma<0$ then there is no 
infinite negative divergence of the dual temperature implying that the columnar 
elastic glass is stable in a finite range of temperatures.

Using the exact scaling relation in $d=3$, Eq. (\ref{exact}) allows $\gamma$ to 
be reexpressed in terms of the more experimentally accessible $\eta_\kappa$ exponent. 
Doing this one finds 
\begin{equation}
\gamma=2(\eta_\kappa-1)\label{gammapower2}\;.
\end{equation}
Thus for $\eta_\kappa<1$ the columnar elastic glass phase is stable. In $d=3$, the 
estimate of $\eta_\kappa$, found in the earlier $\epsilon=7/2-d$ expansion, is 
$\eta_\kappa= 0.74$, so that a columnar elastic glass is indeed predicted to be stable 
against dislocation unbinding.

\begin{figure}[t]
\includegraphics[width=8cm, height=5cm]{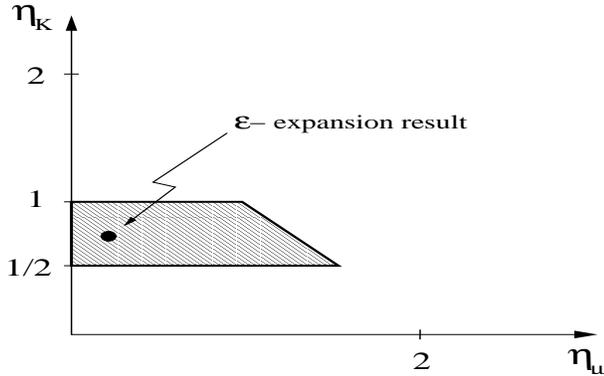}
\caption{
The region of $\eta_\kappa$ and $\eta_\mu$ that satisfy the criteria 
Eqs. (\ref{CBGcrit1})-(\ref{CBGcrit3})  for a long-ranged 
orientationally ordered, stable (at sufficiently low disorder strength) ``columnar 
elastic glass'' phase.
}
\label{ExpntsCrit}
\end{figure}

Thus, in summary, for a topologically ordered phase to exist the anomalous 
exponents must satisfy three distinct criteria, corresponding to a 
destruction of long-ranged {\em translational} order, the existence of 
long-range {\em orientational} order and stability against dislocation 
unbinding. These three criteria are respectively:
\begin{mathletters}
\begin{eqnarray}
\eta_\mu+\eta_\kappa>0 \,, 
\label{CBGcrit1}\\
\eta_\mu+\eta_\kappa<2 \,, 
\label{CBGcrit2}
\\
\eta_\kappa<1 \,,
\label{CBGcrit3}
\end{eqnarray}
\end{mathletters}
and are along with a fourth bound that $\eta_\kappa>1/2$ (which is equivalent, via 
the exact scaling relation in $d=3$, to the condition that $\eta_{\Delta_t} >0$), 
summarized graphically, in Fig. \ref{ExpntsCrit}. The exponents found in the 
$\epsilon=7/2-d$ expansion satisfy all of these bounds; hence a topologically ordered
phase of the SV lattice is possible in three dimensions. We shall term this phase
(which is analogous to the Bragg glass phase of vortices in disordered
type II superconductors\cite{GLD}) a ``columnar elastic glass" (CEG) phase.

\section{Magnetization curve and other phenomenology}
\label{Exp_consq}

We expect that for length scales longer than $\xi^{NL}$ the strong,
power-law anomalous elasticity will manifest itself in the behavior of
all physical observables of the SV solid, such as, {\em e.g.}, the width of
the structure function peak $S({\bf q})$ measured in neutron
scattering and the behavior in its tails. Anomalous elasticity also
implies that the stress ($\sigma$)-strain ($\epsilon_c$) relation
becomes non-Hookean, {\em {\em i.e.}}, nonlinear, even for arbitrarily small
applied stress $\sigma$.  (Note here, that $\sigma$ and $\epsilon_c$
should not be confused with the Poisson ratio, $\sigma_p$ and
$\epsilon$-expansion parameter, respectively). In this section we will
discuss the expected behavior of the magnetization curve and the
structure function.

\subsection{Nonlinear Magnetization Curve}

The origin of the non-Hookean stress-strain behavior can be seen most
clearly for a purely compressive stress,
$\sigma_{ij}=\sigma\delta_{ij}$, (which, as we shall see shortly,
arises from an externally applied magnetic field ${\bf H}$ along the
flux lines). Such a stress adds a term proportional to
\begin{eqnarray}
-\sigma\nabla\cdot{\bf u} \,,
\end{eqnarray}
and a term proportional to
\begin{eqnarray}
\sigma(\partial_z {\bf u})^2
\end{eqnarray}
(which explicitly breaks rotational invariance) to the original
Hamiltonian. The actual values of the dimensionless constants of
proportionality are not important and for simplicity we set them equal
to unity. In Fourier space, the second, symmetry breaking term becomes
\begin{eqnarray}
\sigma q_z^2 |{\bf u}({\bf q})|^2 \,,
\end{eqnarray}
and begins to dominate over the
$\kappa({\bf q}) q_z^4|{\bf u}({\bf q})|^2$ term once
\begin{eqnarray}
\sigma q_z^2\ge \kappa({\bf q})q_z^4 \; .
\end{eqnarray}
This clearly happens for $q_z$'s smaller than a critical $q_c$ given
by
\begin{eqnarray}
\sigma q_c^2=\kappa(q_z=q_c,{\bf q}_{\perp}=0)q_c^4 \;.
\end{eqnarray}
Using equation (\ref{kappa-qz}) for $\kappa({\bf q})$ in this
expression and solving for
$q_c$, we find
\begin{eqnarray}
q_c(\sigma)=\left(\frac{\sigma}{\kappa}\right)^{1/\delta}
(\xi^{NL}_z)^{\eta_{\kappa}/\delta} \,,
\label{q_c}
\end{eqnarray}
where we defined
\begin{eqnarray}
\delta = 2-\eta_{\kappa} \,.
\end{eqnarray}
The important thing to note about this result is that it ({\em
singularly}) depends on the applied stress $\sigma$.

Now, for $q_z>q_c(\sigma)$ the new, stress-induced rotational symmetry
breaking term $\sigma q_z^2|{\bf u}({\bf q})|^2$ in the Hamiltonian is
negligible (subdominant) relative to the vortex curvature energy
$\kappa({\bf q}) q_z^4|{\bf u}({\bf q})|^2$.  However, on longer
length scales, $q_z\ll q_c(\sigma)$, the $\sigma$-term dominates,
suppressing the SV solid fluctuations and hence cutting off the
anomalous elasticity. Therefore, the elastic moduli saturate, for $q_z
\ll q_c(\sigma)$, at their values at $q_z=q_c(\sigma)$, $q_{\perp}=0$,
given by
\begin{mathletters}
\begin{eqnarray}
\mu({\bf q} \to 0) &\longrightarrow& \mu(q_z=q_c,{\bf q}_{\perp}= 0) \,,
\\
\lambda({\bf q}\to 0) &\rightarrow& \lambda(q_z=q_c,{\bf q}_{\perp}= 0) \,,
\end{eqnarray}
\end{mathletters}
and, hence, both are proportional to $q_c^{\eta_\mu}$.  Since $q_c$ is
a function of the applied stress $\sigma$, as shown in equation
(\ref{q_c}), this implies that the long wavelength limit of the $2d$
bulk modulus:
\begin{eqnarray}
C(\sigma) & = & \mu(q_c)+\lambda(q_c),
\nonumber\\
& = & \mu (1+x_*)\;[q_c(\sigma)\xi^{NL}_z]^{\eta_\mu} \,,
\nonumber\\
&\approx& \mu[q_c(\sigma)\xi^{NL}_z]^{\eta_\mu} \,,
\nonumber\\
& = &\mu\left(\frac{\sigma}{\sigma_{NL}}\right)^\beta \,,
\end{eqnarray}
with $\sigma\ll\sigma_{NL} =
\kappa(\xi^{NL}_z)^{-2+\eta_\kappa-\eta_\kappa/\eta_\mu}$,
and
\begin{eqnarray}
\beta=\frac{\eta_\mu}{2-\eta_\kappa} \,.
\end{eqnarray}

To obtain the response of the system to a uniform, isotropic
compressive stress within the plane transverse to the spontaneous vortex
axis, the $2d$ bulk modulus $C$ times the compressive strain
$\epsilon_c\equiv\nabla\cdot{\bf u}$ must be balanced against the applied
compressive stress, {\em {\em i.e.}}, $C(\sigma)\epsilon_c = \sigma$.  Hence, we
find that for a stress $\sigma\le\sigma_{NL}$, the SV glass exhibits
nonlinear (non-Hookean) strain-stress relation:
\begin{eqnarray}
\epsilon_c(\sigma) & = &\sigma/C(\sigma) \,,
\nonumber\\
& \propto & \sigma^{\alpha} \,,
\label{stressstrain}
\end{eqnarray}
with
\begin{mathletters}
\begin{eqnarray}
\alpha & = & 1-\beta\nonumber,\\
&=&1 - \frac{\eta_{\mu}}{2-\eta_{\kappa}} \,,\\
&\approx & 0.72\,,
\end{eqnarray}
\end{mathletters}
where in the last equality, the numerical $3d$ values of $\eta_\mu$ and
$\eta_\kappa$ obtained from the $\epsilon$-expansion have been used.

To see that the non-Hookean stress-strain relation,
Eq. (\ref{stressstrain}), leads to a nonlinear relation between $B$ and
$H$, note that the presence of an external field, ${\bf H}$,
necessitates the inclusion of an additional $-{\bf B}\cdot{\bf H}$
term in the Hamiltonian. A field ${\bf H}$ along the $\hat{\bf z}$
direction (the direction of the initial magnetic induction ${\bf
B}(H=0)$), will therefore induce an increase in the magnetic
induction, $\delta B(H) \equiv B(H)-B(H=0)$.  In the vortex lattice
state, however, since each flux line carries {exactly} one flux
quantum, $\phi_0$, the only way to increase $B$ is to move the vortex
lines closer together, {\em i.e.}, compress the lattice. This implies that
the external field leads to a compressive strain term, proportional to
$-H{\nabla}\cdot{\bf u}$. The external field ${\bf H}$ also explicitly
breaks rotational symmetry, leading to the addition of a term $\sim
H|\partial_z{\bf u}|^2$. Thus, the application of a field ${\bf H}$
leads to a compressional strain (proportional to $\delta B$) and to
explicit rotational symmetry breaking, just as a uniform compressive
stress does. The two cases are exactly analogous, with $\sigma$
replaced by\cite{Note} $H$, and the non-Hookean stress-strain relation,
Eq. (\ref{stressstrain}), immediately translates into a nonlinear
$\delta B(H)-H$ relation of exactly the same form; {\em i.e.}, into
\begin{eqnarray}
\delta B(H) \propto H^\alpha \,,
\end{eqnarray}
with the universal exponent $\alpha = 0.72\pm 0.04$. This relation
is shown in Fig. \ref{BvsH}.

Now, of course, in a real crystalline ferromagnet, there are always
crystalline symmetry breaking fields, which pick out a preferred
direction for the magnetization, thereby explicitly breaking the
rotational invariance.  These symmetry breaking fields lead to a
non-zero tilt modulus, {\em i.e.}, to a term
\begin{eqnarray}
\frac{1}{2}V_\times|\partial_z{\bf u}|^2
\end{eqnarray}
in the Hamiltonian, which
cuts off the anomalous elasticity even in the absence of an external
field ${\bf H}$. Consequently, we expect the predictions of our
rotationally invariant theory to break down beyond a crossover length
scale
\begin{eqnarray}
\xi^\times_z=(\frac{\kappa}{V_\times})^{1/\delta}
(\xi^{NL}_z )^{-\eta_{\kappa}/\delta} \,.
\end{eqnarray}
If the crystal symmetry breaking
fields are weak, $V_\times$ will be small, and our anomalous elastic theory
will be valid for a wide range of length scales $L_{z,\perp}$
satisfying
\begin{eqnarray}
\xi^{NL}_{z,\perp} \ll L_{z,\perp}\ll \xi^\times_{z,\perp} \,.
\end{eqnarray}
For longer length scales the soft anisotropic elasticity of the SV
lattice will cross over to the conventional ``tension'' elasticity of
ordinary vortex lattices.  This crossover in length scales manifests
itself in a crossover value $H_{\times}$ for the applied magnetic field $H$
below which the non-Hookean law, Eq. (\ref{nonhook}), breaks down, to
be replaced by a conventional, Hookean relation $\delta B \propto H$,
as illustrated in Fig. \ref{BvsH}. The value of the crossover field
$H_{\times}$ is proportional to the crystal symmetry breaking field $V_\times$.

The order of magnitude of  $H_\times$ is determined by
$V_\times$, which in turn can be estimated from the width of the hysteresis loops for the
magnetic superconductor in the spontaneous vortex lattice state as
$V_\times\sim M_0H_0$, where $M_0$ is the remnant (spontaneous) magnetization and
$H_0$ is the coercive field. While we were not able to find any hysteresis data of ferromagnetic 
superconductors in the literature, we hope that our work will stimulate such experiments  
as well as other attempts to estimate $H_\times$, such as microscopic models and numerical methods.

The presence of the crystal symmetry breaking fields also makes it
experimentally possible to apply the external magnetic field opposite
to the direction of the spontaneous magnetization. While the true
ground state clearly has a ``spontaneous '' magnetization which is
parallel to the externally applied field, this anti-parallel
configuration is naively metastable if $|{\bf H}| < V_{\times}$. Writing
${\bf H} = -H {\bf\hat{z}}$, we readily see that, when $H$ reaches the
metastability limit $V_\times$, the $\frac{1}{2}V_\times|\partial_z{\bf u}|^2$
term is exactly canceled by the $-\frac{1}{2}H|\partial_z{\bf u}|^2$
term arising from the ${\bf H} \cdot {\bf M}$ interaction. Thus, at
precisely this metastable limit, the flux lattice becomes ``soft'' in
{\it almost} exactly the sense we have been discussing throughout this
paper. Therefore, we expect anomalous elasticity out to arbitrarily
long length scales at this metastable point, even in the presence of
crystal symmetry breaking fields.

Of course, the exact metastability limit can never be reached; as it is
approached, the metastable barrier gets smaller and smaller, and before it
can be reduced to zero the system will thermally tunnel over the barrier.
Thus, it might appear that the point of exact ``softness'' can never be
reached, and, hence, that there is no regime with anomalous elasticity out to arbitrarily
long length scales.

However, this is not necessarily the case. Recently, it has been
found\cite{JohnLeiming} that if the crystal symmetry breaking field has the
right mixture of higher spherical harmonics in its dependence on the
orientation of the flux lattice, it is possible for the metastable state to
remain metastable {\it right up to the point of perfect ``softness''}.
In a sense, the metastability actually persists beyond this point: in fact,
as we move beyond the point of perfect softness, the coefficient (call
it $D(H)$) of $|\partial_z {\bf u}|^2 $ in the Hamiltonian becomes negative, inducing an
instability {\it not} to the true ground state with the flux lattice tipped
$180^o$ relative  to the metastable state, but, rather, to a local minimum
very near the original metastable state which moves {\it continuously} away
from the old metastable position as we continue to make $D(H)$ more
negative. Indeed, this looks very much like the tilting of the molecules
away from the layer normals in a smectic A to smectic C transition.
We will discuss this fascinating and novel transition more in the
Sec. \ref{Conclusion} below.

It should also be noted that many of the ferromagnetic superconductor
materials are characterized by strong easy-plane anisotropy and only
much weaker in-plane crystal-symmetry
breaking fields. Our theory can easily accommodate such systems by
confining fluctuations of the spontaneous vortex lattice to lie in a
plane (that we define to be the $xz$ plane) by setting $u_y=0$. The
symmetry breaking field $V_\times$ then describes the remnant in-plane
($xz$)  anisotropy. The remnant anisotropy and associated
$V_\times$ should be quite small in such layered easy plane materials. 
In this case the system is in the universality class of the ``hybrid columnar 
Bragg glass'' phase described in Ref. \cite{SRT}, which mixes the properties
of an ``$m=1$ smectic elastic glass"\cite{JSRT_smectic} with those of a conventional Bragg glass.

\subsection{The Spontaneous Vortex Lattice Structure Function}

Even though the neutron scattering peaks of the SV lattice will be
broad (due to the absence of long-ranged translational order), they
will still contain a great deal of information about the anomalous
elasticity of the CEG phase. Since, as
demonstrated in Appendix \ref{TopOrderApp} (see also ref. \onlinecite{SaundersThesis}) 
the CEG phase possesses long-ranged orientational order, {\em single
domain}, rather then powdered averaged, neutron scattering should be
possible.\cite{EquilFootnote} Thus, the anisotropic scaling
information which is usually lost in a powder averaged x-ray
scattering experiment would be retained, allowing detailed tests of
the quantitative predictions for $\eta_{\kappa}$, $\eta_\mu$ and
$\eta_{\Delta_t}$.

The neutron scattering will have broadened spots in the $\perp$ plane
centered at the reciprocal lattice vectors. The intensity of a Bragg
spot centered at reciprocal lattice vector ${\bf Q}$ is:
\begin{eqnarray}
I({\bf q}) &=& \overline{\langle|\rho_{\bf Q}({\bf q})|\rangle^2}\;,
\label{xray1}
\end{eqnarray}
where $\rho_{\bf Q}({\bf q})$ is the Fourier transform of piece of the
spatially varying density of vortex lines modulated along ${\bf
Q}$. For any lattice ${\bf u}$ can be defined via
\begin{eqnarray}
\rho_{\bf Q}({\bf r})=|\rho_{\bf Q}|e^{i{\bf
Q}\cdot\left({\bf r}+{\bf u}({\bf r}\right))} \,,
\end{eqnarray}
and so the scattering intensity of Eq. (\ref{xray1}) can be expressed as:
\begin{eqnarray}
I({\bf q}) &=& |\rho_{\bf Q}|^2\overline{\langle\int_{\bf r}|e^{i{\bf
Q}\cdot({\bf u}({\bf r})-{\bf u}({\bf 0}))}e^{i({\bf Q}-{\bf q})\cdot{\bf
r}}|\rangle} \,,
\nonumber\\
&=& |\rho_{\bf Q}|^2 \int_{\bf r}e^{i({\bf Q}-{\bf q})\cdot{\bf
r}}e^{-\overline{\langle({\bf Q}\cdot({\bf u}({\bf r})-{\bf u}({\bf
0}))^2\rangle}/2}\;,
\label{xray2b}
\end{eqnarray}
where in going from the first to the second line the property
of a Gaussian distributed variable $x$, $\langle e^{i x} \rangle =
e^{-\langle x^2
\rangle/2}$, has been used. The effect of the ${\bf u}$ dependent 
exponential is to broaden the peak at ${\bf Q}$. This exponential can
be expressed in terms of the correlation function 
$C_{\alpha\beta}({\bf r})=\langle[u_\alpha({\bf r})-u_\alpha(0)][u_\beta({\bf r})- u_\beta(0)]\rangle$:
\begin{eqnarray}
e^{-\overline{\langle({\bf Q}\cdot({\bf u}({\bf r})-{\bf u}({\bf
0}))^2\rangle}/2}=e^{-Q_\alpha Q_\beta C_{\alpha\beta}({\bf r})/2}
\;.\label{xray4}
\end{eqnarray}
From Eq. (\ref{xray4}) the widths in the ${\hat q_z}$ and ${\hat
q_\perp}$ directions of the broadened peaks can be obtained. The peak
widths in the ${\hat q_z}$ and ${\hat q_\perp}$ directions are defined
as the inverse of the values of $z$ and $r_\perp$, respectively, at
which the exponent in Eq. (\ref{xray4}) is $\approx 1/2$. Using the expressions
for $C_{s}({\bf r}_{\perp},0)$ and $C_{s}({\bf 0}_\perp,z)$, given by
Eqs. (\ref{CDelta}) (with the appropriate wavevector anomalous elastic
constants and disorder variance) the peak widths $(\xi_z^X)^{-1}$ and
$(\xi_\perp^X)^{-1}$ can be obtained by solving:
\begin{mathletters}
\begin{eqnarray}
Q^{-2}&=&\left({\mu \over \kappa}\right)
\left({\xi_z^X\over\xi^{NL}_z}\right)^{\eta_\kappa+\eta_\mu} \,,
\label{xray5z} 
\\
Q^{-2}&=&\left({\mu \over \kappa}\right)
\left({\xi_\perp^X\over\xi^{NL}_\perp}\right)^{(\eta_\kappa+\eta_\mu)/\zeta}\,,
\label{xray5perp}
\end{eqnarray}
\end{mathletters}
which gives
\begin{mathletters}
\begin{eqnarray}
\xi_z^X&=&\xi^{NL}_z\left({\kappa \over
G^2\mu}\right)^{1/(\eta_\kappa+\eta_\mu)}\,,
\label{xray6z} 
\\
\xi_\perp^X&=&\xi^{NL}_\perp\left({\kappa \over
G^2\mu}\right)^{\zeta/(\eta_\kappa+\eta_\mu)}\label{xray6perp}\;.
\end{eqnarray}
\end{mathletters}
The temperature dependence of $\xi^X_z$ and $\xi^X_\perp$ could be
used to determine the exponents $\eta_{\kappa}$, $\eta_\mu$ and
$\eta_{\Delta_t}$ since the {\em bulk} $\kappa(T)$, $\mu(T)$ in
Eqs. (\ref{xray5z}), (\ref{xray5perp}) have temperature dependences that
can be extracted from data on bulk materials.

It is also possible to extract information from the scattering in the
``tails'' of the peak. By ``tails'' one means at wavevectors ${\bf
q}={\bf Q}+\delta {\bf q}$ such that:
\begin{eqnarray}
  (\xi^X_\alpha)^{-1} \ll | \delta q_\alpha | \ll (\xi^{NL})^{-1}\;,
\label{tails}
\end{eqnarray}
where $\alpha$ is either $\perp$ or $z$. In this regime, the scattering
\begin{eqnarray}
I(\delta{\bf q}) &=& |\rho_{\bf Q}|^2 \int_{\bf r}e^{-i\delta{\bf q}\cdot{\bf
r}}e^{-\overline{\langle({\bf Q}\cdot({\bf u}({\bf r})-{\bf u}({\bf
0}))^2\rangle}/2}\;,
\label{xray7}
\end{eqnarray}
is dominated by $r < (\delta q)^{-1}$. If $|\delta q_\alpha | \gg
(\xi^X_\alpha)^{-1}$ (the first part of the condition Eq. (\ref{tails}))
then the exponent $\overline{\langle({\bf Q}\cdot({\bf u}({\bf
r})-{\bf u}({\bf 0}))^2\rangle}$ is still small, because
$\xi^X_\alpha$ is by definition the length one must go to before
$\overline{\langle({\bf Q}\cdot({\bf u}({\bf r})-{\bf u}({\bf
0}))^2\rangle}$ becomes appreciable. This means that the exponential
can be expanded to give:
\begin{eqnarray}
I(\delta{\bf q}) &\sim& \int_{\bf r}e^{-i\delta{\bf q}\cdot{\bf r}}\;
\overline{\langle({\bf Q}\cdot({\bf u}({\bf r})-{\bf u}({\bf
0}))^2\rangle}\;,
\label{xray8}
\end{eqnarray}
which is just the Fourier transform of the quantity between angular
brackets at wavevector $\delta {\bf q}$. If $|\delta q_\alpha | \ll
(\xi^{NL})^{-1}$ (the second part of the condition Eq. (\ref{tails})),
then at this wavevector the fluctuations will be anomalous and the
intensity will be given by:
\begin{eqnarray}
I({\bf q})& \sim & { \Delta_t (\delta {\bf q}) \delta q_z^2 \over \big[\mu(\delta
{\bf q}) \delta q_\perp^2 + \kappa ( \delta {\bf q}) \delta q_z^4\big]^2} \;,
\label{tails2}
\end{eqnarray}
where the longitudinal and transverse pieces of $\overline{\langle u_i({\bf q}) u_j({\bf -q}) \rangle}$ 
have, for simplicity, been set
equal, since at $|\delta q_\alpha | \ll \xi_{NL}^{-1}$, the
longitudinal and transverse propagators are essentially the same
(since $\lambda(\delta{\bf q})/\mu(\delta{\bf q})\rightarrow$
constant). Thus, by examining the tails of the broadened peaks one
could directly observe the anomalous elasticity.

\section{Future Theoretical  Work}
\label{Conclusion}

There are a number of interesting questions that remain
to be investigated experimentally. One of them is the problem touched upon
in section \ref{Exp_consq} of applying a magnetic field in the opposite
direction to the spontaneous magnetization to ``cancel off'' the effects of
the crystal field. The state thereby produced, though only metastable,
could, in principle, be {\it exactly}, rather than approximately, soft.
This exact anomalous elasticity is probably not,
however, controlled by the columnar elastic glass (CEG) fixed point we have
been discussing up to now. The reason for this is that the CEG fixed point
described a model with exact rotation invariance, to all orders in the
rotation angle $\theta$.  In contrast, at the metastable fixed point of a
model with crystal symmetry breaking fields, we have simply tuned $H$ to
cancel off the leading order in $\theta$ (i.e., the $|\partial_z{\bf
u}|^2$) term in the Hamiltonian. The full model however, is not
rotation invariant to all orders in $\theta$, since, for large
$\theta$, the crystal symmetry breaking field has a different
dependence on $\theta$ than the ${\bf H} \cdot {\bf M}$ term (e.g.,
$\cos(\theta)$ {\em vs.}  $\cos(4\theta)$ for an underlying cubic crystal). This lower symmetry of the
``metastable limit'' model allows, e.g., the coefficients of the cubic
$({\nabla}\cdot{\bf u})|\partial_z{\bf u}|^2$ and quartic
$|\partial_z{\bf u}|^4$ nonlinearities in the Hamiltonian to be
different from one another and from those of the quadratic
$|{\nabla}{\bf u}|^2$ terms.  This should be contrasted with the
columnar elastic glass Hamiltonian, in which these cubic and quartic
terms arise only from the square of the nonlinear strain tensor
$v_{ij}$, and must therefore have the same coefficients (up to known
factors of order unity) as each other and the quadratic terms.

A recent study of this problem\cite{JohnLeiming} has found that the departure of
these cubic and quartic coefficients away from equality with the quadratic
terms is a relevant perturbation away from the rotationally invariant
fixed point to a new non-rotationally invariant fixed point.
This new fixed point exhibits anomalous elasticity with different
universal exponents than those given above for the rotationally
invariant problem. Further details will be given in a forthcoming
publication.\cite{JohnLeiming}

We have focussed in this paper exclusively on static, equilibrium
properties of (putative) spontaneous vortex lattices in ferromagnetic 
superconductors in presence of positional and random tilt disorder. 
Since, in the presence of tilt disorder
these lattices are glassy, their dynamics should exhibit all of the
interesting slow phenomenology associated with glasses. The interplay
of this slow dynamics with the anomalous elasticity we have
studied here should make the dynamic behavior even more fascinating.
In particular, the dynamics of depinning of these flux lattices, which
determines their voltage-current $(I-V)$ characteristics, should be quite
novel.

\medskip

{\em Note added in proof.} It has recently been shown \cite{TBKT} that
the elastic theory developed here for spontaneous flux lattices in $s$-wave
superconductors also applies, without modification, to superconductors of
arbitrary (e.g., $p$-wave) symmetry.

\acknowledgements
A.M.E. wishes to thank the Department of Physics, University of Florida, for 
financial support. L.R. acknowledges support by the NSF under 
grants DMR-0213918, DMR-0321848, and by the David and Lucile Packard Foundation.
J.T. thanks the Aspen Center for Physics, and the Kavli Institute for Theoretical Physics, 
Santa Barbara, for their hospitality while a portion of this work was being
completed, and D. Belitz and S. Tewari for many valuable discussions.

\appendix

\section{Large scale behavior of correlation functions}
\label{App_vmag_replicas}

In this Appendix, we give some technical details on the solution of the
saddle-point equation (\ref{eq-sigma-1}) of the replica formalism
of section \ref{vmag_replicas}, which we rewrite
here for definiteness:
\begin{equation}
\sigma(u)\int\frac{d^{d_\perp}{\bf q}_\perp}{(2\pi)^{d_\perp}}
\int_{-\infty}^\infty \frac{dq_z}{2\pi}
\frac{TQ_0^2}{\big[\kappa q_z^4 + Kq_\perp^2 +[\sigma](u)\big]^2} = 1\,,
\label{eq-sigma-1new}
\end{equation}
with $[\sigma](u) = u\sigma(u) -\int_0^udv\;\sigma(v)$.
Using the fact that
\begin{eqnarray}
\int_{-\infty}^\infty \frac{dx}{(x^4+a^4)^2} =
\frac{3\pi\sqrt{2}}{a^7} \quad ,\quad a > 0
\end{eqnarray}
we obtain
\begin{equation}
\int_{-\infty}^\infty \frac{dq_z}{\big[\kappa q_z^4 + Kq_\perp^2 +
[\sigma](v)\big]^2} =
\frac{3\pi\sqrt{2}}{8\kappa^{1/4}}\,\frac{1}{\big(Kq_\perp^2 +
[\sigma](v)\big)^{7\over 4}},
\end{equation}
and hence, equation (\ref{eq-sigma-1}) becomes
\begin{eqnarray}
\frac{3\sqrt{2}TQ_0^2\sigma(u)}{16\kappa^{1/4}}
\int\frac{d^{d_\perp}{\bf q}_\perp}{(2\pi)^{d_\perp}}
\frac{1}{\big(Kq_\perp^2 +[\sigma](u)\big)^{7/4}} = 1.
\label{eq-sigma-2}
\end{eqnarray}
Performing the change of variables
\begin{eqnarray}
{\bf q}_\perp={\bf q}\;\sqrt{\frac{[\sigma]}{K}} \,,
\nonumber
\end{eqnarray}
equation (\ref{eq-sigma-2}) becomes
\begin{eqnarray}
\frac{3\sqrt{2}c_{d_\perp}}{16\kappa^{1/4}K^{d_\perp/2}}
TQ_0^2\sigma(u) [\sigma(u)]^{\frac{2d_\perp-7}{4}} = 1 \,,
\label{eq-sigma-3}
\end{eqnarray}
where $c_{d_\perp}$ is a numerical constant given by:
\begin{eqnarray}
c_{d_\perp} = \int\frac{d^{d_\perp}{\bf
q}_\perp}{(2\pi)^{d_\perp}}\frac{1}{(q_\perp^2+1)^{7\over 4}} \,,
\end{eqnarray}
where, in this last equation, the ultraviolet cutoff $\Lambda=2\pi/a$ was sent to
infinity. Now, equation (\ref{eq-sigma-3}) can be rewritten in the form:
\begin{eqnarray}
[\sigma(u)]^{\frac{7-2d_\perp}{4}} =
\frac{3\sqrt{2}c_{d_\perp}}{16\kappa^{1/4}K^{d_\perp/2}}
\,TQ_0^2\sigma(u)\,.
\end{eqnarray}
Taking the derivative of this last expression with respect to $u$,
and taking into account the fact that $[\sigma]'(u)=u\sigma'(u)$,
we obtain:
\begin{eqnarray}
\frac{7-2d_\perp}{4}[\sigma(u)]^{\frac{3-2d_{\perp}}{4}} =
\Big(\frac{3\sqrt{2}c_{d_\perp}TQ_0^2}{16\kappa^{1/4}
K^{d_\perp/2}}\Big)\;\frac{1}{u}\,.
\end{eqnarray}
Solving for $[\sigma](u)$, we obtain:
\begin{eqnarray}
[\sigma](u) = \Big(\frac{u}{u_0}\Big)^{\frac{2}{\theta}} ,
\label{result-sigma}
\end{eqnarray}
where we called $\theta=d_{\perp}-3/2$, and where $u_0$ is given by:
\begin{eqnarray}
u_0 =\frac{3\sqrt{2}c_{d_\perp}TQ_0^2}{4(7-2d_{\perp})
\kappa^{1/4}K^{d_\perp/2}} \,.
\end{eqnarray}
In terms of $[\sigma](u)$, the elastic propagator in the presence
of disorder $\tilde G({\bf q})$ is given by:
\begin{eqnarray}
\tilde G({\bf q}) & = & \frac{1}{Kq_\perp^2 + \kappa q_z^4}
\Big(1 + \int_0^1 \frac{du}{u^2}
\frac{[\sigma](u)}{Kq_\perp^2 + \kappa q_z^4 + [\sigma](u)}\Big),
\nonumber\\
& = & \frac{1}{Kq_\perp^2 + \kappa q_z^4}
\Big(1 + \int_0^1 \frac{du}{u^2}
\frac{(u/u_0)^{2/\theta}}{Kq_\perp^2 + \kappa q_z^4
+ (u/u_0)^{\frac{2}{\theta}}}\Big),
\nonumber\\
\label{eq-G-1}
\end{eqnarray}
where, in going from the first to the second line we replaced
$[\sigma](u)$ by its expression, equation
(\ref{result-sigma}). Now, using the change of variables
$x=u^{2/\theta}$, we can write:
\end{multicols}
\begin{eqnarray}
\int_0^1 \frac{du}{u^2}\frac{(u/u_0)^{2/\theta}}{Kq_\perp^2 +
\kappa q_z^4 + (u/u_0)^{2/\theta}} & = &
\frac{\theta}{2}\int_0^1 dx \;\frac{x^{-\theta/2}}{x +
u_0^{2/\theta}(Kq_\perp^2+\kappa q_z^4)}
\nonumber\\
&=&\frac{\theta}{2-\theta}\frac{1}{u_0^{2/\theta}
(Kq_\perp^2+\kappa q_z^4)}\;
{}_2F^1[1,1-\frac{\theta}{2},2-\frac{\theta}{2},
-\frac{1}{u_0^{2/\theta}(Kq_\perp^2+\kappa q_z^4)}]\,,
\end{eqnarray}
\begin{multicols}{2}\noindent
where ${}_2F^1$ is a hypergeometric function.
Replacing this last expression back into equation (\ref{eq-G-1}),
and using the following limiting behavior when $a\to 0$:
\begin{equation}
{}_2F^1[1,1-\frac{\theta}{2},2-\frac{\theta}{2},-\frac{1}{a}] =
\Gamma(2-\frac{\theta}{2})\Gamma(\frac{\theta}{2})\,
a^{1-\frac{\theta}{2}} + {\cal O}(a)
\end{equation}
we obtain the following long wavelength behavior of the
correlation function:
\begin{equation}
\tilde G({\bf q}) \simeq
\frac{\theta}{2-\theta}\;
\;\frac{\Gamma(2-\frac{\theta}{2})\,\Gamma(\theta/2)}{u_0(K q_\perp^2 + \kappa q_z^4)^{1+\theta/2}}\,.
\end{equation}
We now are in a position to calculate the large scale behavior of
the correlation function $C^{p}({\bf r})=\langle[{\bf u}({\bf
r})-{\bf u}(0)]^2\rangle$, which is now given by:
\end{multicols}
\begin{eqnarray}
C^{p}({\bf r}) & = & \int\frac{d^{d_\perp}{\bf 
q}_\perp\,dq_z}{(2\pi)^{d_{\perp}+1}}\;(1-\mbox{e}^{i{\bf q}\cdot{\bf r}})
\,\tilde G({\bf q}) \,,
\nonumber\\
& = & \frac{2T}{u_0}\frac{\theta}{2-\theta}
\;\Gamma(2-\theta/2)\Gamma(\theta/2)
\;\int\frac{d^{d_\perp}{\bf q}_\perp\,dq_z}{(2\pi)^{d_{\perp}+1}}\;
\frac{1-\mbox{e}^{i{\bf q}\cdot{\bf r}}}
{(Kq_\perp^2+\kappa q_z^4)^{1+\theta/2}} \,.
\end{eqnarray}
As we did in the last paragraph, we shall evaluate
$C^{p}({\bf r}_\perp)$ and $C^{p}(z)$ separately. We have:
\begin{eqnarray}
C^{p}({\bf r}_\perp) & = & \frac{2T}{u_0\kappa^{1+\theta/2}}
\frac{\theta}{2-\theta}\;\Gamma(2-\theta/2)\Gamma(\theta/2)
\int\frac{d^{d_\perp}{\bf q}_\perp}{(2\pi)^{d_\perp}}\;
\int_{-\infty}^\infty\frac{dq_z}{2\pi}
\frac{1-\mbox{e}^{i{\bf q}\cdot{\bf r}}}
{(q_z^4 + Kq_\perp^2/\kappa)^{1+\theta/2}} \,,
\nonumber\\
& = & \frac{2T}{\pi u_0\kappa^{1+\theta/2}}
\Big(\frac{\theta}{2-\theta}\Big)
\;\Gamma(2-\frac{\theta}{2})\Gamma(\frac{\theta}{2})
\frac{\Gamma(\frac{5}{4})
\Gamma(\frac{3}{4}+\frac{\theta}{2})}{\Gamma(1+\frac{\theta}{2})}\;
\Big(\frac{\kappa}{K}\Big)^{\frac{3+2\theta}{4}}
\int\frac{d^{d_\perp}{\bf q}_\perp}{(2\pi)^{d_\perp}}
\;\frac{1-\mbox{e}^{i{\bf q}_\perp\cdot{\bf r}_\perp}}
{q_\perp^{(3+2\theta)/2}} \,,
\nonumber
\end{eqnarray}
\begin{multicols}{2}\noindent
where, in going from the first to the second line we used the fact that
\begin{eqnarray}
\int_{-\infty}^\infty\frac{dx}{(x^4+a)^{1+\frac{\theta}{2}}}
=\frac{2\Gamma(5/4)\Gamma(\frac{3}{4}+\frac{\theta}{2})}
{\Gamma(1+\theta/2)} \; a ^{-(\frac{3}{4}+\frac{\theta}{2})}.
\end{eqnarray}
For a spontaneous vortex lattice in three dimensions,
$d_{\perp}=2$, $\theta=d_{\perp}-3/2=1/2$, and we obtain:
\begin{eqnarray}
C^{p}(r_\perp) & = & \frac{2T\Gamma(7/4)\Gamma(1/4)}{3\pi
u_0\kappa^{1/4}K}
\int_0^\Lambda\frac{dq_\perp}{2\pi}\;
\frac{1-\mbox{J}_0(q_\perp r_\perp)}{q_\perp} \,,
\nonumber\\
& \simeq & \frac{T\sqrt{2}}{2 u_0\kappa^{1/4}K}
\;\ln(\Lambda r_\perp) \,,
\end{eqnarray}
where, in going from the first to the second line, we used the
fact that $\Gamma(7/4)\Gamma(1/4)=3\pi\sqrt{2}/4$.

\medskip

We now turn our attention to the calculation of $C^{p}(z)$. We have:
\end{multicols}
\begin{eqnarray}
C^{p}(z) & = & 2T \int\frac{d^{d_\perp}{\bf q}dq_z}
{(2\pi)^{d_{\perp}+1}}\,(1-\mbox{e}^{iq_zz})\tilde G({\bf q}) \,,
\nonumber\\
& = & \frac{T}{\pi u_0K^{1+\theta/2}}\Big(\frac{\theta}{2-\theta}\Big)
\,\Gamma(2-\theta/2)\Gamma(\theta/2)\;
\int_{-\infty}^\infty dq_z\;(1-\mbox{e}^{iq_zz})K_{d_{\perp}}\int_0^\Lambda
dq_\perp\;\frac{q_\perp^{d_\perp-1}}
{(q_\perp^2+\kappa q_z^4/K)^{1+\theta/2}} \,,
\label{eq-C-z}
\end{eqnarray}
where $K_{d}=(2\pi^{d/2}/\Gamma(d/2))$ is the total solid angle
(surface area of a unit sphere) in $d$
dimensions. Now, using the fact that
\begin{eqnarray}
\int_0^\Lambda dq_\perp
\;\frac{q_\perp^{d_{\perp}-1}}{(q_\perp^2+\kappa q_z^4/K)^{1+\theta/2}} =
\frac{\Lambda^{d_\perp}}{d_{\perp}}
\Big(\frac{\kappa}{K}\,q_z^4\Big)^{-(1+\theta/2)}\;
{}_2F^1[\frac{d_{\perp}}{2},1+\frac{\theta}{2},
1+\frac{d_{\perp}}{2},-\frac{\Lambda^2K}{\kappa q_z^4}] \,,
\end{eqnarray}
and the limiting behavior when $q_z\to 0$:
\begin{eqnarray}
{}_2F^1[\frac{d_{\perp}}{2},1+\frac{\theta}{2},
1+\frac{d_{\perp}}{2},-\frac{\Lambda^2K}{\kappa q_z^4}] \approx
\frac{\Gamma(1+d_{\perp}/2)\Gamma(1+\frac{\theta}{2}
-\frac{d_{\perp}}{2})}{\Gamma(1+\theta/2)}\;
\Big(\frac{\kappa q_z^4}{K\Lambda^2}\Big)^{d_{\perp}\over 2} ,
\end{eqnarray}
\begin{multicols}{2}\noindent
we obtain, in three dimensions ($d_\perp=2$):
\begin{eqnarray}
\int_0^\Lambda dq_\perp\;\frac{q_\perp^{d_{\perp}-1}}
{(q_\perp^2+\kappa q_z^4/K)^{1+\theta/2}} =
2\Big(\frac{K}{\kappa}\Big)^{1/4}\;\frac{1}{q_z} .
\end{eqnarray}
Replacing this last expression into equation (\ref{eq-C-z}), we
obtain (here we use the fact that $K_2=1/(2\pi)$):
\begin{eqnarray}
C^{p}(z) & = & \frac{2T\Gamma(7/4)\Gamma(1/4)}{3\pi^2 u_0
K\kappa^{1/4}}\int_0^{\Lambda_z} dq_z
\;\frac{1-\cos(q_zz)}{q_z}\,,
\nonumber\\
& = & \frac{T\sqrt{2}}{2\pi u_0 K\kappa^{1/4}}
\;\int_0^{\Lambda_zz} dq_z \;\frac{1-\cos(x)}{x} \,,
\end{eqnarray}
where we introduced the cut-off $\Lambda_z\simeq 1/\xi$ for the
integration over $q_z$, and where, in going from the first to the
second line, we used the change of variables $x=q_zz$. Now, from
the definition of the cosine-integral function:\cite{Abramowitz}
\begin{eqnarray}
\mbox{Ci}(x) = \ln|x| + \gamma + \int_0^x\frac{\cos t -1}{t}\;dt \,,
\end{eqnarray}
where $\gamma\simeq 0.577\ldots$ is Euler's constant, and the fact
that $\mbox{Ci}(x)\to 0$ when $x\to\infty$, we see that
$\int_0^{\Lambda_zz}dx\;(1-\cos x)/x \simeq \ln(\Lambda_z|z|) +
\gamma$ for $\Lambda_z|z|\gg 1$, and hence the long distance
behavior of $C^{p}(z)$ in three dimensions is given by:
\begin{eqnarray}
C^{p}(z) \simeq \frac{T\sqrt{2}}{2\pi u_0K\kappa^{1/4}}\;\Big[\ln(\Lambda|z|) 
+ \gamma\Big]
\end{eqnarray}
as claimed in the text.

\section{Elastic propagator in the presence of disorder}
\label{App_elastic_propagator}

In this Appendix, we show how to derive the expression of the propagator
$G_{\alpha\beta}^{ab}({\bf q})$ of
the theory defined by the Hamiltonian $H_{0n}$ of equation 
(\ref{Heff0}),
\begin{eqnarray}
H_{0n} = \sum_{a,b}\int_{\bf q}\;\frac{1}{2}\,
\Gamma_{\alpha\beta}^{ab}({\bf q})
\,u_\alpha^a({\bf q})u_\beta^b(-{\bf q}) \,,
\end{eqnarray}
where
\begin{eqnarray}
\Gamma_{\alpha\beta}^{ab}({\bf q}) = \Gamma_{\alpha\beta}({\bf q})\delta_{ab} -
\frac{{\Delta_t}}{T}q_z^2\delta_{\alpha\beta} \,.
\end{eqnarray}
Finding $G_{\alpha\beta}^{ab}({\bf q})$ amounts to finding the 
inverse of the matrix
$\Gamma_{\alpha\beta}^{ab}({\bf q})$ such that:
\begin{eqnarray}
G_{\alpha\beta}^{ab}({\bf q}){\Gamma}_{\beta\gamma}^{bc}({\bf q}) =
\Gamma_{\alpha\beta}^{ab}({\bf q})G_{\beta\gamma}^{bc}({\bf q}) = 
\delta_{\alpha\gamma}\,\delta_{bc} \,.
\end{eqnarray}
It is easy to verify that, in the limit $n\to 0$, 
$G_{\alpha\beta}^{ab}({\bf q})$ is given by:
\begin{eqnarray}
G_{\alpha\beta}^{ab}({\bf q}) = 
T\big(\Gamma^{-1}\big)_{\alpha\beta}({\bf q})\,\delta_{ab} +
{\Delta_t} q_z^2\big[\big(\Gamma^{-1}\big)^2\big]_{\alpha\beta}({\bf q}) \,.
\end{eqnarray}
But, since
\begin{eqnarray}
\Gamma_{\alpha\beta}({\bf q}) = \Gamma_L({\bf 
q})P^L_{\alpha\beta}({\bf q}_\perp) +
\Gamma_T({\bf q})P^T_{\alpha\beta}({\bf q}_\perp) \,,
\end{eqnarray}
we see that
\begin{eqnarray}
\big(\Gamma^{-1}\big)_{\alpha\beta}({\bf q}) = \Gamma^{-1}_L({\bf 
q})P^L_{\alpha\beta}({\bf q}_\perp) +
\Gamma^{-1}_T({\bf q})P^T_{\alpha\beta}({\bf q}_\perp) \,,
\end{eqnarray}
and
\begin{eqnarray}
\big[(\Gamma^{-1})^2\big]_{\alpha\beta}({\bf q}) & = &
\big(\Gamma^{-1}\big)_{\alpha\gamma}({\bf 
q})\big(\Gamma^{-1}\big)_{\gamma\beta}({\bf q}) \,,
\nonumber\\
& = &
\big[\Gamma^{-1}_L({\bf q})P^L_{\alpha\gamma}({\bf q}_\perp) +
\Gamma^{-1}_T({\bf q})P^T_{\alpha\gamma}({\bf q}_\perp)\big]\times
\nonumber\\
&\times&\big[\Gamma^{-1}_L({\bf q})P^L_{\gamma\beta}({\bf q}_\perp) +
\Gamma^{-1}_T({\bf q})P^T_{\gamma\beta}({\bf q}_\perp)\big]\,,
\nonumber\\
& = &  \Gamma^{-2}_L({\bf q})P^L_{\alpha\beta}({\bf q}_\perp) +
\Gamma^{-2}_T({\bf q})P^T_{\alpha\beta}({\bf q}_\perp) \,,
\nonumber\\
\end{eqnarray}
where, in going from the second to the third line, we used the 
following results from projection
operators algebra:
\begin{mathletters}
\begin{eqnarray}
P^L_{\alpha\gamma}({\bf q}_\perp)P^L_{\gamma\beta}({\bf q}_\perp) &=& 
P^L_{\alpha\beta}({\bf q}_\perp) \,,
\label{LLalg}\\
P^T_{\alpha\gamma}({\bf q}_\perp)P^T_{\gamma\beta}({\bf q}_\perp) &=& 
P^T_{\alpha\beta}({\bf q}_\perp) \,,
\label{TTalg}\\
P^L_{\alpha\gamma}({\bf q}_\perp)P^T_{\gamma\beta}({\bf q}_\perp) &=& 0 \,.
\label{LTalg}
\end{eqnarray}
\end{mathletters}
We therefore obtain for $G_{\alpha\beta}^{ab}({\bf q})$, in the limit $n\to 0$:
\begin{eqnarray}
G_{\alpha\beta}^{ab}({\bf q}) = G_L^{ab}({\bf q})P^L_{\alpha\beta}({\bf q}_\perp) +
G_T^{ab}({\bf q})P^T_{\alpha\beta}({\bf q}_\perp) \,,
\end{eqnarray}
where
\begin{mathletters}
\begin{eqnarray}
G_L^{ab}({\bf q}) & = & T\Gamma_L^{-1}({\bf q})\,\delta_{ab} + 
{\Delta_t} q_z^2\Gamma_L^{-2}({\bf q}) \,,
\\
G_T^{ab}({\bf q}) & = & T\Gamma_T^{-1}({\bf q})\,\delta_{ab} + 
{\Delta_t} q_z^2\Gamma_T^{-2}({\bf q}) \,.
\end{eqnarray}
\end{mathletters}
We therefore can write:
\begin{eqnarray}
\langle u_\alpha^a({\bf q})u_\beta^b({\bf q}')\rangle_0 = 
(2\pi)^d\delta({\bf q}+{\bf q}')\,
G_{\alpha\beta}^{ab}({\bf q}) \,,
\end{eqnarray}
where the subscript $0$ in $\langle\cdots\rangle_0$ indicates that 
the average is taken with the
statistical weight $\exp(-\beta H_{0n})/Z_0$, with 
$Z_0=\mbox{Tr}\big(\exp(-\beta H_{0n})\big)$.

\section{Perturbation theory in the elastic nonlinearities}
\label{App_vmag_PT1}

In this Appendix, we fill in a few details on the perturbative
calculation of Sec. \ref{PT}. We shall start by looking at the
perturbative corrections to $\lambda$ and $\mu$, before addressing the
relatively more involved corrections to $\kappa$ and $\Delta$.

\subsection{Correction to $\lambda$ and $\mu$}

Starting from equation (\ref{def-Hint}), and using Wick's theorem, we
find that the part of the connected average $\langle
H_{int}^2\rangle_{0>}^c$ which corrects $\lambda$ and $\mu$ is given
by:
\end{multicols}
\begin{eqnarray}
\langle H_{int}^2\rangle_{0>}^c [\lambda,\mu] & \equiv &
\sum_{a,b}\int d{\bf r}\,d{\bf r}'\; \Big[\;
\mu^2 \partial_\alpha u_\beta^a({\bf r})\partial_\gamma
u_\delta^b({\bf r}')
\big\langle\,\partial_z u_\alpha^a({\bf r})\partial_z u_\beta^a({\bf r})
\partial_z u_\gamma^b({\bf r}')\partial_z
u_\delta^b({\bf r}')\,\big\rangle_{0>}^c +
\nonumber\\
& + & \frac{\mu\lambda}{2}\,\partial_\alpha u_\beta^a({\bf
r})\partial_\gamma u_\gamma^b({\bf r}')
\,\big\langle\,\partial_z u_\alpha^a({\bf r})
\partial_zu_\beta^a({\bf r})
\,\big(\partial_z u_\delta^b({\bf r}')\big)^2\,\big\rangle_{0>}^c +
\nonumber\\
& + & \frac{\lambda\mu}{2}\,\partial_\alpha u_\alpha^a({\bf r})
\partial_\gamma u_\delta^b({\bf r}')
\,\big\langle\,\big(\partial_z u_\beta^a({\bf r})\big)^2
\partial_z u_\gamma^b({\bf r}')\partial_z u_\delta^b({\bf r}')
\,\big\rangle_{0>}^c +
\nonumber\\
& + & \frac{\lambda^2}{4}\,\partial_\alpha u_\alpha^a({\bf r})
\partial_\gamma u_\gamma^b({\bf r}')
\,\big\langle\,\big(\partial_z u_\beta^a({\bf r})\big)^2
\,\big(\partial_z u_\delta^b({\bf r}')\big)^2\,\big\rangle_{0>}^c
\;\Big] \,,
\end{eqnarray}
where we have used the symbol $\equiv$ to emphasize the fact
that we only show the part which corrects $\lambda$ and $\mu$.
Evaluating the above averages using Wick's theorem, we find:
\begin{eqnarray}
-\frac{1}{2T}\langle H_{int}^2\rangle_{0>}^c & = &
\frac{1}{2}\int_{\bf q}^< u_i^{a<}({\bf 
q})\,\delta\Gamma_{ij}^{ab}({\bf q})u_j^{b<}(-{\bf q}) \,,
\end{eqnarray}
with the kernel:
\begin{eqnarray}
\delta\Gamma_{ij}^{ab}({\bf q}_\perp) & = &
-\frac{1}{T}\int_{\bf q'}^> q_z'^4
\times\Big\{
\frac{\lambda^2}{2}\,q_iq_j\,G_{\beta\delta}^{ab}({\bf q'})
\,G_{\beta\delta}^{ab}({\bf q}+{\bf q'}) +
\mu^2 q_\alpha q_\gamma
\,\big(G_{\alpha\gamma}^{ab}({\bf q'})\,G_{ij}^{ab}({\bf q}+{\bf q'})
+ G_{\alpha j}^{ab}({\bf q'})\,G_{i\gamma}^{ab}({\bf q}+{\bf q'})
\big) +
\nonumber\\
& + & \mu\lambda q_iq_\gamma\,G_{\beta j}^{ab}({\bf q'})
\,G_{\beta\gamma}^{ab}({\bf q}+{\bf q'})
+ \lambda\mu q_\alpha q_j\,G_{\alpha\delta}^{ab}({\bf q'})
\,G_{i\delta}^{ab}({\bf q}+{\bf q'}) \Big\} \,,
\label{dGamma_perp1}
\nonumber\\
& = & -\frac{1}{T}\int_{\bf q'} q_z'^4
\times\Big\{
\frac{\lambda^2}{2}\,q_iq_j\,G_{\beta\delta}^{ab}({\bf q'})
\,G_{\beta\delta}^{ab}({\bf q'}) +
\mu^2\, q_\alpha q_\gamma
\,\big(G_{\alpha\gamma}^{ab}({\bf q'})\,G_{ij}^{ab}({\bf q'})
+ G_{\alpha j}^{ab}({\bf q'})\,G_{i\gamma}^{ab}({\bf q'})
\big) +
\nonumber\\
& + & \mu\lambda\, q_iq_\gamma\,G_{\beta j}^{ab}({\bf q'})
\,G_{\beta\gamma}^{ab}({\bf q'})
+ \lambda\mu\, q_\alpha q_j\,G_{\alpha\delta}^{ab}({\bf q'})
\,G_{i\delta}^{ab}({\bf q'}) \Big\} \,,
\label{dGamma_perp2}
\end{eqnarray}
where $G_{\alpha\beta}^{ab}({\bf q})$ is the propagator of
the theory defined by the Hamiltonian $\tilde{H}_{0n}$ of Eq.
(\ref{Heff0}), which can be written in terms of
transverse and longitudinal parts in the manner:
\begin{eqnarray}
G_{\alpha\beta}^{ab}({\bf q}) = G_L^{ab}({\bf q})
P^L_{\alpha\beta}({\bf q}_\perp) +
G_T^{ab}({\bf q})P^T_{\alpha\beta}({\bf q}_\perp) \,.
\label{decompG}
\end{eqnarray}
where (in the $n\to 0$ limit)
\begin{mathletters}
\begin{eqnarray}
G_L^{ab}({\bf q}) & = & T\Gamma_L^{-1}({\bf q})\,\delta_{ab}
+ {\Delta_t} q_z^2\Gamma_L^{-2}({\bf q}) \,,
\\
G_T^{ab}({\bf q}) & = & T\Gamma_T^{-1}({\bf q})\,\delta_{ab}
+ {\Delta_t} q_z^2\Gamma_T^{-2}({\bf q}) \,.
\end{eqnarray}
\end{mathletters}
In going from equation
(\ref{dGamma_perp1}) to (\ref{dGamma_perp2}), we approximated
$G_{\rho\sigma}^{ab}({\bf q}+{\bf q'})\simeq
G_{\rho\sigma}^{ab}({\bf q'})$, since we only need terms of
order $q_\perp^2$ in $\delta\Gamma_{ij}^{ab}({\bf q}_\perp)$.
Now, it is easy to verify, using the algebra of projection operators
(Eqs. (\ref{LLalg})-(\ref{LTalg}) of Appendix \ref{App_elastic_propagator}),
that
\begin{mathletters}
\begin{eqnarray}
G_{\beta\delta}^{ab}({\bf q'})\,G_{\beta\delta}^{ab}({\bf q'})
& = & \big(G_L^{ab}({\bf q}')\big)^2 +
(d_\perp-1)\big(G_T^{ab}({\bf q}')\big)^2 \,,
\\
G_{\beta j}^{ab}({\bf q'})\,G_{\beta\gamma}^{ab}({\bf q'}) & = &
\big(G_L^{ab}({\bf q}')\big)^2 P^L_{j\gamma}({\bf q}_\perp')
+ \big(G_T^{ab}({\bf q}')\big)^2P^T_{j\gamma}({\bf q}_\perp') \,,
\\
G_{\alpha\delta}^{ab}({\bf q'})\,G_{i\delta}^{ab}({\bf q'}) & = &
\big(G_L^{ab}({\bf q}')\big)^2 P^L_{i\alpha}({\bf q}_\perp')
+ \big(G_T^{ab}({\bf q}')\big)^2P^T_{i\alpha}({\bf q}_\perp') \,.
\end{eqnarray}
\end{mathletters}
The expression of $\delta\Gamma_{ij}^{ab}({\bf q}_\perp)$
therefore becomes:
\begin{eqnarray}
\delta\Gamma_{ij}^{ab}({\bf q}_\perp) & = &
-\frac{1}{T}\int_{\bf q'} q_z'^4
\times\Big\{
q_iq_j\Big(\frac{\lambda^2}{2}+\frac{2\lambda\mu}{d_\perp}\Big)
\big[\,\big(G_{L}^{ab}({\bf q'})\big)^2 +
(d_\perp-1)\,\big(G_T^{ab}({\bf q'})\big)^2
\big] +
\nonumber\\
& + & \mu^2\,q_\alpha q_\gamma
\,\big(G_{\alpha\gamma}^{ab}({\bf q'})\,G_{ij}^{ab}({\bf q'})
+ G_{\alpha j}^{ab}({\bf q'})\,G_{i\gamma}^{ab}({\bf q'})
\big)\Big\} \,,
\label{dGamma_perp4}
\end{eqnarray}
where we used the following rotational
averages of single projection operators
(Appendix \ref{App_rotavg}):
\begin{mathletters}
\begin{eqnarray}
\langle P^L_{\alpha\beta}({\bf q}_\perp)\rangle & = &
\frac{\delta_{\alpha\beta}}{d_\perp} \,,
\\
\langle P^T_{\alpha\beta}({\bf q}_\perp)\rangle & = &
\frac{d_\perp-1}{d_\perp}\,\delta_{\alpha\beta} \,.
\end{eqnarray}
\end{mathletters}
Now, in the second line on the rhs of equation
(\ref{dGamma_perp4}), using the decomposition (\ref{decompG})),
we obtain
\begin{eqnarray}
G_{\alpha\gamma}^{ab}({\bf q}')G_{ij}^{ab}({\bf q}') & = &
\big(G_L^{ab}({\bf q}')\big)^2P^L_{\alpha\gamma}({\bf q}_\perp')
P^L_{ij}({\bf q}_\perp') + \big(G_L^{ab}({\bf q}')\big)^2
P^L_{\alpha\gamma}({\bf q}_\perp')P^L_{ij}({\bf q}_\perp') +
\nonumber\\
& + & G_L^{ab}({\bf q}')G_T^{ab}({\bf q}')
\,\big(P^L_{\alpha\gamma}({\bf q}_\perp')P^T_{ij}({\bf q}_\perp')
+ P^T_{\alpha\gamma}({\bf q}_\perp')P^L_{ij}({\bf q}_\perp')\big) \,.
\end{eqnarray}
Performing the integration over polar angles, we obtain
\begin{eqnarray}
\mu^2\,q_\alpha q_\gamma
\int_{\bf q'} q_z'^4\big(G_{\alpha\gamma}^{ab}({\bf
q}')G_{ij}^{ab}({\bf q}') &+&
G_{\alpha j}^{ab}({\bf q}')G_{i\gamma}^{ab}({\bf q}')\big) =
\nonumber\\
& = & \frac{\mu^2q_\perp^2\delta_{ij}}{d_\perp(d_\perp+2)}
\int_{\bf q'} q_z'^4\;\Big[
2\big(G_L^{ab}({\bf q}')\big)^2 +
(d_\perp^2-2)\big(G_T^{ab}({\bf q}')\big)^2
+ 2d_\perp G_L^{ab}({\bf q}')G_T^{ab}({\bf q}')
\Big] +
\nonumber\\
&+& \frac{\mu^2q_iq_j}{d_\perp(d_\perp+2)}\int_{\bf q'}
q_z'^4\;\Big[
4\big(G_L^{ab}({\bf q}')\big)^2 + d_\perp^2
\big(G_T^{ab}({\bf q}')\big)^2
+ 2(d_\perp-2) G_L^{ab}({\bf q}')G_T^{ab}({\bf q}')
\Big]\,.
\nonumber
\end{eqnarray}
Now, using the fact that
\begin{eqnarray}
q_iq_j & = & q_\perp^2 P^L_{ij}({\bf q}_\perp) \,,\nonumber
\\
\delta_{ij} & = & P^L_{ij}({\bf q}_\perp) +
P^T_{ij}({\bf q}_\perp) \,,
\nonumber
\end{eqnarray}
we obtain that $\delta\Gamma_{ij}^{ab}({\bf q}_\perp)$ can be written in
the form:
\begin{eqnarray}
\delta\Gamma_{ij}^{ab}({\bf q}_\perp) =
\delta\Gamma_L^{ab}({\bf q}_\perp)\,q_\perp^2
\,P^L_{ij}({\bf q}_\perp)
+ \delta\Gamma_T^{ab}({\bf q}_\perp)\,q_\perp^2
\,P^T_{ij}({\bf q}_\perp) \,,
\end{eqnarray}
with:
\begin{mathletters}
\begin{eqnarray}
\delta\Gamma_L^{ab}({\bf q}_\perp) & = & -\frac{1}{T}
\Big(\frac{\lambda^2}{2}+\frac{2\lambda\mu}{d_\perp}\Big)
\int_{\bf q'} q_z'^4\Big[
\big(G_L^{ab}({\bf q}')\big)^2 +
(d_\perp-1)\big(G_T^{ab}({\bf q}')\big)^2
\Big] +\nonumber\\
& - & \frac{\mu^2}{Td_\perp(d_\perp+2)}\;\int_{\bf q'}q_z'^4\;\Big[
6\big(G_L^{ab}({\bf q}')\big)^2 + (2d_\perp^2-2)
\big(G_T^{ab}({\bf q}')\big)^2
+  4(d_\perp-1) G_L^{ab}({\bf q}')G_T^{ab}({\bf q}')
\Big] \,,
\label{delGammaL}
\\
\delta\Gamma_T^{ab}({\bf q}_\perp) & = &
- \frac{\mu^2}{Td_\perp(d_\perp+2)}\;\int_{\bf q'}q_z'^4\;\Big[
2\big(G_L^{ab}({\bf q}')\big)^2 + (d_\perp^2-2)\big(G_T^{ab}({\bf q}')\big)^2
+ 2d_\perp G_L^{ab}({\bf q}')G_T^{ab}({\bf q}')
\Big] \,.
\label{delGammaT}
\end{eqnarray}
\end{mathletters}
\begin{multicols}{2}
Now, $\delta\Gamma_T^{ab}({\bf q}_\perp)$ is nothing but the
correction to the transverse part of the elastic tensor $B_T$,
{\em i.e.} $\delta\Gamma_T^{ab}({\bf q}_\perp)=\delta\mu$. On the other hand,
$\delta\Gamma_L^{ab}({\bf q}_\perp)$ is the correction to the
longitudinal part $B_L=\lambda +2\mu$. Hence:
\begin{eqnarray}
\delta\lambda & = & \delta(\lambda+2\mu) - 2\delta\mu \nonumber\\
& = & \delta\Gamma_L^{ab}({\bf q}_\perp) - 2
\delta\Gamma_T^{ab}({\bf q}_\perp)
\end{eqnarray}

Performing the above ${\bf q}$-integrations, we find that both
corrections $\delta\mu$ and $\delta\lambda$ behave like
\begin{eqnarray}
\delta\mu \sim \delta\lambda \sim
\mu^{\frac{3-d}{2}}\kappa^{\frac{d-7}{2}}{\Delta_t}\; L_z^{7-2d}
\label{del-mu}
\end{eqnarray}
where $L_z$ is the size of the system along the direction of the
flux lines (which comes in as a result of imposing an infrared cut-off
$1/L_z$ on $q_z$ integrations), which is the result
(\ref{new-del-mu}) quoted in the text.

In the RG analysis of Sec. \ref{RGanalysis}, the integrals in Eqs.
(\ref{delGammaL})-(\ref{delGammaT})
are evaluated within the momentum shell defined by Eqs. 
(\ref{shell1})-(\ref{shell2}), and this
leads to:
\begin{mathletters}
\begin{eqnarray}
\delta D = g D  F_D(x),
\\
\delta\mu = g \mu  F_\mu(x),
\end{eqnarray}
\end{mathletters}
where $x=\lambda/\mu$, and where the functions $F_D$ and $F_\mu$
are the functions defined in Eqs. (\ref{Flambda})-(\ref{Fmu}) of the text.

\subsection{Corrections to $\kappa$ and ${\Delta_t}$}
\label{App_vmag_PTkappa}

We now turn our attention to the calculation of the
perturbative corrections to the curvature modulus $\kappa$ and to the
disorder strength ${\Delta_t}$. These corrections will come from the
following terms of the connected average $\langle
H_{int,n}^2\rangle_{0>}^c$,
\end{multicols}
\begin{eqnarray}
\langle H_{int,n}^2\rangle_{0>}^c[\kappa,{\Delta_t}] &\equiv&
\sum_{a,b}\int d{\bf r}d{\bf r}' \Big\{
\mu^2\partial_z u_\alpha^{a<}({\bf r})
\partial_z u_\gamma^{b<}({\bf r}')
\big\langle \partial_\alpha u_\beta^{a>}({\bf r}) \partial_z
u_\beta^{a>}({\bf r})
\partial_\gamma u_\delta^{b>}({\bf r}')\partial_z
u_\delta^{b>}({\bf r}') \big\rangle_{0>}^c
\nonumber\\
&+& \mu^2 \partial_z
u_\alpha^{a<}({\bf r}) \partial_z u_\delta^{b<}({\bf r}')
\big\langle\partial_\alpha u_\beta^{a>}({\bf r}) \partial_z
u_\beta^{a>}({\bf r})
\partial_\gamma u_\delta^{b>}({\bf r}')\partial_z
u_\gamma^{b>}({\bf r}') \big\rangle_{0>}^c
\nonumber\\
&+& \mu^2 \partial_z u_\beta^{a<}({\bf
r}) \partial_z u_\gamma^{b<}({\bf r}')
\big\langle \partial_\alpha u_\beta^{a>}({\bf r}) \partial_z
u_\alpha^{a>}({\bf r})
\partial_\gamma u_\delta^{b>}({\bf r}')\partial_z
u_\delta^{b>}({\bf r}') \big\rangle_{0>}^c
\nonumber\\
&+& \mu^2\partial_z
u_\beta^{a<}({\bf r}) \partial_z u_\delta^{b<}({\bf r}')
\big\langle \partial_\alpha u_\beta^{a>}({\bf r}) \partial_z
u_\alpha^{a>}({\bf r})
\partial_\gamma u_\delta^{b>}({\bf r}')\partial_z
u_\gamma^{b>}({\bf r}') \big\rangle_{0>}^c
\nonumber\\
&+& \mu\lambda\partial_z
u_\alpha^{a<}({\bf r}) \partial_z u_\delta^{b<}({\bf r}')
\big\langle \partial_\alpha u_\beta^{a>}({\bf r}) \partial_z
u_\beta^{a>}({\bf r})
\partial_\gamma u_\gamma^{b>}({\bf r}')\partial_z
u_\delta^{b>}({\bf r}') \big\rangle_{0>}^c
\nonumber\\
&+& \mu\lambda\partial_z
u_\beta^{a<}({\bf r}) \partial_z u_\delta^{b<}({\bf r}')
\big\langle \partial_\alpha u_\beta^{a>}({\bf r})
\partial_z u_\alpha^{a>}({\bf r})
\partial_\gamma u_\gamma^{b>}({\bf r}')\partial_z
u_\delta^{b>}({\bf r}') \big\rangle_{0>}^c
\nonumber\\
&+&\lambda^2\partial_z
u_\beta^{a<}({\bf r}) \partial_z u_\delta^{b<}({\bf r}')
\big\langle \partial_\alpha u_\alpha^{a>}({\bf r})
\partial_z u_\beta^{a>}({\bf r})
\partial_\gamma u_\gamma^{b>}({\bf r}')
\partial_z u_\delta^{b>}({\bf r}') \big\rangle_{0>}^c \Big\}\,.
\label{kappa_Delta}
\end{eqnarray}
All the connected averages on the right hand side of the above
equation are of the general form
$\big\langle \partial_\alpha u_\beta^{a>}({\bf r}) \partial_z
u_\rho^{a>}({\bf r}) \partial_\gamma u_\delta^{b>}({\bf
r}')\partial_z u_\sigma^{b>}({\bf r}') \big\rangle_{0>}^c$.
Using Wick's theorem, we can write:
\begin{eqnarray}
\big\langle \partial_\alpha u_\beta^{a>}({\bf r})
\partial_z u_\rho^{a>}({\bf r})
\partial_\gamma u_\delta^{b>}({\bf r}') \partial_z
u_\sigma^{b>}({\bf r}') \big\rangle_{0>}^c & = &
\int_{{\bf q},{\bf q}'}^> \Big[
q_\alpha q_\gamma {q'}_z^2 G_{\beta\delta}^{ab}({\bf
q}) G_{\rho\sigma}^{ab}({\bf q}') +
q_\alpha q_z {q'}_z {q'}_\gamma G_{\beta\sigma}^{ab}({\bf q})
G_{\rho\delta}^{ab}({\bf q}')
\Big]\, \mbox{e}^{i({\bf q}+{\bf q}')\cdot({\bf r}-{\bf r}')} \,.
\nonumber
\end{eqnarray}
Integration over ${\bf r}$ and ${\bf r}'$ leads to the result
\begin{eqnarray}
\int d{\bf r}d{\bf r}'\partial_z u_\mu^{a<}({\bf r})
&{\partial_z u_\nu^{b<}({\bf r}')}&
\big\langle \partial_\alpha u_\beta^{a>}({\bf r}) \partial_z
u_\rho^{a>}({\bf r})
\partial_\gamma u_\delta^{b>}({\bf r}')\partial_z
u_\sigma^{b>}({\bf r}') \big\rangle_{0>}^c =
\int_{\bf q}^> q_z^2 u_\mu^{a<}({\bf q}) u_\nu^{b<}(-{\bf q})
\times
\nonumber\\
&\times&\int_{{\bf q}'}^> \Big[
{q}_{\alpha} {q}_\gamma' (q_z + {q}_z')^2
G_{\beta\delta}^{ab}({\bf q}')
G_{\rho\sigma}^{ab}(-{\bf q}-{\bf q}') +
{q}_{\alpha} {q}_z' (q_z + {q}_z')(q_\gamma+{q}_\gamma')
G_{\beta\sigma}^{ab}({\bf q}')G_{\rho\delta}^{ab}(-{\bf q}-{\bf q}')
\Big] \,.
\nonumber\\
\label{GenRule}
\end{eqnarray}
\begin{multicols}{2}
Using the above general result, we calculate the seven terms appearing
on the right hand side of equation (\ref{kappa_Delta}), and then
Taylor expand the result in $q_z$ near $q_z=0$.  Tedious but
straightforward calculations (which are best carried out using a
symbolic math processor such as {\sc Mathematica}) lead to the result:
\begin{eqnarray}
-\frac{1}{2T}\big\langle H_{int}^2 \rangle_{0>} & = &
\int_{\bf q}^> u_i^{a<}({\bf q})\delta
\Gamma_{ij}^{ab}({\bf q})u_j^{b<}(-{\bf q}) \,,
\end{eqnarray}
where $\delta\Gamma_{ij}^{ab}(q_z)$ is given by
\begin{eqnarray}
\delta\Gamma_{ij}^{ab}(q_z) = \delta\kappa q_z^4\delta_{ab} -
\delta\Big(\frac{{\Delta_t}}{T}\Big)q_z^2\delta_{ij} \,.
\end{eqnarray}
Again, it can be shown that evaluating the ${\bf q}'$ integrals in
Eq. (\ref{GenRule}) with an IR cut-off $1/L_z$ on $q_z$ integrations
leads to same conclusion as in the previous paragraph, namely that
$\delta\kappa$ and $\delta(\Delta_t/T)$ both diverge with the system
size in the fashion
\begin{equation}
\delta\kappa \sim \delta(\Delta_t/T) \sim L_z^{7-2d}\,.
\end{equation}
On the other hand, if the ${\bf q}'$ integrals on the {\em rhs} of
Eq. (\ref{GenRule}) are evaluated within the momentum shell of
Eqs. (\ref{shell1})-(\ref{shell2}), then it follows that:
\begin{mathletters}
\begin{eqnarray}
\delta\kappa & = & \kappa g F_\kappa(x) \,,
\label{pert-kappa}\\
\delta\Big(\frac{{\Delta_t}}{T}\Big) & = & {\Delta_t} g F_{\Delta_t}(x) \,,
\label{pert-Delta}
\end{eqnarray}
\end{mathletters}
where the functions $F_\kappa$ and $F_{\Delta_t}$ are the functions defined
in Eqs. (\ref{Fkappa})-(\ref{FDelta}) of the text.

\section{Rotational averages of projection operators}
\label{App_rotavg}

In this Appendix, for completeness we evaluate angular averages of
projection operators, necessary for the RG computations in the main
text and Appendix \ref{App_vmag_PT1}.  More explicitly, we want to express the
following $d_\perp$-dimensional integrals:
\begin{mathletters}
\begin{eqnarray}
I_{\alpha\beta}^L[f] & = & \int_{\bf q_\perp} f(q_\perp^2)\,P^L_{\alpha\beta}({\bf q}_\perp),
\\
I_{\alpha\beta}^T[f] & = & \int_{\bf q_\perp} f(q_\perp^2)\,P^T_{\alpha\beta}({\bf q}_\perp),
\\
I_{\alpha\beta\gamma\delta}^{LL}[f] & = & \int_{\bf q_\perp} f(q_\perp^2)
\, P^L_{\alpha\beta}({\bf q}_\perp)P^L_{\gamma\delta}({\bf q}_\perp),
\\
I_{\alpha\beta\gamma\delta}^{TT}[f] & = & \int_{\bf q_\perp} f(q_\perp^2)
\, P^T_{\alpha\beta}({\bf q}_\perp)P^L_{\gamma\delta}({\bf q}_\perp),
\\
I_{\alpha\beta\gamma\delta}^{LT}[f] & = & \int_{\bf q_\perp} f(q_\perp^2)
\, P^L_{\alpha\beta}({\bf q}_\perp)P^T_{\gamma\delta}({\bf q}_\perp),
\end{eqnarray}
\end{mathletters}
where $f(q_\perp^2)$ is an arbitrary function of $q_\perp^2$, in terms of
the spherically symmetric integral
\begin{equation}
I[f] = \int_{\bf q_\perp} f(q_\perp^2) \,.
\end{equation}
We start with the integrals $I_{\alpha\beta}^L[f]$ and
$I_{\alpha\beta}^T[f]$ which can easily be evaluated. We have (in what
follows, $\hat{q}_\alpha$ denotes the quantity $q_\alpha/q$):
\begin{eqnarray}
I^L_{\alpha\beta}[f] & = & \int_{{\bf q}_\perp} f(q_\perp^2) \hat{q}_\alpha {\hat q}_\beta, 
\nonumber\\
& = & \delta_{\alpha\beta}\int_{{\bf q}_\perp} f(q_\perp^2) \frac{q_\alpha^2}{q^2} ,
\end{eqnarray}
where no summation is implied on the index $\alpha$ in the second
line.  Using the rotational symmetry of $f$, we can write:
\begin{eqnarray}
I^L_{\alpha\beta}[f] & = & \frac{\delta_{\alpha\beta}}{d_\perp}
\int_{{\bf q}_\perp} f(q_\perp^2) \frac{1}{q_\perp^2}\,(q_1^2 + \cdots + q_{d_\perp}^2) \,,
\nonumber\\
& = & \frac{\delta_{\alpha\beta}}{d_\perp} \int_{{\bf q}_\perp} f(q_\perp^2) \,,
\end{eqnarray}
{\em  {\em i.e.}}
\begin{eqnarray}
I^L_{\alpha\beta}[f] = \frac{\delta_{\alpha\beta}}{d_\perp}\;I[f]\,.
\end{eqnarray}
Similarly, we have for $I^T_{\alpha\beta}$:
\begin{eqnarray}
I^T_{\alpha\beta}[f] & = & \int_{{\bf q}_\perp} f(q_\perp^2) 
\big[\delta_{\alpha\beta} - P^L_{\alpha\beta}({\bf q}_\perp)\big]\,,
\nonumber\\
& = & \delta_{\alpha\beta}(1-\frac{1}{d_\perp})\int_{{\bf q}_\perp} f(q_\perp^2) \,,
\end{eqnarray}
so
\begin{eqnarray}
I^T_{\alpha\beta}[f] = \frac{d_\perp-1}{d_\perp}\,\delta_{\alpha\beta}\;I[f] \,.
\end{eqnarray}

\medskip

We now turn our attention to the integral
$I^{LL}_{\alpha\beta,\gamma\delta}[f]$. We first consider the case
where the indices are all equal, {\em {\em i.e.}} we want to calculate the
integral:
\begin{eqnarray}
I^{LL}_{\alpha\alpha,\alpha\alpha}[f] & = & \int_{{\bf q}_\perp} {\hat q}_\alpha^4\; f(q_\perp^2) \,,
\nonumber\\
& = & \int_{{\bf q}_\perp} \frac{q_\alpha^4}{q_\perp^4}\; f(q_\perp^2) \,.
\label{integral-4alpha's}
\end{eqnarray}
Due to the rotational symmetry of the function $f$, this integral will
be the same for all possible values of the index $\alpha=1,\cdots,d_\perp$.  
The decomposition of the vector ${\bf q}_\perp$
in spherical coordinates in $d_\perp$ dimensions is given by
\begin{eqnarray}
\left\{
    \begin{array}{ll}
&{q_1} = 
q\,\sin\theta_{d_\perp-1}\sin\theta_{d_\perp-2}\cdots\sin\theta_2\cos\theta_1 
\\
&{q_2} = 
q\,\sin\theta_{d_\perp-1}\sin\theta_{d_\perp-2}\cdots\sin\theta_2\sin\theta_1 
\\
&{q_3} = 
q\,\sin\theta_{d_\perp-1}\sin\theta_{d_\perp-2}\cdots\sin\theta_3\cos\theta_2 
\\
&{q_4} = 
q\,\sin\theta_{d_\perp-1}\sin\theta_{d_\perp-2}\cdots\sin\theta_4\cos\theta_3 
\\
&{\cdots} \nonumber\\
&{q_{d_\perp-1}} = q\, \sin\theta_{d_\perp-1}\cos\theta_{d_\perp-2} \\
&{q_{d_\perp}} = q\,\cos\theta_{d_\perp-1}
    \end{array}
\right.
\end{eqnarray}
and hence we see that the most convenient choice of $\alpha$ for
calculating the integral (\ref{integral-4alpha's}) is $\alpha =
d_\perp$.  Taking $q_\alpha=q_{d_\perp}$ results in the following
expression
\begin{eqnarray}
I^{LL}_{\alpha\alpha,\alpha\alpha}[f] = \int\frac{d^{d_\perp}{\bf q}_\perp}{(2\pi)^{d_\perp}}\; 
\cos^4\theta_{d_\perp-1}\; f(q_\perp^2) \,.
\end{eqnarray}
Using the fact that
\end{multicols}
\begin{eqnarray}
d^{d_\perp}{\bf q}_\perp = q^{d_\perp-1}
\sin^{d_\perp-2}\theta_{d_\perp-1}\,d\theta_{d_\perp-1}\,\sin^{d_\perp-3}\theta_{d_\perp-2}\,d\theta_{d_\perp-2}\,\cdots
\sin\theta_2\,d\theta_2\,d\theta_1 \,,
\end{eqnarray}
(where $0 \le \theta_k <\pi$ for $k\neq 1$, and $0\le \theta_1\le 2\pi$), we obtain:
\begin{eqnarray}
I^{LL}_{\alpha\alpha,\alpha\alpha}[f] & = & 
\frac{1}{(2\pi)^{d_\perp}}\int q_\perp^{{d_\perp}-1}f(q_\perp^2)\; dq_\perp\,
\int_0^\pi 
\sin^{d_\perp-2}\theta_{d_\perp-1}\cos^4\theta_{d_\perp-1}\,d\theta_{d_\perp-1}
\cdots \int_0^\pi 
\sin\theta_2\,d\theta_2\int_0^{2\pi}d\theta_{1} \,,
\nonumber\\
& = & C_{d_\perp} \int_{{\bf q}_\perp} f(q_\perp^2)
\end{eqnarray}
\begin{multicols}{2}
where
\begin{eqnarray}
C_{d_\perp} = \frac{\int_0^\pi \sin^{d_\perp-2}\theta\,
\cos^4\theta\;d\theta}{\int_0^\pi \sin^{d_\perp-2}\theta\;d\theta} \,.
\end{eqnarray}
The integrals in the numerator and denominator of $C_{d_\perp}$ can be
evaluated analytically.  {\sc Mathematica} for example gives:
\begin{eqnarray}
C_{d_\perp} = \frac{\Gamma(d_\perp/2)}{\sqrt{\pi}}\Big[\,
&-&\frac{3\sqrt{\pi}}{\Gamma(d_\perp/2)} + 2^{d_\perp+1}\Big(\;
\frac{3\Gamma(\frac{3+d_\perp}{2})}{\Gamma(1+d_\perp)}
\nonumber\\
&-& \frac{4\Gamma(\frac{5+d_\perp}{2})}{\Gamma(2+d_\perp)}
+ \frac{2\Gamma(\frac{7+d_\perp}{2})}{\Gamma(3+d_\perp)}
\;\Big)\,
\Big] \,.
\end{eqnarray}
Expanding this last expression using the following results for the
Gamma function:
\begin{mathletters}
\begin{eqnarray}
\Gamma(1+z) & = & z\,\Gamma(z) \,,
\\
\sqrt{\pi}\Gamma(2z) & = & 2^{2z-1} \Gamma(z)\Gamma(z+\frac{1}{2}) \,,
\end{eqnarray}
\end{mathletters}
leads, after a few manipulations, to the very simple result:
\begin{eqnarray}
C_{d_\perp} = \frac{3}{d_\perp(d_\perp+2)}  \,.
\label{res-Cd}
\end{eqnarray}
Hence:
\begin{eqnarray}
I^{LL}_{\alpha\alpha,\alpha\alpha}[f] = \frac{3}{d_\perp(d_\perp+2)}\, I[f] \,.
\label{ILL-allequal}
\end{eqnarray}

\medskip

We now consider the case where the indices $\alpha$, $\beta$, $\gamma$
and $\delta$ in $I^{LL}_{\alpha\beta,\gamma\delta}$ are not { all}
equal. In that case, we can write:
\begin{eqnarray}
I^{LL}_{\alpha\beta,\gamma\delta} & = & \int_{{\bf q}_\perp} 
f(q_\perp^2)P^L_{\alpha\beta}({\bf q}_\perp)P^L_{\gamma\delta}({\bf q}_\perp) \,,
\nonumber\\
& = & \int_{{\bf q}_\perp} f(q_\perp^2) {\hat q}_{\alpha}{\hat q}_{\beta}{\hat q}_{\gamma}
{\hat q}_{\delta} \,,
\nonumber\\
& = & \delta_{\alpha\beta}\delta_{\gamma\delta}\int_{{\bf q}_\perp} f(q_\perp^2) 
{\hat q}_\alpha^2{\hat q}_\gamma^2 +
\delta_{\alpha\gamma}\delta_{\beta\delta}\int_{{\bf q}_\perp} f(q_\perp^2) {\hat 
q}_\alpha^2{\hat q}_\beta^2 
\nonumber\\
& + & \delta_{\alpha\delta}\delta_{\beta\gamma}\int_{{\bf q}_\perp} f(q_\perp^2) 
{\hat q}_\alpha^2{\hat q}_\beta^2 \,,
\label{ILL}
\end{eqnarray}
where the two remaining indices in each integral are distinct from
each other.  Let us evaluate one such integral. We have:
\begin{eqnarray}
\int_{{\bf q}_\perp} f(q^2) {\hat q}_\alpha^2{\hat q}_\beta^2 & = & \int_{{\bf q}_\perp} 
f(q_\perp^2) \frac{q_\alpha^2q_\beta^2}{q_\perp^4} \,,
\nonumber\\
& = & \frac{1}{d_\perp-1}\int_{{\bf q}_\perp} 
\frac{1}{q_\perp^4}\;f(q_\perp^2)\,q_\beta^2\sum_{\alpha(\neq\beta)}q_\alpha^2 \,,
\nonumber\\
& = & \frac{1}{d_\perp-1}\int_{{\bf q}_\perp} 
\frac{q_\beta^2}{q_\perp^4}\;f(q_\perp^2)\,(q_\perp^2-q_\beta^2) \,,
\nonumber\\
& = & \frac{1}{d_\perp-1}\big[\int_{{\bf q}_\perp} \frac{q_\beta^2}{q_\perp^2}f(q_\perp^2) -
\int_{{\bf q}_\perp}\frac{q_\beta^4}{q_\perp^4}f(q_\perp^2)\big].
\nonumber\\
\end{eqnarray}
But:
\begin{equation}
\int_{{\bf q}_\perp} f(q_\perp^2)\;\frac{q_\beta^2}{q_\perp^2} = \int_{{\bf q}_\perp} 
f(q_\perp^2)\;{\hat q}_\beta^2 =
\frac{1}{d_\perp} \int_{{\bf q}_\perp} f(q_\perp^2) \,,
\end{equation}
and
\begin{equation}
\int_{{\bf q}_\perp} f(q_\perp^2)\;\frac{q_\beta^4}{q_\perp^4} = \int_{{\bf q}_\perp} 
f(q_\perp^2)\;{\hat q}_\beta^4 =
C_{d_\perp}\int_{{\bf q}_\perp}f(q_\perp^2) \,,
\end{equation}
so we obtain:
\begin{eqnarray}
\int_{{\bf q}_\perp} f(q_\perp^2) {\hat q}_\alpha^2{\hat q}_\beta^2 & = &
\frac{1}{d_\perp-1}\big[\frac{1}{d_\perp} - C_{d_\perp}
\big]\;\int_{{\bf q}_\perp} f(q_\perp^2) \,,
\nonumber\\
& = & \frac{1-d_\perp C_{d_\perp}}{d_\perp(d_\perp-1)}\int_{{\bf q}_\perp} f(q_\perp^2) \,.
\end{eqnarray}
Using the result (\ref{res-Cd}) for $C_{d_\perp}$, we finally obtain:
\begin{eqnarray}
\int_{{\bf q}_\perp} f(q_\perp^2) {\hat q}_\alpha^2{\hat q}_\beta^2 =
\frac{1}{d_\perp(d_\perp+2)}\int_{{\bf q}_\perp} f(q_\perp^2) \,,
\end{eqnarray}
and hence, Eq. (\ref{ILL}) gives ($\alpha$, $\beta$, $\gamma$ and
$\delta$ not all equal):
\begin{equation}
I^{LL}_{\alpha\beta,\gamma\delta} = 
\frac{1}{d_\perp(d_\perp+2)}\big(\delta_{\alpha\beta}\delta_{\gamma\delta}+
\delta_{\alpha\gamma}\delta_{\beta\delta}+\delta_{\alpha\delta}\delta_{\beta\gamma}\big)\;I[f]\,. 
\label{rotLL}
\end{equation}
It is not difficult to see that equation (\ref{ILL-allequal}) is a
special case of this last equation.  Hence, the final result for
$I^{LL}_{\alpha\beta,\gamma\delta}[f]$ is just equation (\ref{rotLL}),
which describes all possible combinations for indices.

\medskip

Now, let us find $I^{LT}_{\alpha\beta,\gamma\delta}[f]$. We have:
\begin{eqnarray}
I^{LT}_{\alpha\beta,\gamma\delta}[f] & = & \int_{{\bf q}_\perp} 
f(q_\perp^2)\,P_{\alpha\beta}^L({\bf q}_\perp)
P_{\gamma\delta}^T({\bf q}_\perp) \,,
\nonumber\\
& = & \int_{{\bf q}_\perp} f(q_\perp^2)\,P_{\alpha\beta}^L({\bf q}_\perp)
\big[\,\delta_{\gamma\delta}-P_{\gamma\delta}^L({\bf q}_\perp)
\big] \,,
\nonumber\\
& = & \delta_{\gamma\delta}\int_{{\bf q}_\perp} 
f(q_\perp^2)\,P_{\alpha\beta}^L({\bf q}_\perp) - I^{LL}_{\alpha\beta,\gamma\delta} \,,
\nonumber\\
& = &
\frac{1}{d_\perp}\delta_{\alpha\beta}\delta_{\gamma\delta} I[f]
- \frac{1}{d_\perp(d_\perp+2)}\times
\nonumber\\
&\times&\big(\delta_{\alpha\beta}\delta_{\gamma\delta}+
\delta_{\alpha\gamma}\delta_{\beta\delta}
+\delta_{\alpha\delta}\delta_{\beta\gamma}\big)\,\Big\}\;I[f] \,,
\end{eqnarray}
hence
\begin{eqnarray}
I^{LT}_{\alpha\beta,\gamma\delta}[f] & = &
\Big\{\,\frac{d_\perp+1}{d_\perp(d_\perp+2)}
\delta_{\alpha\beta}\delta_{\gamma\delta} 
\nonumber\\
& - &\frac{1}{d_\perp(d_\perp+2)}\big(
\delta_{\alpha\gamma}\delta_{\beta\delta}
+\delta_{\alpha\delta}\delta_{\beta\gamma}\big)\,\Big\}\;I[f] \,.
\label{rotLT}
\end{eqnarray}
The last rotational average that we need to calculate is
$I^{TT}_{\alpha\beta,\gamma\delta}[f]$. We have:
\begin{eqnarray}
I^{TT}_{\alpha\beta,\gamma\delta}[f] & = & \int_{{\bf q}_\perp}
f(q_\perp^2)\,P_{\alpha\beta}^T({\bf q}_\perp)P_{\gamma\delta}^T({\bf q}_\perp),
\nonumber\\
& = & \int_{{\bf q}_\perp} \!\!
f(q_\perp^2)\big[\,\delta_{\alpha\beta}-P_{\alpha\beta}^L({\bf q}_\perp)\big]\,
\big[\,\delta_{\gamma\delta} - P_{\gamma\delta}^L({\bf q}_\perp)\big] ,
\nonumber\\
& = & \delta_{\gamma\delta}\int_{{\bf q}_\perp} f(q_\perp^2)\,\big[\,
\delta_{\alpha\beta}\delta_{\gamma\delta}
- \delta_{\alpha\beta}P_{\gamma\delta}^L({\bf q}_\perp)
\nonumber\\
& - & \delta_{\gamma\delta}P_{\alpha\beta}^L({\bf q}_\perp) +
\,P_{\alpha\beta}^L({\bf q}_\perp)P_{\gamma\delta}^L({\bf q}_\perp) \,\big] \,,
\nonumber\\
& = & \delta_{\alpha\beta}\delta_{\gamma\delta}\,I[f]
-\frac{2}{d_\perp}\;\delta_{\alpha\beta}\delta_{\gamma\delta}\,I[f]
\nonumber\\
& + &
\frac{1}{d_\perp(d_\perp+2)}
\big(\delta_{\alpha\beta}\delta_{\gamma\delta}+
\nonumber\\
&+&\delta_{\alpha\gamma}\delta_{\beta\delta}+\delta_{\alpha\delta}
\delta_{\beta\gamma}\big)\,I[f] \,,
\end{eqnarray}
hence
\begin{eqnarray}
I^{TT}_{\alpha\beta,\gamma\delta}[f] &=& \Big\{\,
\frac{d_\perp^2-3}{d_\perp(d_\perp+2)}\;
\delta_{\alpha\beta}\delta_{\gamma\delta} +
\nonumber\\
&+&\frac{1}{d_\perp(d_\perp+2)}\big(
\delta_{\alpha\gamma}\delta_{\beta\delta}
+\delta_{\alpha\delta}\delta_{\beta\gamma}\big)
\Big\}\,I[f] \,.
\label{rotTT}
\end{eqnarray}

\end{multicols}


\begin{references}


\bibitem{Early-exp} O. Fisher, A. Treyvand, R. Chevrel and
M. Sergent, Solid State Commun. {\bf 17}, 721 (1975);
R.N. Shelton, R.W. McCallum and H. Adrain, Phys. Lett. {\bf 56}A,
213 (1976); W. A. Fertig, D.C. Johnson, L.E.
DeLong, R.W. McCallum, M.B. Maple and B.T. Matthias,
Phys. Rev. Lett. {\bf 38}, 987 (1977).

\bibitem{Stassis-et-al} C. Stassis, M. Bullock, J. Zarestky,
P. Canfield and A.I. Goldman, Phys. Rev. B {\bf 55}, R8678 (1997).

\bibitem{PhysicsToday} P. Canfield et al. Phys. Today {\bf 51},
No. 10, 40 (1998).


\bibitem{experiments} U. Yaron et al., Nature {\bf 386}, 236 (1996).


\bibitem{Canfield-et-al} P.C. Canfield, S.L. Bud'ko and B.K. Cho,
Physica C {\bf 262}, 249 (1996).

\bibitem{exp1} J. Tallon {\em et al.}, IEEE Trans. Appl. Supercond. 
{\bf 9}, 1696 (1999).

\bibitem{exp2} S.S. Saxena {\em et al.}, Nature {\bf 406}, 587 (2000);
A. Huxley {\em et al.}, Phys. Rev. B {\bf 63}, 144519 (2001).

\bibitem{exp3} C. Pfleiderer {\em et al.}, Nature {\bf 412}, 58 (2001).

\bibitem{exp4} D. Aoki {\em et al.}, Nature {\bf 413}, 613 (2001).

\bibitem{Blount} E.I. Blount and C.M. Varma, Phys. Rev. Lett. {\bf
42}, 1079, 1979.

\bibitem{Tachiki1} M. Tachiki, H. Matsumoto and H. Umezawa,
Phys. Rev. B{\bf 20}, 1915 (1979).

\bibitem{Kuper} C.G. Kuper, M. Revzen and A. Ron,
Phys. Rev. Lett.{\bf 44}, 1545 (1980).

\bibitem{Tachiki2} M. Tachiki, H. Matsumoto, T. Koyama and
H. Umezawa, Solid State Comm.{\bf 34}, 19 (1980).

\bibitem{Greenside} H.S. Greenside. E.I. Blount and C.M. Varma,
Phys. Rev. Lett.{\bf 46}, 49 (1981).

\bibitem{Ng1} T.K. Ng and C.M. Varma, Phys. Rev. Lett.{\bf 78},
330 (1997).

\bibitem{Ng2} T.K. Ng and C.M. Varma, Phys. Rev. Lett.{\bf 78},
3745 (1997).

\bibitem{Kawano-et-al} H. Kawano-Furukawa, E. Habuta, T. Nagata,
M. Nagao, H. Yoshizawa, N. Furukawa, H. Takeya and K. Kadowaki, 
preprint cond-mat/0106273.



\bibitem{anisotropy} As discussed in detail in Sec.\ref{Exp_consq}, 
in real materials rotational invariance is explicitly broken by a
crystal field anisotropy. As we show there, this cuts off the physics
discussed here at a length scale that is long if the anisotropy is
weak, diverging with a vanishing crystal pinning field. This proviso
is not special to our theory and is a property of any crystalline
ferromagnet.  The fully rotationally invariant theory might be best
applicable to finely-powdered ferromagnetic-superconductor samples.

\bibitem{Larkin} A.I. Larkin and Yu.N. Ovchinnikov,
Sov. Phys. JETP {\bf 38}, 854 (1974); J. Low Temp. Phys. {\bf 34}, 409 (1979).

\bibitem{Brandt} E.H. Brandt,
J. Low Temp. Phys. {\bf 26}, 709,(1977); J. Low Temp. Phys. {\bf 26},
735 (1977); J. Low Temp. Phys. {\bf 28}, 263 (1977); J. Low
Temp. Phys. {\bf 28}, 291 (1977).

\bibitem{SaundersThesis} Karl Saunders, PhD Thesis (University of 
Oregon, 2001).

\bibitem{softsolids} A class of ``soft'' solids in which (as a result of
symmetry-enforced strict vanishing of a harmonic elastic constant)
nonlinear elastic terms are qualitatively important, leading to
anomalous universal elasticity includes 
smectic\cite{GP,Radzihovsky-Toner,JSRT_smectic}
and columnar liquid crystals\cite{RT_columnar,SaundersThesis}, a flat phase of
thermal and disordered polymerized membranes\cite{membranes}, and
nematic elastomers\cite{review_elastomers,LMRX,XRdisorderPRL}.


\bibitem{GP} G. Grinstein and R.A. Pelcovits, 
Phys. Rev. Lett. {\bf 47}, 856 (1981).

\bibitem{membranes} D.R. Nelson and L.Peliti, J. Phys. (France) {\bf 48}, 1085 (1987);
J.A. Aronovitz and T.C. Lubensky, Phys. Rev. Lett. {\bf 60}, 2634 (1988);
P. Le Doussal and L. Radzihovsky, {\em ibid.} {\bf 69}, 1209 (1992).


\bibitem{review_elastomers} X. Xing, R. Mukhopadhyay, T. C. Lubensky and
Radzihovsky, Phys. Rev. E 68, 021108 (2003).


\bibitem{LMRX} T.C. Lubensky, R. Mukhapadhyay, L. Radzihovsky and X. Xing,
Phys. Rev. E {\bf 66}, 011702 (2002).

\bibitem{XRdisorderPRL} X. Xing and L. Radzihovsky, Phys. Rev. Lett., 
{\bf 90}, 168301 (2003); Europhys. Lett. {\bf 61}, 769 (2003);
O. Stenull and T.C. Lubensky, {\em ibid.} {\em 61}, 776 (2003).


\bibitem{deGennes} P.G.~de Gennes, {\em Superconductivity of Metals
and Alloys}, Addison-Wesley, 1966.

\bibitem{Radzihovsky-Toner} L. Radzihovsky and J. Toner,
Phys. Rev. Lett. {\bf 78}, 4414 (1997); Phys. Rev. Lett. {\bf 79},
4214 (1997); Phys. Rev. B {\bf 60}, 206 (1999).

\bibitem{MSCprl} L. Radzihovsky, A.M. Ettouhami, K. Saunders and J. Toner,
Phys. Rev. Lett. {\bf 87}, 027001 (2001).

\bibitem{JSRT_smectic}  B.Jacobsen, K. Saunders, L. Radzihovsky and 
J. Toner, Phys. Rev. Lett. {\bf 83}, 1363 (1999).


\bibitem{JohnLeiming} L. Chen and J. Toner, preprint nlin.SI/0407066 (unpublished).

\bibitem{RT_columnar}  K. Saunders, L. Radzihovsky and J. Toner, 
Phys. Rev. Lett. {\bf 85}, 4309 (2000).

\bibitem{GLD} T. Giamarchi and P. Le Doussal, Phys. Rev. B {\bf 52}, 
1242 (1995).

\bibitem{FFLO} The Ginzburg-Landau functional, Eq. (\ref{FGL}), focuses
on a uniform singlet superconducting order
parameter, thereby from the start omitting the possibility of a
Fulde-Ferrell-Larkin-Ovchinikov state expected to be induced at
sufficiently large magnetization, corresponding to Zeeman energy on order
of the quasi-particle gap, so as to make it favorable to break up a
fraction of Cooper pairs into spin-polarized electrons.

\bibitem{Blatter-et-al} G. Blatter, M.V.~Feigel'man,
V.B. Geshkenbein, A.I. Larkin and V.M.~Vinokur, Rev. Mod. Phys. {\bf 66}, 
1125 (1994).

\bibitem{NattermanReview} T. Nattermann and S. Scheidl, Adv. Phys. 
{\bf 49}, 607 (2000).

\bibitem{Natterman} J. Villain and J.F. Fernandez, Z. Phys. B {\bf 
54}, 416 (1986);
T. Nattermann, Phys. Rev. Lett. {\bf 64}, 2454 (1990);
S.E. Korshunov, Phys. Rev. B, {\bf 48} 3969 (1993); S. Bogner, T. 
Emig and T. Nattermann,
Phys. Rev. B {\bf 63},174501 (2001).

\bibitem{Mezard} M. M\'ezard and G. Parisi, J. Phys. (France) I {\bf 
1}, 809 (1991).

\bibitem{Remark1} The reader should be warned that this
variational replica calculation is somewhat an academic one, in the
sense that it is valid only, strictly speaking, for $7/2<d<9/2$ where
nonlinear elasticity and tilt disorder can be ignored. Including these
will result in a significantly different and richer behavior for the
relative displacement correlator $\overline{\langle[{\bf u}({\bf
r})-{\bf u}({\bf 0})]^2\rangle}$, as we shall explore in the rest of
this paper.

\bibitem{commentVariational} The effects of positional pinning on lengths
scales longer than the Larkin length can be most carefully treated via
a functional renormalization group method\cite{DSFisher,GLD} that
takes into account multiple minima of the random potential.  However,
here, for simplicity, we employ a less rigorous variational
approach\cite{Mezard,GLD}, that in conventional vortex lattices leads
to results in agreement with predictions of the functional
renormalization group.\cite{GLD,Natterman} The only qualitatively
important result that we are after here is the logarithmic scaling of
phonon correlation functions for purely positional disorder treated
within harmonic elastic approximation. This is expected by general
symmetry arguments and is indeed reproduced by replica variational
approach that we take here.

\bibitem{Anderson} S.F. Edwards and P.W. Anderson, J. Phys. (France) 
{\bf 5}, 965 (1975).


\bibitem{commentTilt} One might worry that dominance by the tilt
disorder does not survive in the presence of both positional and tilt
disorder, and anharmonic elasticity. However, as was shown in a
closely related randomly pinned smectic liquid crystal
system\cite{Radzihovsky-Toner}, here too, tilt disorder continues to
dominate even if feedback effects of positional disorder are taken
into account.


\bibitem{commentStrain} As discussed in more detail in
Refs.\onlinecite{LMRX,XR} the elasticity of a d-dimensional isotropic
solid embedded in a d-dimensional isotropic environment can be
formulated equally well in terms of the (Lagrangian) left Cauchy-Green
tensor $v_{\alpha\beta}=\frac{1}{2}(\partial_i R_\alpha\partial_i
R_\beta-\delta_{\alpha\beta})$ or the (more common, Lagrangian) right
Cauchy-Green tensor $u_{ij}=\frac{1}{2}(\partial_i R_\alpha\partial_j
R_\alpha-\delta_{ij})$, with the former a scalar in the reference
space and a rank-2 tensor in the embedding space, and the latter a
rank-2 tensor in the reference space and a scalar in the embedding
space. However, phenomena in which an external field breaks spatial
symmetry of the embedding space (as the pinning disorder does in the
problem at hand), quite clearly can only be formulated in terms of the
{\em left} Cauchy-Green strain tensor. Furthermore, the elasticity of
solids with a dimension $d_\perp$ of the elastic displacements field
$u_\alpha$ smaller than the reference space dimension $d$ (as for
example in a line-vortex solid or a columnar liquid crystal phase,
with $d_\perp=2$ and $d=3$) is also most easily formulated in terms of
the left Cauchy-Green strain tensor $v_{\alpha\beta}$.


\bibitem{XR} X. Xing and L. Radzihovsky, Europhysics Letters {\bf 
61}, 769 (2003); and
unpublished.

\bibitem{commentReplica} We stress that here replica trick is used purely as a
technical diagrammatic device, which allows an efficient way of
excluding {\em annealed} disorder contributions, that can be
equally-well (but less efficiently) done without the replica trick.

\bibitem{linearH} It is easy to see that the first-order cumulant
$\langle H_{int}\rangle_0$ does not yield any corrections to the
parameters of the original Hamiltonian $H_{0n}$. These corrections
arise starting only at the second-order cumulant $-\langle
H_{int}^2\rangle_{0>}^c/2T$. For details see Appendix \ref{App_vmag_PT1}.

\bibitem{Wilson} K.G. Wilson and J. Kogut, Phys. Rep. {\bf 12}, 75 (1974).

\bibitem{comment_phi} The underlying rotational invariance of the spontaneous
solid ensures that graphical (perturbative) corrections leave the
nonlinear strain tensor $v_{\alpha\beta}$, Eq. (\ref{def-strain}), invariant,
independent of the arbitrary choice of the rescaling exponents, $\phi$
and $\omega$. It is a matter of convenience to have the arbitrary
rescaling to also preserve the form of the nonlinear strain tensor
$v_{\alpha\beta}$, by choosing $\phi=2-\omega$, Eq. (\ref{omega}).

\bibitem{Nelson_Rudnick} J. Rudnick and D.R. Nelson, Phys. Rev. B 
{\bf 13}, 2208 (1976).

\bibitem{DSFisher} D. S. Fisher, Phys. Rev. Lett. {\bf 56}, 1964 (1986).

\bibitem{GH} M.J.P. Gingras and D.A. Huse, Phys. Rev. B {\bf 53}, 15183 (1996).

\bibitem{KT} J. M. Kosterlitz and D. J. Thouless, J. Phys. C {\bf 6},
1181 (1973); V. L. Berezinskii, Zh. Eksp. Teor. Fiz. {\bf 59}, 907
(1970) [Sov. Phys. JETP {\bf 32}, 493 (1971)]; B. I. Halperin and
D. R. Nelson, Phys. Rev. Lett. {\bf 41}, 121 (1978); D. R. Nelson and
B. I. Halperin, Phys. Rev. B {\bf 19}, 2457 (1979); A. P. Young,
Phys. Rev. B {\bf 19}, 1855 (1979).

\bibitem{Rokhsar} P. E. Lammert, D. Rokhsar and J. Toner,
Phys. Rev. Lett.  {\bf 70}, 1650 (1993); Phys. Rev. E {\bf 52}, 1778
(1995).
 

\bibitem{Toner} J. Toner, Phys. Rev. B {\bf R26}, 462 (1982).

\bibitem{HLM} B.I. Halperin, T. Lubensky and S.K. Ma, Phys. Rev. Lett. {\bf 32}, 292 (1974);
J. Chen, T.C. Lubensky and D.R. Nelson, Phys. Rev. B {\bf 17}, 4274 (1978);
B.I. Halperin and T.C. Lubensky, Solid State Comm. {\bf 14}, 997 (1973); L. Radzihovsky,
Europhys. Lett. {\bf 29}, 227 (1995).


\bibitem{Note} More precisely,  $\sigma \sim H B(H)/4\pi = H(H_0 + H)/4\pi$, 
where $H_0$ is the coercive field.

\bibitem{SRT} K. Saunders, L. Radzihovsky and J. Toner, Phys. Rev. Lett. {\bf 87}, 27001 (2001).

\bibitem{EquilFootnote} Because, as is usually the case with glassy
systems, we expect equilibration into the orientationally-ordered
ground state to be slow, such equilibration will  therefore require cooling under field alignment.

\bibitem{TBKT} S. Tewari, D. Belitz, T.R. Kirkpatrick and J. Toner, 
Phys. Rev. Lett. {\bf 93}, 177002 (2004).


\bibitem{Abramowitz} {\em Handbook of Mathematical Functions},
M. Abramowitz and I.A.~Stegun (editors); Dover, 1965.




                                                                                       

\end{references}
\end{document}